\documentclass[a4paper,12pt,oneside]{article}
\usepackage[latin1]{inputenc}
\usepackage{amsmath}
\usepackage{amsfonts}
\usepackage{amssymb}
\usepackage{graphicx}
\usepackage{footnote}
\usepackage{float}
\usepackage{subcaption}
\DeclareGraphicsExtensions{.bmp,.png,.pdf,.jpg,.eps}
\usepackage{physics}
\usepackage{mathrsfs}

\usepackage{bigints}

\usepackage[colorlinks]{hyperref}
\hypersetup{
    colorlinks=true,
    linkcolor=blue,
    filecolor=magenta,  
    urlcolor=[rgb]{0.00,0.00,0.50},    
    citecolor=red,
    }

\usepackage{url}

\advance \textheight by \topskip
\usepackage{amsmath}
\usepackage[english]{babel}
\makeatletter
 \@addtoreset{equation}{section}
\makeatother
\usepackage[latin1]{inputenc}
\usepackage{amsmath}
\numberwithin{equation}{section}

\advance \textheight by \topskip
\usepackage{amsmath}
\usepackage[english]{babel}
\DeclareMathAlphabet\mathbfcal{OMS}{cmsy}{b}{n}
\DeclareMathAlphabet{\boldmathe}{T1}{cmr}{bx}{it}
\newcommand{\mbf}[1]{\boldmathe{#1}}

\def\vA{\mbf{A}}

\def\vG{\mbf{G}}

\def\ve{\mbf{e}}

\def\vr{\mbf{r}}

\def\be{\begin{equation}}
\def\ee{\end{equation}}

\def\R{\mathbb R}
\def\Q{\mathbb Q}

\def\be{\begin{equation}}
\def\ee{\end{equation}}

\def\R{\mathbb R}

\def\vA{\mbf{A}}

\def\vG{\mbf{G}}

\def\ve{\mbf{e}}

\def\vr{\mbf{r}}

\def\vv{\mbf{v}}

\def\be{\begin{equation}}
\def\ee{\end{equation}}

\def\Z{\mathbb Z}

\def\R{\mathbb R}

\usepackage{todonotes}

\begin{document}

\title{
{\bf 
Dynamics, symmetries, anomaly  and vortices
in a rotating cosmic  string background
}}

\author{{\bf  Luis Inzunza and Mikhail S. Plyushchay} 
 \\
[8pt]
{\small \textit{Departamento de F\'{\i}sica,
Universidad de Santiago de Chile, Casilla 307, Santiago,
Chile  }}\\
[4pt]
 \sl{\small{E-mails:   
\textcolor{blue}{luis.inzunza@usach.cl},
\textcolor{blue}{mikhail.plyushchay@usach.cl}
}}
}
\date{}
\maketitle

\begin{abstract}

Non-relativistic  
conformally invariant systems  in a rotating cosmic string (conical) spacetime
 are analyzed at the classical and quantum levels
 by means of 
the  gravitoelectromagnetic 
interpretation of the background.
Solutions of the  
equations of motion are 
found
by 
employing a local canonical transformation, 
that leads to
their  natural 
interpretation 
 in terms of  Riemann surfaces. 
The 
cone parameter  $\alpha$ and the angular velocity $\Omega$ 
of the background
 determine
 the existence of
hidden 
symmetries.
Globally defined 
higher order
integrals associated with perihelion of 
geodesic orbits appear 
at rational values of
 $\alpha$.
For the harmonic oscillator system with  frequency $\omega$,  the integrals 
responsible for the trajectory closure arise  only 
for rational values  of
$\alpha$ and $|\gamma|=|\Omega/\omega|$, with $|\gamma|=1$ 
corresponding to  the Landau problem. 
We face
a  quantum anomaly problem  since
the 
hidden symmetry operators 
can only be constructed
when 
 $\alpha$ is 
 integer.
 Such
 operators 
   are non-local in the case of the free particle.
  For the
harmonic oscillator, 
 the symmetry
 generators are obtained 
 with the help of
 the conformal bridge transformation.  
 We also study a multi-particle
  version of the harmonic oscillator system with $|\gamma|=1$ using the mean-field theory
  and find  
  that the emerging vortex  
  structure
   respects
    a singular point 
  of
  the background.

\vskip.5cm\noindent
\end{abstract}

\section{Introduction}
In physics, a quantum anomaly occurs when classical symmetries fail at the quantum level. 
The well known examples of this  are the so-called  Adler-Bell-Jackiw anomaly in
electrodynamics 
\cite{Adler,BellJackiw}, where the conservation of axial vector current is violated in quantum theory,
and the 
gauge anomalies, where quantum corrections do not preserve classical gauge invariance.
\cite{anomaly}.  
In this article, we face a type of quantum anomaly
similar to what was observed in  \cite{anomaly2}, 
which is more related to the structure of 
the spacetime background than quantum corrections, 
and where the conserved operators commute with the
Hamiltonian, but their action can produce non-physical eigenstates.

The phenomenon 
of a quantum anomaly produced by geometric constraints is not too strange since it is natural 
to expect that spacetime has 
to influence on  
the dynamics of the systems that live on it. 
The very well known  
examples where the geometry 
effects 
essentially  
on
 the symmetry properties
 of
the particle systems are
provided by   the integrable dynamics of  a test 
 particle in the Kerr black hole due to 
the Carter integral \cite{Carter}, and  
the enhanced supersymmetry that characterizes the motion  of a spinning particle 
in a Kerr-Newman black hole \cite{GRH}. 
In these two cases we are dealing  
with hidden symmetries and 
the so-called Killing and Killing-Yano tensors associated with them 
\cite{Cariglia2,Frolov,Frolov2,CarFKK,CarigliaRev,Frolov3}.
Other effects related with geometry are   
the Lense-Thirring effect produced by gravitoelectromagnetic fields \cite{Bardeen,Volovik} 
and the production of additional  particles in quantum field theories in curved spaces \cite{curvedQFT1,curvedQFT2}. 

The specific issue of the ``geometrical quantum anomaly" that we refer above 
was observed by us earlier when we analyzed the problem of 
 canonical quantization of non-relativistic dynamics in the cosmic string spacetime 
 \cite{InzPly7}, which
is  characterized by a geometrical
parameter $\alpha$  that encodes the information on the conical singularity 
of the background
\cite{Kibble,Vilen,Vilenkin,Dow}. Various 
authors 
explored this topic in the past \cite{tHooft,DesJack,Kay,Furtado,Coelho}, but their research avoids going deeper 
into quantum symmetries, since they claim that  there are difficulties in defining self-adjoint 
operators correctly. In our case, we were able to construct the 
well defined hidden symmetry generators
and gave a non-linear algebraic description of conformally invariant quantum systems \cite{AFF} living in this 
spacetime only for natural values of $\alpha$. This goes in contrast with the classical level 
(and for this reason we have 
a quantum anomaly), where the well defined on phase space 
hidden integrals of motion can be constructed for any rational value of $\alpha$. 

In our study, the so-called conformal bridge transformation \cite{InzPlyWipf1,PTrev} played a crucial role. 
This relatively new technique relates asymptotically free conformally invariant models 
with their harmonically confined versions. 
Therefore, if one system has an anomaly issue, 
then the other also faces 
 this problem \cite{InzPly7}. 
On the other hand, the (super)conformally 
invariant systems  are the simplest integrable models that 
can be studied in  
different geometric backgrounds, and which 
 are essential with their own right 
due to appearance  in various applications in physics. For  some recent 
interesting results see 
 \cite{Conformal1,Conformal2,Conformal3,Conformal4} 
 and references therein.

One may expect that a way to cure the geometric anomaly
 in our case should  again be related to geometry.
With this in mind, we  address here  the idea to include a homogeneous rotation 
of angular velocity 
$\Omega$ 
about the cone's symmetry axis 
 in the corresponding 
spacetime metric as a possible solution to the anomaly issue. 
As is well known, such a rotation  
 induces gravitoelectromagnetic fields  \cite{Bardeen,Volovik}.
Historically,  the importance 
of studying rotating reference frames  comes from the simple fact that the Earth's 
rotation strongly influences the experiments carried out on it.
In quantum case,  the rotation  results, in particular,  in appearance of the  
Sagnac effect \cite{AnaSu}, the impossibility of forming rigid flows \cite{GulRo},
 and modifications 
in the Hall effect \cite{Hall1,Hall2}. 
Since rotation also introduces a kind of singularity at the point in which 
its axis intersects the  two-dimensional surface, it is  somewhat natural to expect that its 
inclusion,  at least, should 
provide  some new insight in the context of the quantum anomaly problem.
In addition, by switching on of  a rotationally-invariant harmonic trap with its  
independent frequency parameter, one could also hope to dynamically solve the anomaly problem.

Although the formulated idea has not worked as it was expected,
we have obtained several interesting results 
that apply to both single-particle and many-particle systems in conical backgrounds, 
and this article aims to report them.

Let us  describe the organization of the article and 
briefly summarize the results. 
Sec. \ref{SecFreeRotatingD} investigates the geodesic motion in a homogeneously rotating conical 
background, which we interpret as a cosmic string
with a gravitoelectromagnetic field. 
We describe there a particular type of an integral of motion which  plays a role similar to that of  the  
Laplace-Runge-Lentz vector in the Kepler problem \cite{CarigliaRev,Pauli}. 
At the classical level, this quantity 
is a well defined phase space function for arbitrary values of $\Omega$
but only when  $\alpha$ is rational.
It has a well defined non-local quantum analog only for integer values of  $\alpha$. 
In Sec. \ref{SecERIHOClasic}, we supply the system with a harmonic trap potential with frequency $\omega$ 
and study its classical dynamics. As a result, we obtain 
an analog of the Euclidean exotic rotationally invariant
 harmonic oscillator (ERIHO) system \cite{ERIHO}, but now in the cosmic string background. 
 This system has closed orbits and  integrals of motion associated with
 a hidden symmetry 
 only when both parameters  $|\gamma|=|\Omega/\omega|$ and $\alpha$
  take rational values. 
 In particular  case $|\gamma|=1$, the  system 
 is equivalent to the Landau problem in a  space with a topological defect
 \cite{DeAMarques,Muniz}. 
 In Sec. \ref{SecERIHOQuantum}, we consider the quantum version
  of the system using the conformal bridge transformation 
 \cite{InzPlyWipf1,PTrev}.  
 There we show that the quantum analogs of the hidden symmetry 
 generators are well defined operators that correctly reflect the 
 degeneracy of energy levels 
only when $\alpha$ takes integer values.  
 In Sec. \ref{Vessel}, we consider  the mean-field theory approximation to 
 study the Bose-Einstein condensation of the harmonically trapped systems in a
  conical vessel in their Landau problem phases corresponding 
  to $\gamma=\pm 1$.
  Similar problem for the Euclidean case was considered in \cite{Cooper,ALF}. 
 As a result, we show that the appearing 
 vortex structure respects a  singular point of the 
 rotating cosmic string background.  
 In Sec. \ref{SecDisOut}, 
 the  discussion of the obtained results and outlook are presented.
 In Appendix \ref{AppCantime} we show 
 how the nontrivial integrals found in Sec. \ref{SecFreeRotatingD} 
 can be obtained from  
 the non-relativistic  two-dimensional free particle in the inertial reference frame
  ($\Omega=0$).

\section{Dynamics in rigidly rotating spacetimes}
\label{SecFreeRotatingD}
The weak field approximation
of general relativity is characterized by
appearance of electric- and magnetic-like 
fields. 
 They are  usually called the gravitoelectric and gravitomagnetic fields
because  their action on 
 massive test particles  
is  similar 
to the action of external electric and magnetic  fields on charged particles. 
An indirect  verification 
of gravitoelectromagnetism is provided by 
 relativistic jets \cite{Penrose1,Williams} appearing 
in the study of rotating black holes, and in  the observed 
 excess  of energy and luminosity produced by quasars and active galactic nuclei, 
 which are explained by the  Lense-Thirring effect \cite{Bardeen,Volovik}.
 Recently, gravitomagnetism has also been used to explain 
 the rotation curves of galaxies as an alternative to dark matter
 \cite{Ludwig}.

The dynamics of test particles in rigidly rotating spacetime backgrounds 
 is  essentially affected by the appearance of 
  gravitoelectromagnetic fields
   at the classical and quantum levels.
 For special values of the parameters of 
 such and similar  systems,
 classical dynamics of test particles can be completely integrable
 due to appearance of hidden symmetries, 
 which also 
 reveal themselves in peculiar properties of the
 corresponding quantum systems \cite{Carter,GRH,CarFKK,CarigliaRev,Frolov3,InzPly7,ERIHO}.
This section  aims to investigate such effects for geodesic motion 
 in a  rotating conical background.

 \subsection{Classical picture}
 
The simplest case in which  gravitoelectromagnetic  
effects can be observed corresponds to a 
$(3+1)$-dimensional Minkowski vacuum subjected to a uniform rotation.
The metric of this spacetime with the angular velocity $\Omega$  oriented along 
the $z$  axis  is given by
\be
\label{rotating metric}
ds^2=-c^2dt^2+ (d\rho^2+\rho^2(d\varphi+\Omega dt)^2+dz^2)\,.
\ee
Lagrangian of a non-relativistic test particle that experiences geodesic motion  
in this background
is obtained from the relativistic action 
\begin{eqnarray}
\label{Rlimit}
&-mc\int \sqrt{-ds^2}=-mc^2\bigintsss\sqrt{1-\frac{1}{c^2}
\left(\left(\frac{d\rho}{dt}\right)^2+\rho^2\left(\frac{d\varphi}{dt}+\Omega\right)^2+
\left(\frac{dz}{dt}\right)^2\right)}\,\,dt&
\end{eqnarray}
by taking the limit $c\rightarrow \infty$. It 
is given by 
\begin{eqnarray}
&
L=L_{\Omega}+\frac{m}{2}\dot{z}^{2}\,,\qquad 
L_{\Omega}=\frac{m}{2}(\dot{\rho}^2+\rho^2(\dot{\varphi}+\Omega)^2)\,,
&
\end{eqnarray}
where the dot denotes 
time derivative. 
This is a Lagrangian 
of a free particle in a rotating reference frame in cylindrical coordinates. 
It can be obtained in an alternative way from the non-relativistic free particle  system
in inertial reference frame 
by means of a time dependent canonical transformation, see Appendix \ref{AppCantime}.
Written in Cartesian coordinates  
$x^1=\rho\cos \varphi $ and  $x^2=\rho\sin \varphi $, 
 Lagrangian $L_{\Omega}$ takes the form 
\begin{eqnarray}
&
\label{flatLagrangian}
L_{\Omega}=\frac{m}{2}\dot{x}^i\dot{x}^i+\frac{m}{2}\Omega^2 x^ix^i
+\frac{q_{{}_{GM}}}{c}\dot{x}^iA^i\,,\qquad
A^i=\frac{\Omega c}{2}\epsilon^{ij}x^j\,.
&
\end{eqnarray}
Here, $A^i$ can be interpreted as a two-dimensional vector potential of a 
gravitomagnetic field applied to  a particle with gravitomagnetic charge $q_{{}_{GM}}=-2m$. 
As a result,  two-dimensional Lagrangian (\ref{flatLagrangian}) 
corresponds to that of  the Landau problem  in symmetric gauge with magnetic field
$B_{G}=\epsilon_{ij}\partial_i A_j=-\Omega c$,  supplemented with 
an isotropic quadratic  centrifugal  potential $-\frac{1}{2}m\Omega^2 x^ix^i$.
The latter can be reinterpreted as
the external gravitoelectric potential applied to a particle with gravitoelectric 
charge $q_{{}_{GE}}=-m$ \footnote{
The ratio of the gravitomagnetic charge to the gravitoelectric charge  
equals two because, unlike electromagnetism,   gravity is a spin-2 field \cite{Mashhoon}.}.
The  canonical   Hamiltonian for the system  (\ref{flatLagrangian})  is
\begin{eqnarray}
&
\label{FreeClasicH}
H_{\Omega}=\frac{1}{2m} p_ip_i-\Omega p_\varphi\,,\qquad
p_\varphi=x_1p_2-x_2p_1\,.
&
\end{eqnarray}
The peculiarity of the system (\ref{flatLagrangian}) 
is that the additional repulsive centrifugal
potential term exactly balances  the confining effect of the 
magnetic field as is seen  from 
 the solution of  the classical equations of motion,
\begin{eqnarray}
&x_+(t)=x_1+ix_2=\rho e^{i\varphi}=e^{-i\Omega t}\left(Bt+C\right)\,,&
\end{eqnarray}
where $B$ and $C$ are complex constants of the dimensions of $\dot{x}$ and $x$,
respectively, 
and the bracket term corresponds to a free dynamics of a particle
in a plane.

A conical singularity can be introduced  by reducing
metric (\ref{rotating metric}) to 
a conical surface
$z(\rho)=\rho \cot{\beta}$, that results in  the (2+1)-dimensional spacetime 
with  a  metric 
\begin{eqnarray}
\label{ConeGeo1}
&
ds^2=-c^2dt^2+\alpha^2d\rho^2+\rho^2(d\varphi+\alpha\Omega dt)^2\,,&\\&
0<\rho<\infty\,,\qquad 0\leq \varphi\leq2\pi\,,\qquad \alpha^2=1+\cot^2{\beta}>1\,.
\label{ConeGeo2}&
\end{eqnarray}
To simplify future calculations,  we have rescaled $\Omega\rightarrow \alpha\Omega$.
When $\Omega=0$, this metric can be associated with  a $(2+1)$-dimensional 
cosmic string spacetime with the identification $\alpha=(1-4\mu c^{-2} G )^{-1}$, 
where $G$ is the Newton constant and  $\mu$ is the linear mass density 
of the string \cite{InzPly7}. If we allow the parameter $\alpha$
to take  
values in the interval  $(0,\,1)$, metric (\ref{ConeGeo1}) 
 can be associated, on the one hand,  with  a geometric background 
describing radial dislocations in superfluids \cite{Volovik}, and, on the other hand, 
with metric of an  anti-gravitating   cosmic string
 with negative mass density 
 (which corresponds to a wormhole's background) \cite{Volovik,CFMVBL}.
 
Note here that in the case of   $\Omega\not=0$, 
 $\alpha>1$
one can consider 
  metric 
(\ref{ConeGeo1}),  (\ref{ConeGeo2})
as that corresponding  
to a rotating conical  vessel of an infinitesimal thickness. 
This  interpretation will be employed in Section  \ref{Vessel}.

In the spacetime (\ref{ConeGeo1}), the non-relativistic geodesic  
Lagrangian 
is given by 
\begin{eqnarray}
&
\label{roteconela}
L^{(\alpha)}_\Omega=\frac{m}{2}\left(\alpha^2\dot{\rho}^2+\rho^2(\dot{\varphi}+\alpha\Omega)^2\right)\,.
&
\end{eqnarray}
It is obtained in the same way as 
for the flat case by substituting $ds^2$ in (\ref{Rlimit}) for the metric
(\ref{ConeGeo1}), (\ref{ConeGeo2})
\footnote{One can also start from the spacetime  metric 
 of the spinning cosmic string background, but the non-relativistic 
limit has to be taken in a special way. We return to this point in the last section.}. 
In terms of the conical metric in Cartesian coordinates, 
\begin{eqnarray}
&
\label{conicalmetricCar}
g_{ij}=\delta_{ij} +(\alpha^2-1)\frac{1}{(x^1)^2+(x^2)^2}\left(\begin{array}{cc}
(x^1)^2 & x^1x^2\\
x^1x^2 & (x^2)^2
\end{array}\right)\,,
&
\end{eqnarray}
 Lagrangian (\ref{roteconela}) takes the form  
 \begin{eqnarray}
 &
 \label{cartesianLag}
 L_\Omega^{(\alpha)}=
\frac{m}{2}g_{ij}\dot{x}^i\dot{x}^j+\frac{m}{2}\Omega^2 g_{ij}x^i x^j+
\frac{q_{{}_{GM}}}{c}g_{ij}\dot{x}^iA^{(\alpha)j}\,,\qquad
A^{(\alpha)i}=\frac{\alpha \Omega  c}{2}(x^2,-x^1)\,. 
&
 \end{eqnarray}
For $\alpha>1$, $A^{(\alpha)i}$ can be interpreted as a vector potential 
corresponding to the Landau problem  in the symmetric gauge 
with  magnetic field 
$B^{(\alpha)}=-\alpha B_G= 
\alpha \Omega c$, which 
 is perpendicular  to the surface of the cone. 
 
Classical Hamiltonian of the system (\ref{roteconela}) is
\begin{eqnarray}
\label{Homegaalpha}&
H_\Omega^{(\alpha)}=
\frac{1}{2m}\left(\frac{1}{\alpha^2}p_\rho^2+\frac{1}{\rho^2}p_\varphi^2\right)
-\alpha \Omega p_\varphi\,,&
\end{eqnarray}
from where one 
sees  that the repulsive quadratic potential appearing in (\ref{cartesianLag})
again 
compensates the confinement effect of the magnetic field.
 It is clear that the system is rotationally invariant, $p_\varphi=const$,  and so, 
 has the integral of motion 
\begin{eqnarray}\label{H0alpha}
&
H_0^{(\alpha)}= \frac{1}{2m}\left(\frac{1}{\alpha^2}p_\rho^2+\frac{1}{\rho^2}p_\varphi^2\right).
&
\end{eqnarray}
The latter
 is the Hamiltonian of a free particle in a conical geometry in
 the
 inertial reference frame.

In the case of arbitrary values of 
the  
 parameter $\alpha$,  there are also formal 
dynamical integrals of motion~\footnote{Such explicitly depending on time
integrals 
satisfy the evolution equation $\frac{dA}{dt}=\{A,H_\Omega^{(\alpha)}\}+
\frac{\partial A}{\partial t}=0$.}      
 \begin{eqnarray}\label{P+-X+-}
 &P_\pm^{(\alpha)}=e^{\pm i\alpha\Omega t}\Pi_\pm^{(\alpha)}\,,\qquad
G_\pm^{(\alpha)}= me^{\pm i\alpha\Omega t}\,  \Xi_\pm^{(\alpha)}\,,&
 \end{eqnarray}
 where 
\begin{eqnarray}
&\label{ConeGenerators}
 \Pi_\pm^{(\alpha)}=
\left(\frac{p_\rho}{\alpha} \pm i
\frac{p_\varphi}{\rho}
 \right)e^{\pm i\frac{\varphi}{\alpha}}\,,\qquad
\Xi_\pm^{(\alpha)}= \alpha \rho e^{\pm i\frac{\varphi}{\alpha}}-\frac{t}{m}\Pi_\pm^{(\alpha)}&
\end{eqnarray} 
 are the 
 analogues of the canonical momenta and the Galilean boost generators 
  in the 
 conical geometry \cite{InzPly7}.
 They are globally well defined phase space  functions  in the flat case
$\alpha=1$, see Appendix \ref{AppCantime}, as well as for integer values of 
$\alpha$ greater than one.
Integrals (\ref{P+-X+-}) 
generate the two-dimensional Heisenberg algebra,
 $\{G_\pm^{(\alpha)},P_\mp^{(\alpha)}\}=2 m$, 
 $\{G_\pm^{(\alpha)},P_\pm^{(\alpha)}\}=0$, 
 $\{G_+^{(\alpha)},G_-^{(\alpha)}\}=\{P_+^{(\alpha)},P_-^{(\alpha)}\}=0$.
 From them, one can  construct 
 the well defined in the  phase space, rotationally invariant
 generators of dilatations, $ D^{(\alpha)} $,  and  
 special conformal 
transformations, $ K^{(\alpha)} $, 
\begin{eqnarray}\label{Halpha0}
&
D^{(\alpha)}=\frac{1}{4 
m}(G_+^{(\alpha)}P_-^{(\alpha)}+ G_-^{(\alpha)}P_+^{(\alpha)})=\frac{1}{2}\rho p_\rho-H^{(\alpha)}_0t\,,&\\ 
&K^{(\alpha)}=\frac{1}{2 m}G_+^{(\alpha)}G_-^{(\alpha)}=\frac{1}{2}m\alpha^2 \rho^2-2D^{(\alpha)}t-H^{(\alpha)}_0t^2\,.&
\label{Halpha1}
\end{eqnarray}
Quadratic in $P_\pm^{(\alpha)}$ and 
$G_\pm^{(\alpha)}$ dynamical integrals $D^{(\alpha)}$
and $ K^{(\alpha)} $
 together with the Hamiltonian $H_\Omega^{(\alpha)}=\frac{1}{2m}
 P_+^{(\alpha)}P_-^{(\alpha)}-\alpha \Omega p_\varphi$ 
satisfy  the centrally extended $\mathfrak{sl}(2,\R)$ conformal algebra 
\begin{eqnarray}
&
\label{sl2RZ}
\{D^{(\alpha)},H^{(\alpha)}_\Omega\}=H^{(\alpha)}_\Omega
+Z^{(\alpha)}\,,\quad
\{D^{(\alpha)},K^{(\alpha)}\}=-K^{(\alpha)}\,,
\quad
\{K^{(\alpha)},H^{(\alpha)}_\Omega\}=2D^{(\alpha)}\,,\quad
&
\end{eqnarray}
in which $Z^{(\alpha)}=\alpha\Omega p_\varphi=i\frac{\Omega}{2m}
(G_+^{(\alpha)}P_-^{(\alpha)}-G_-^{(\alpha)}P_+^{(\alpha)})$ plays a role of the central charge.
The Casimir of this algebra is 
\begin{eqnarray}&\label{Casimir0}
\mathcal{C} 
=(D^{(\alpha)})^2-(H_\Omega^{(\alpha)} +Z^{(\alpha)})
K^{(\alpha)}+\frac{1}{4\Omega^2}(Z^{(\alpha)})^2=0\,.&
\end{eqnarray}
The central extension here is rather formal
since the central charge can be absorbed
 by changing the generator $H^{(\alpha)}_\Omega$ for
$H^{(\alpha)}_0=H^{(\alpha)}_\Omega+Z^{(\alpha)}=\frac{1}{2m}P_+^{(\alpha)}P_-^{(\alpha)}$, 
that transforms 
(\ref{sl2RZ}) into 
$\mathfrak{gl}(2,\R)\cong \mathfrak{sl}(2,\R)\oplus \mathfrak{u}(1)$ 
algebra, where 
$\mathfrak{sl}(2,\R)$ is the usual conformal
algebra generated by $H^{(\alpha)}_0$, 
$D^{(\alpha)}$ and $K^{(\alpha)}$, and $ \mathfrak{u}(1)$ 
is generated by $p_\varphi$.
The value of the $\mathfrak{sl}(2,\R)$ Casimir
$C=(D^{(\alpha)})^2-H_0^{(\alpha)} 
K^{(\alpha)}$
 reduces in this case to
$-\frac{\alpha^2}{4}p_\varphi^2$.

From 
the dynamical integral 
(\ref{Halpha1}) 
and Casimir (\ref{Casimir0})
one gets  that $\rho^2$ as a function of time is given by the 
quadratic polynomial 
\begin{eqnarray}
\label{rho2}
\rho^2(t)=a(t-t_*)^2+b^2
\end{eqnarray}
with $a=2H^{(\alpha)}_0/(m\alpha^2)$, 
$b^2=p_\varphi^2/(2mH^{(\alpha)}_0)$
and 
\be\label{t*}
t_*=-D^{(\alpha)}/H^{(\alpha)}_0\,.
\ee
 One finds that 
at  the moment of time
$t= t_*$,  the variable  $\rho$ takes the minimum value, 
\begin{eqnarray}
&
\rho(t_*)=b=(2m H_{0}^{(\alpha)})^{-\frac{1}{2}}\vert p_\varphi\vert 
:=\rho_*\,,
&
\end{eqnarray}
that corresponds to the `perihelion'  of the trajectory.

To find the angle evolution and 
identify the  trajectories of the particle,
we consider the transformation
\begin{eqnarray}
&\label{canonalpha}
\rho\rightarrow \alpha\rho\,,\qquad
\varphi\rightarrow \alpha^{-1}\varphi\,.
&
\end{eqnarray}
Under this rescaling of the variables, 
Lagrangian (\ref{roteconela}) is transformed  into $L_{\Omega}$ in 
(\ref{flatLagrangian}), and therefore, 
\emph{locally} the geodesic motion in metric (\ref{ConeGeo1}) is given by 
$x_+(t)=x^1(t)+ix^2(t)=\rho (t)e^{i\varphi(t)}$, where now
\begin{eqnarray}
&\label{tajalpha}
\rho (t)=\alpha^{-1} | B t+C|\,,\qquad
e^{i\varphi(t)}=e^{-i\alpha\Omega t} |B t+C|^{-\alpha}(B t+C)^{\alpha}\,,
&
\end{eqnarray}
and so, 
\begin{eqnarray}
&\label{zalpha}
x_+(t)=
\alpha^{-1} | w|^{1-\alpha}(w(t))^{\alpha}\,,\qquad w(t)=e^{-i\Omega t}(B t+C)\,.
&
\end{eqnarray}
Parametrizing  complex integration constants as 
$
B=V e^{i\vartheta_1}$ and $
C=R e^{i\vartheta_2}
$, and employing the corresponding Hamiltonian equations of motion of the system,
 one finds that the integral $H_0^{(\alpha)}$, 
 the angular momentum $p_\varphi$, the perihelion of the orbit $\rho_*$
 and 
 the corresponding moment of time $t_*$ 
 reduce to
\begin{eqnarray}
&
H_0^{(\alpha)}=\frac{m}{2}V^{2}\,,\qquad 
p_\varphi= \alpha^{-1}mV R  \sin(\vartheta_1-\vartheta_2)\,,&\\&
\rho_*=\alpha^{-1}R |\sin(\vartheta_1-\vartheta_2)|\,,\qquad 
t_*=-RV^{-1}\cos(\vartheta_1-\vartheta_2)\,,
\end{eqnarray}
while
 the total mechanical energy 
is expressed as
$H_\Omega^{(\alpha)}=\frac{m}{2}V^{2}(1-2\Omega R V^{-1}\sin(\vartheta_1-\vartheta_2))\,.$
From these relations it 
follows
 that when the constants of integration are chosen 
so that $\vartheta_1-\vartheta_2=\pi n$, $n\in \Z$, 
the angular momentum is zero, and with $V\neq 0$ 
the particle falls to the origin of
 the coordinate system
being the vertex of the cone.
 One also notes 
that when the mechanical energy $ H_{\Omega}^{(\alpha)} $ 
takes positive (negative) values, 
the translational (rotational) motion
governed by  $H^{(\alpha)}_0$ ($-\alpha\Omega p_\varphi$)
 dominates. 
Furthermore, when $ V = 0 $, 
we have $H^{(\alpha)}_0=0$, and 
 the 
orbit is a circle of radius 
$ \rho_* $. 

The second equation in  (\ref{tajalpha}) can be written  in an 
equivalent form
 \begin{eqnarray}
 &\label{Traj2}
 e^{i\alpha^{-1}(\varphi-\varphi_*)}=e^{-i\Omega (t-t_*)}\left[
 \frac{1}{\rho_*\rho}\left(
\rho_*^2+i\frac{p_\varphi}{m\alpha}(t-t_*) 
 \right)
 \right]\,,&
 \end{eqnarray}
 where $\varphi_*=\varphi(t_*)$. After replacing here the time  evolution parameter
  by its expression in terms of  the phase space functions,   
\begin{eqnarray}&
\label{t-t*}
 t-t_*= \frac{\rho p_\rho}{2 H_{0}^{(\alpha)}}\,,
 &
\end{eqnarray}
that is directly obtained from 
(\ref{t*})
and the 
 definition of $D^{(\alpha)}$, and  multiplying (\ref{Traj2})
  by $\mp (2mH_0^{(\alpha)})^{1/2}e^{-i\alpha^{-1}\varphi}$, we get
 as a result  
the following 
complex pair $I_\pm^{(\alpha)}$
 of the true (not depending explicitly on time)
 mutually conjugate integrals of motion,
 \begin{eqnarray}
 \label{I1}
 &I_\pm^{(\alpha)}=\exp(\pm i \frac{\Omega\rho p_\rho}{2H_{0}^{(\alpha)}} )\, \Pi_\pm^{(\alpha)}\,,\qquad
&
 \end{eqnarray}
 see Eq. (\ref{ConeGenerators}) and  Appendix \ref{AppCantime}.
 In the case of $\alpha=1$ and $\Omega=0$ these integrals reduce to
the 
complex linear combinations $p_\pm=p_1\pm ip_2$  of the 
momentum vector components
of the free particle in $\R^2$~\footnote{
The
conical background ($\alpha\neq 1$) topologically is like  the punctured  Euclidean plane $\R^2- (0,0)$.}.
Obviously
the quantities 
$
f(H_{0}^{(\alpha)},p_\varphi) I_\pm^{(\alpha)}
$
also are  integrals of motion for
arbitrary choice of the function $f(\xi,\eta)$.
 
The integrals $I_+^{(\alpha)}$  and $I_-^{(\alpha)}$
have the Poisson brackets
$\{I_+^{(\alpha)},I_-^{(\alpha)}\}=2im\alpha\Omega\,,
$
and so, are similar to the complex linear combinations of the integrals
corresponding to a center of  a circular orbit in the Landau problem
in a plane.
 However,
due to the phase factor, these integrals 
are globally well defined functions in the phase space only 
in the cases of $ \alpha = 1/k $ with $ k = 1,2, \ldots $. For general case 
of rational values
$ \alpha = q/k $ with $ q = 1,2, \ldots,$ 
the integrals 
 \begin{eqnarray}
 \label{Ipmq}
&(I_\pm^{(q/k)})^{q}=\exp(\pm i  \frac{q\Omega \rho p_\rho}{2H_{0}^{(q/k)}}) (\Pi_\pm^{(q/k)})^{q}  & 
 \end{eqnarray}
 can be considered instead of (\ref{I1}) since
they do not have the  phase problem. 
Here and in what follows
we assume that  the fraction $q/k$  is irreducible, i.e., $q$ and $k$ are mutually prime.

Another way to obtain the formal integrals (\ref{I1}) is by starting from the integrals of the 
system in the 
reference frame at rest ($\Omega=0$), and applying to them the canonical transformation 
\begin{eqnarray}
\label{Falpha0}
&
\exp(\lambda F^{(\alpha)})\star f=
f
+\sum_{n=1}^\infty 
\frac{\lambda^n}{n!}\{F^{(\alpha)},\{\ldots,\{F^{(\alpha)},f\underbrace{\}\ldots\}\}}_{n}:=T_{F^{(\alpha)}}(\lambda)\star(f) \,,\quad&\\&
 F^{(\alpha)}=\alpha\frac{\rho p_\rho p_\varphi}{2H_0^{(\alpha)}}
 \,, \qquad
 \lambda=-\Omega\,.\label{Falpha}
&
\end{eqnarray} 
The chosen generator $F^{(\alpha)}$ of the 
canonical transformation includes dependence on 
the dilatation dynamical integral 
$D^{(\alpha)}$ taken at $t=0$, see Eq. (\ref{Halpha0}), and so, 
is not an integral of motion.
By means of the Poisson brackets relations 
\begin{eqnarray}
&
\{F^{(\alpha)},H_0^{(\alpha)}\}=\alpha p_\varphi\,,\qquad 
\{F^{(\alpha)},p_\varphi\}=0\,,
\label{Br1}&\\&
\{F^{(\alpha)}, \frac{1}{2} \rho p_\rho\}=F^{(\alpha)}\,,\qquad 
\{\frac{1}{2}\rho p_\rho,\Pi_\pm^{(\alpha)}\}=\frac{1}{2}\Pi_\pm^{(\alpha)}\,,\qquad 
\{p_\varphi,\Pi_\pm^{(\alpha)}\}=\mp i \Pi_\pm^{(\alpha)}\,, \label{Br2}
&
\end{eqnarray}
one gets as a result 
\be
T_{F^{(\alpha)}}(-\Omega)\star 
(H_0^{(\alpha)})=H_\Omega^{(\alpha)}\,,\qquad
T_{F^{(\alpha)}}(-\Omega)\star
((H_0^{(\alpha)})^{-\frac{1}{2}}\Pi_\pm^{(\alpha)})=(H_0^{(\alpha)})^{-\frac{1}{2}}I_\pm^{(\alpha)}\,,
\ee
from were it is  deduced
\begin{eqnarray}
&T_{F^{(\alpha)}}(-\Omega)\star 
\left(((H_0^{(\alpha)}+\alpha\Omega p_\varphi )^2)^{1/4}
(H_0^{(\alpha)})^{-\frac{1}{2}}\Pi_\pm^{(\alpha)}\right)=I_\pm^{(\alpha)}\,.&
\end{eqnarray}

In the case $\alpha=q/k$ we get 
$
T_{F^{(q/k)}}(-\Omega)\star((H_0^{(q/k)})^{-\frac{q}{2}}(\Pi_\pm^{(q/k)})^{q})=
(H_0^{(q/k)})^{-\frac{q}{2}}(I_\pm^{(q/k)})^{q}\,. 
$
This technique will be useful for the quantum analysis in the next
section. 

\begin{figure}[H]
\begin{center}
\begin{subfigure}[c]{0.28\linewidth}
\includegraphics[scale=0.28]{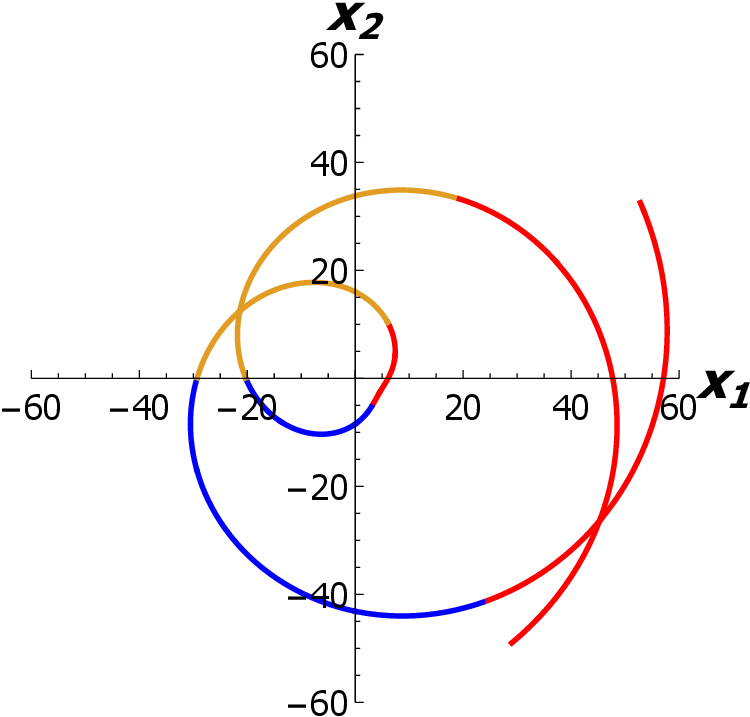}
\caption{\small{$\alpha=1/3$, $\dot{\varphi}_*<0$.}}
\label{freetaja}
\end{subfigure}
\begin{subfigure}[c]{0.28\linewidth}
\includegraphics[scale=0.28]{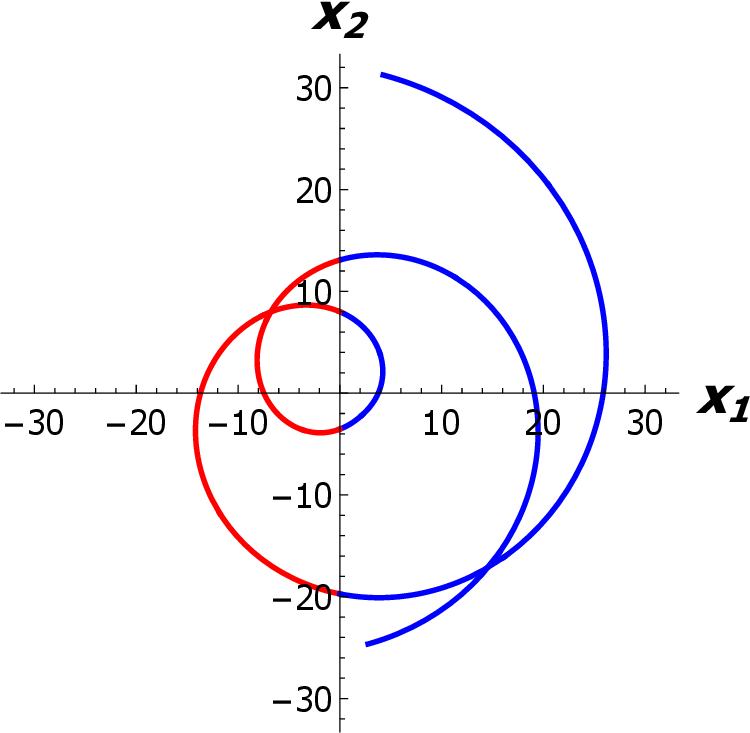}
\caption{\small{$\alpha=1/2$, $\dot{\varphi}_*<0$.}}
\label{freetajb}
\end{subfigure}
\begin{subfigure}[c]{0.28\linewidth}
\includegraphics[scale=0.28]{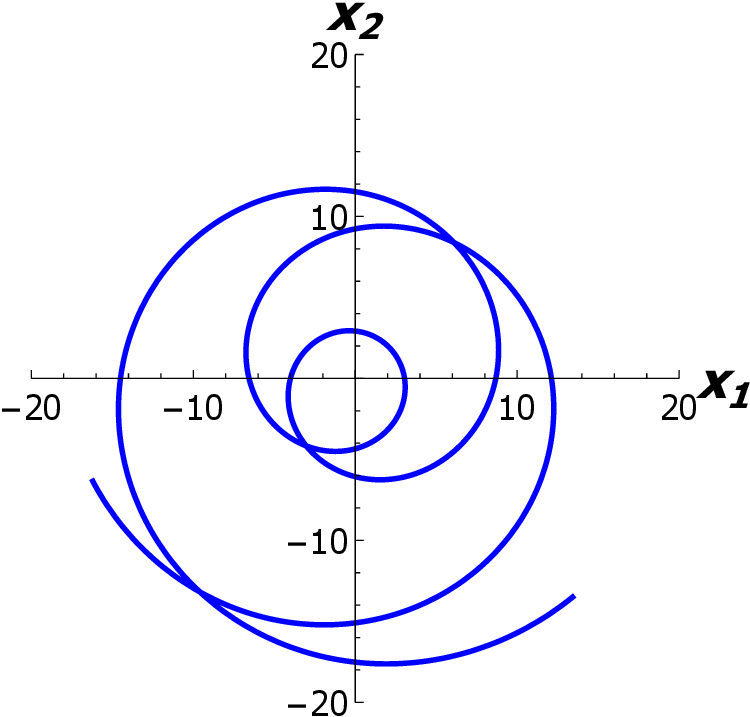}
\caption{\small{$\alpha=1$, $\dot{\varphi}_*<0$.}}
\label{freetajc}
\end{subfigure}
\vskip0.25cm
\begin{subfigure}[c]{0.28\linewidth}
\includegraphics[scale=0.28]{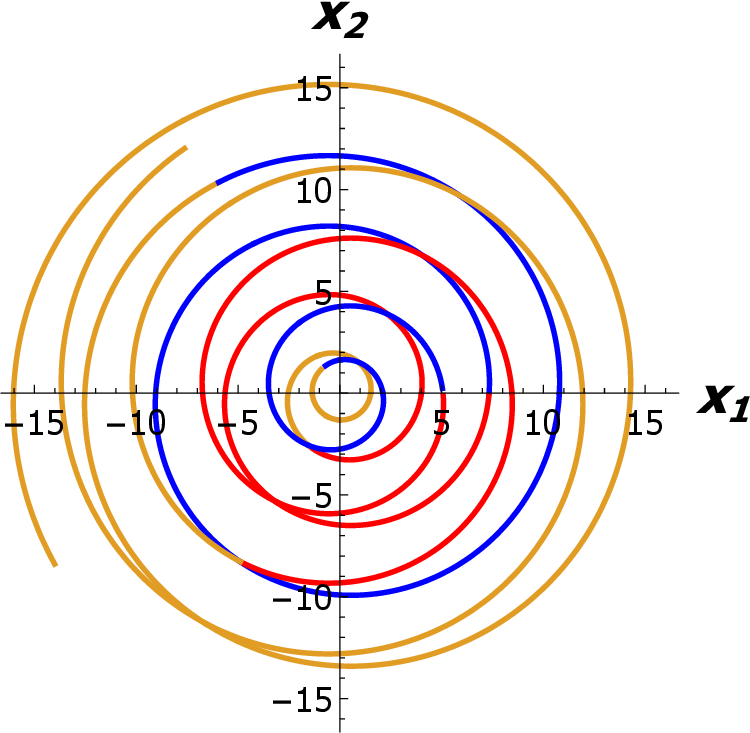}
\caption{\small{ $\alpha=4/3$, $\dot{\varphi}_*<0$.}}
\label{freetajd}
\end{subfigure}
\begin{subfigure}[c]{0.28\linewidth}
\includegraphics[scale=0.28]{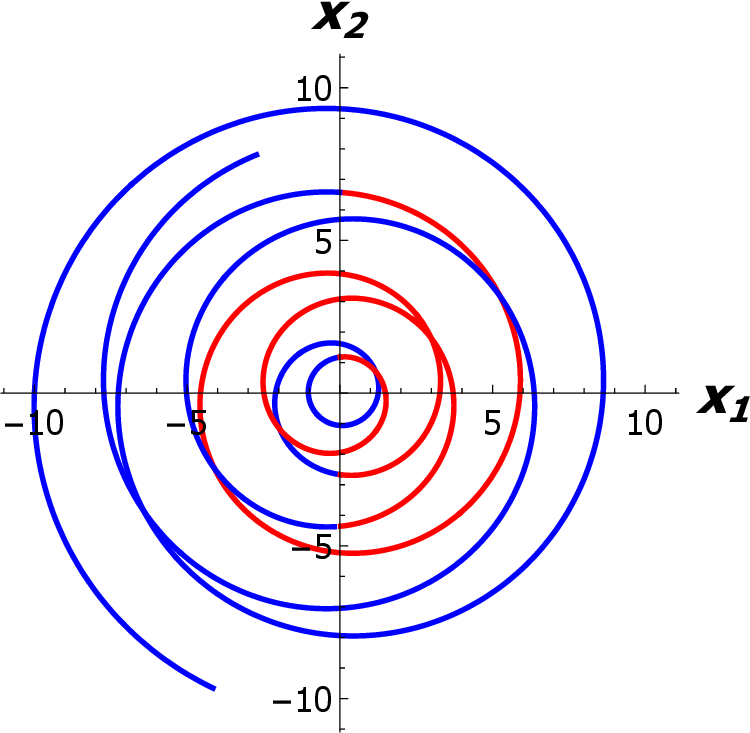}
\caption{\small{$\alpha=3/2$, $\dot{\varphi}_*<0$.}}
\label{freetaje}
\end{subfigure}
\begin{subfigure}[c]{0.28\linewidth}
\includegraphics[scale=0.28]{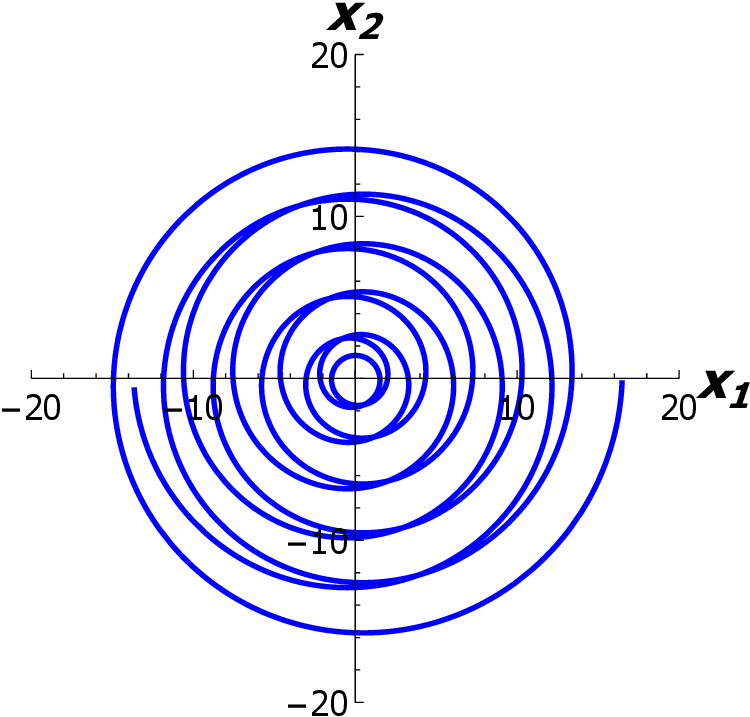}
\caption{\small{$\alpha=2$, $\dot{\varphi}_*<0$.}}
\label{freetajf}
\end{subfigure}

\begin{subfigure}[c]{0.28\linewidth}
\includegraphics[scale=0.28]{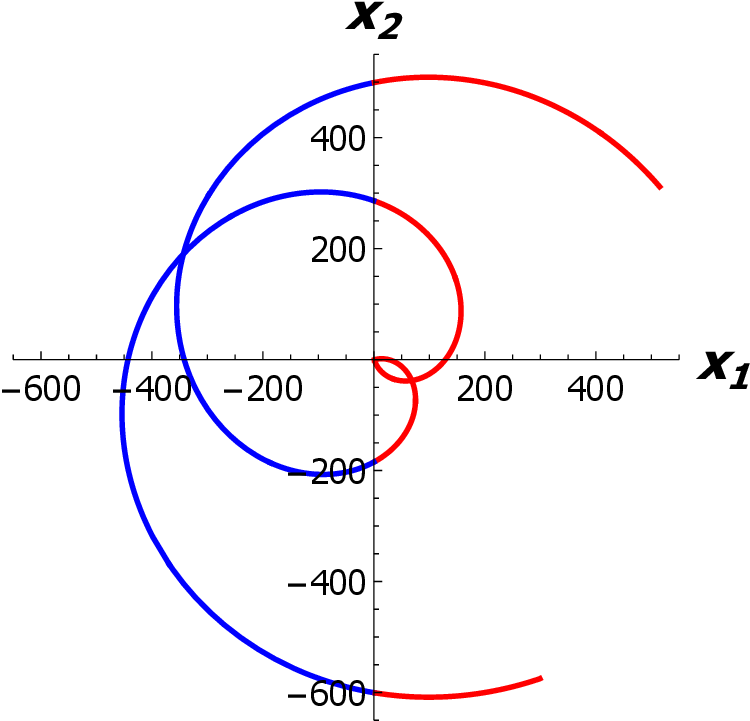}
\caption{\small{$\alpha=1/2$, $\dot{\varphi}_*>0$.}}
\label{freetaj2a}
\end{subfigure}
\begin{subfigure}[c]{0.28\linewidth}
\includegraphics[scale=0.28]{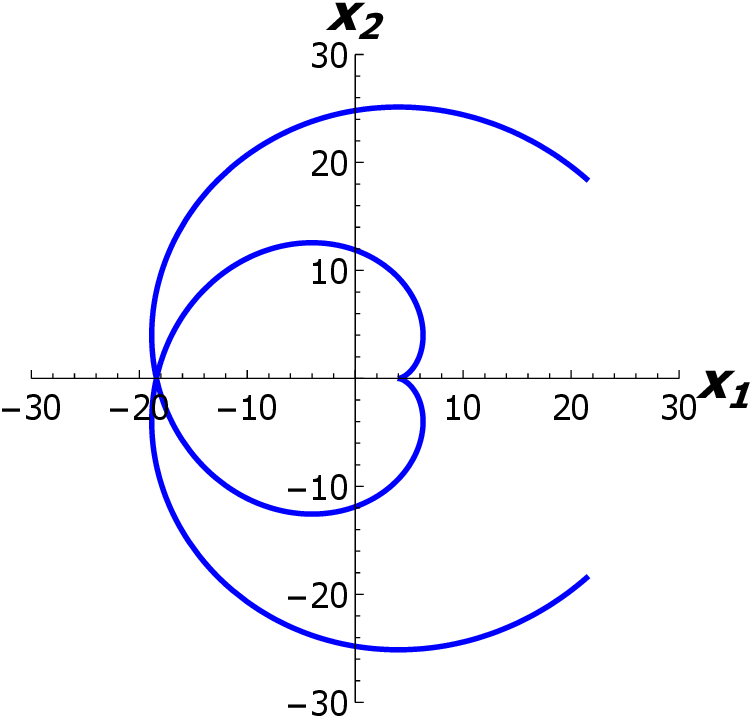}
\caption{\small{$\alpha=1$, $\dot{\varphi}_*=0$.}}
\label{freetaj2b}
\end{subfigure}
\begin{subfigure}[c]{0.28\linewidth}
\includegraphics[scale=0.28]{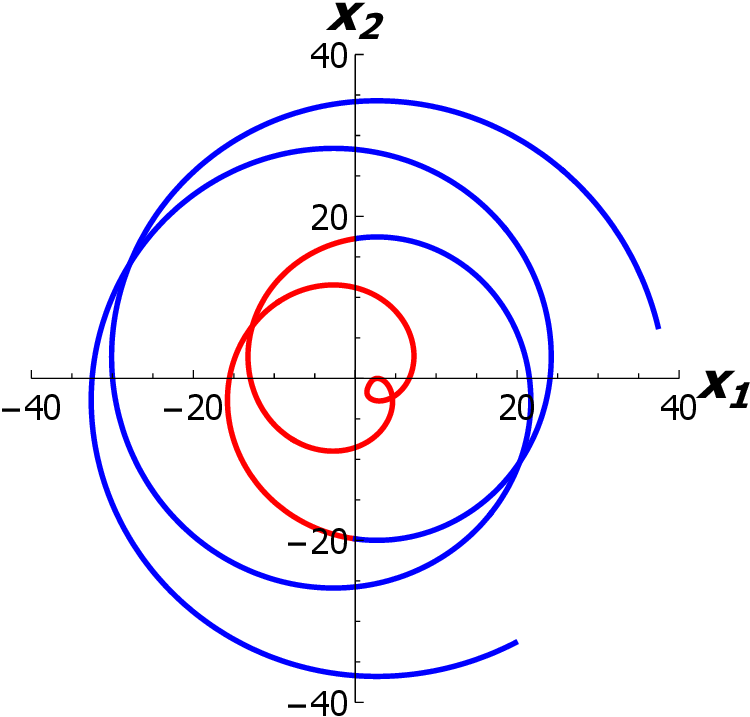}
\caption{\small{$\alpha=3/2$, $\dot{\varphi}_*>0$.}}
\label{freetaj2c}
\end{subfigure}
\caption{\small{Trajectories for different rational values of the 
cone parameter $\alpha$ and 
 $\Omega>0$. 
The colour change in the trajectories indicates the transition from one 
Riemann surface sheet to another. 
In figures \ref{freetaja} and \ref{freetajd} (\ref{freetajb}, \ref{freetaje}, \ref{freetaj2a}
and \ref{freetaj2c})
 the corresponding 
Riemann surface has three (two) sheets, while in figures \ref{freetajc} and \ref{freetajf} the Riemann 
surface is the complex plane itself. 
For
$\dot{\varphi}_*=\dot{\varphi}(t_*)=\alpha(V(R\sin(\vartheta_1-\vartheta_2))^{-1}-\Omega)>0$,
the external angular velocity $\Omega>0$  is subtracted  from the intrinsic angular velocity 
of the particle
at the position of ``perihelion'' of the orbit, 
producing  a small loop that does not enclose the origin.
For $\dot{\varphi}_*<0$, such small loops are not generated.
When  $\dot{\varphi}_*=0$, a cusp appears 
(figure \ref{freetaj2b}). 
}}
\label{Figtraj1}
\end{center}
\end{figure}

To complete
the classical analysis, we 
 visualise  the trajectories  of the particle 
 for different values of $\alpha$.
To 
draw them 
in the $(x^1,x^2)$ plane, 
where $x^1$ and $x^2$ are real and imaginary parts of 
$x_+$ in (\ref{zalpha}),
 it is necessary to take into account some properties associated with the function $f_\alpha(w)=w^{\alpha}$ in the
  $w$ complex 
 plane. First, the function $f_\alpha(w)$ is analytic in the $w$ complex plane  when $\alpha$ is an integer number, 
and then there are no problems to draw the orbit. 
On the other hand, when $\alpha=q/k$ with $q,k=1,2,\ldots$, $f_{q/k}(w)$ is a $k$-valued function,  and its entire domain 
corresponds to a Riemann surface with $k$ sheets.
In this  case, for $k>1$
we have to take care on which 
sheet we are in a definite moment of time. Finally,  $f_\alpha(w)$ 
with irrational $\alpha$ is an infinite-valued function,
whose domain corresponds to a
Riemann surface with infinite number of sheets. In Fig. \ref{Figtraj1} some examples 
of trajectories are presented by following the indicated peculiarities. 

 \subsection{Quantum picture}
 \label{SecQuantumFreeRot}

After quantization we get the Hamiltonian operator 
\begin{eqnarray}
&\label{quantumHconeOm}
\hat{H}_\Omega^{(\alpha)}=\hat{H}_0^{(\alpha)}-\alpha\Omega
\hat{p}_\varphi\,,
&
\end{eqnarray}
which represents  a linear combination of the free particle Hamiltonian 
in a conical background \cite{InzPly7}
and the quantum angular momentum operator, 
\begin{eqnarray}
&\label{QHJandpPhi}
\hat{H}_0^{(\alpha)}= -\frac{\hbar^2}{2m}
\left(\frac{1}{\alpha^2\rho}\frac{\partial}{\partial \rho}\left(\rho\frac{\partial}{\partial \rho}\right) +
\frac{1}{\rho^2}\frac{\partial^2}{\partial\varphi^2}\right)\,,
\qquad 
\hat{p}_\varphi=-i\hbar\frac{\partial}{\partial \varphi}\,.
&
\end{eqnarray}
The eigenstates that simultaneously diagonalize these two operators, 
and therefore are the eigenstates of the system  (\ref{quantumHconeOm}), are 
given by the functions 
\begin{eqnarray}
\label{FreeParticleE}&
\psi_{\kappa,\ell}^{(\alpha)}(r,\varphi)=\sqrt{\frac{\kappa}{2\pi \alpha}}J_{\alpha |\ell| }(\kappa r)e^{i\ell
 \varphi}\,,&
\end{eqnarray}  
where $J_{\beta }(\eta)$ are the Bessel functions of the first kind. 
These states satisfy the eigenvalue 
equations 
\begin{eqnarray}
&
\hat{H}_{0}^{(\alpha)}\psi_{\kappa,\ell}^{(\alpha)}(r,\varphi)=\frac{\hbar^2\kappa^2}{2m\alpha^2}\psi_{\kappa,\ell}^{(\alpha)}(r,\varphi)\,,\qquad
\hat{p}_\varphi \psi_{\kappa,\ell}^{(\alpha)}(r,\varphi)=\hbar\ell \psi_{\kappa,\ell}^{(\alpha)}(r,\varphi)\,, &\\&
\hat{H}_{\Omega}^{(\alpha)}\psi_{\kappa,\ell}^{(\alpha)}=  E_{\kappa,\ell}^{(\alpha)}\psi_{\kappa,\ell}^{(\alpha)}\,,\qquad 
  E_{\kappa,\ell}^{(\alpha)} =\frac{\hbar^2\kappa^2}{2m\alpha^2}-\alpha \Omega\hbar\ell\,,\quad \kappa>0\,, \quad \ell =0,\pm 1,\ldots\,,
&
\end{eqnarray}
as well as the orthogonality relation 
$\braket{\psi_{\kappa,\ell}^{(\alpha)}}{\psi_{\kappa',\ell'}^{(\alpha)}}=\
\delta_{\ell,\ell'}\delta(\kappa-\kappa') $, 
where  
\begin{eqnarray}
\label{scalarp}&
\braket{\Psi_1}{\Psi_2}=
\iint \Psi_1^*\Psi_2  \sqrt{g} \, dx^1dx^2=
\alpha\int_0^{\infty} r dr\int_{0}^{2\pi}d\varphi \Psi_1^*\Psi_2\,&
\end{eqnarray}
is the  
inner product in conical geometry,
$g=\det \,(g_{ij})$.
The formal eigenfunctions  corresponding 
 to 
  $\kappa=0$ and $\ell\not=0$, 
which are  given by $r^{\alpha|\ell|}e^{i\ell\varphi}$
and $r^{-\alpha|\ell|}e^{i\ell\varphi}$,
are excluded from the physical sets of eigenstates since 
they are not  Dirac delta normalizable  with 
respect to the scalar product (\ref{scalarp}). 
The physical  eigenstates with zero energy correspond to all those functions whose quantum numbers 
 satisfy the condition 
$\hbar\kappa^{2}=2m\alpha^3\Omega\ell$,  
as well as the constant function 
when $\kappa=\ell=0$. The latter does not belong to the set (\ref{FreeParticleE}), 
but  can be 
obtained by
a limiting procedure
\begin{eqnarray}
&
\label{limit}
\lim_{\kappa\rightarrow0} \sqrt{\frac{2\pi
\alpha
}{\kappa}}\psi_{\kappa,0}^{(\alpha)}=1\,.
&
\end{eqnarray}
For the sake of definiteness 
 in what follows
we assume $\Omega>0$.

It is notable that unlike  the energy eigenvalues, 
the form of the eigenfunctions 
$ \psi_{\kappa,\ell}^{(\alpha)}(r,\varphi)$
is independent of the value of the external angular velocity $ \Omega $. 
Therefore,
if for certain values of $\alpha$, the system has well defined integral operators 
related to
a degeneration of the spectrum, they 
 should modify coherently
both indices (discrete and continuous) 
in a corresponding wave function. 
In this way from the condition  $E_{\kappa,\ell}^{(\alpha)}=
E_{\tilde{\kappa},\ell\pm 1}^{(\alpha)}$
we obtain
\begin{eqnarray}
\label{kappapm}&
\tilde{\kappa}=\kappa_\pm=\kappa\sqrt{1\pm \frac{2m\alpha^3 \Omega}{\hbar\kappa^2}}\,.&
\end{eqnarray}
This 
relation   
implies that when $\hbar\kappa^2<2m\alpha^3\Omega$, 
 $\kappa_-$
becomes an imaginary number, and therefore, 
physical integral operators that produce  the transformation $\psi_{\kappa,\ell}^{(\alpha)}
\rightarrow \psi_{\kappa_-,\ell- 1}^{(\alpha)} $ 
must annihilate all physical eigenstates with  $\hbar \kappa^2\leq2m\alpha^3\Omega$.
No restrictions appear for the eigenvalue $\ell$  here. 

In order  to try 
to construct such symmetry integral operators,  we introduce
the operators
\be
\hat{W}_-^{(\alpha)} =e^{-i\frac{\Omega}{\hbar} \hat{F}}\,,\qquad 
\hat{W}_+^{(\alpha)}=(\hat{W}_-^{(\alpha)})^{-1}=e^{i\frac{\Omega}{\hbar} \hat{F}}\,.
\ee
The quantum version $\hat{F}$ of the classical quantity 
 (\ref{Falpha}) 
we fix 
so that
the commutation relations 
\begin{eqnarray}
&
[\hat{F}^{(\alpha)},\hat{H}_0^{(\alpha)}]=i\hbar \alpha \hat{p}_\varphi\,,\quad 
[\hat{F}^{(\alpha)},\hat{p}_\varphi]=0\,
&
\end{eqnarray}
will be satisfied  in correspondence with (\ref{Br1}).
This 
 implies the relations
\be
\label{comW}
[\hat{H}_0^{(\alpha)},\hat{W}_\pm^{(\alpha)}]=\pm \Omega \alpha\hat{p}_\varphi W_\pm^{(\alpha)} \,,\qquad
[\hat{p}_\varphi,\hat{W}_\pm^{(\alpha)}]=0\,.
\ee
The appearance of the quantity $(H_{0}^{(\alpha)})^{-1}$ in the classical definition 
of $F$ means
 that some non-trivial quantization prescription must be employed
to escape a singularity in the corresponding quantum operators.
By taking the symmetric ordering
\footnote{A commutator of the form $[(\hat{H}_0^{(\alpha)})^{-1},\hat{O}]$, with
an arbitrary well defined operator $\hat{O}$,  is singular 
when it is applied to a constant  wave function. 
For this reason 
the constant wave function has  to be excluded from the 
domain of the operator (\ref{SymOrderF}).
},
 \be
\label{SymOrderF}
\hat{F}^{(\alpha)}=\frac{\alpha}{4}\left[\hat{p}_\varphi\left( (\hat{H}_0^{(\alpha)})^{-1} \hat{D}_0^{(\alpha)}+
\hat{D}_0^{(\alpha)}(\hat{H}_0^{(\alpha)})^{-1}\right)+\left( (\hat{H}_0^{(\alpha)})^{-1} \hat{D}_0^{(\alpha)}+
\hat{D}_0^{(\alpha)}(\hat{H}_0^{(\alpha)})^{-1}\right)\hat{p}_\varphi\right]\,,\quad
\ee
we produce the Hermitian operator, $\hat{F}^{(\alpha)}=\hat{F}^{(\alpha)}{}^\dagger$,
where $\hat{D}_0^{(\alpha)}=\frac{-i\hbar}{2}(\rho \frac{\partial}{\partial \rho}+1)$ coincides with the dilatation operator
taken at $t=0$. In this construction 
$\hat{W}_\pm^{(\alpha)}$ formally are unitary and their
 action is well defined on eigenstates of the form (\ref{FreeParticleE}).
 However,
 the 
  action of the operator (\ref{SymOrderF}) on  the constant wave function is not defined. 
   As this last eigenstate corresponds to a free particle at 
rest in the inertial
 reference frame,  with zero eigenvalue of $\hat{H}_0^{(\alpha)}$ 
 and 
 zero angular momentum, 
it seems reasonable to treat it  
  separately for the moment.
With these considerations, operator (\ref{SymOrderF}) is 
simplified, 
\be
\label{SymOrderF+}
\hat{F}^{(\alpha)}=\frac{\alpha}{2}(\hat{H}_0^{(\alpha)})^{-1} ( 2\hat{D}_0^{(\alpha)}-i\hbar)\hat{p}_\varphi\,.
\ee
Then, the unitary  operator 
$\hat{W}_+^{(\alpha)}$ and its inverse, 
$\hat{W}_-^{(\alpha)}=\hat{W}_+^{(\alpha)}{}^\dagger$,
can be employed to produce the following unitary transformation,
\be
\label{UniTrans}
\hat{W}_-^{(\alpha)}
\hat{H}_0^{(\alpha)}
\hat{W}_+^{(\alpha)}=\hat{H}_\Omega^{(\alpha)}\,,\qquad
\hat{W}_-^{(\alpha)}
\hat{p}_\varphi
\hat{W}_+^{(\alpha)}=\hat{p}_\varphi\,,
\ee
and as a consequence,  
the transformed eigenstates 
$\Psi_{\kappa,\ell}^{(\alpha)}=\hat{W}_-^{(\alpha)}\psi_{\kappa,\ell}^{(\alpha)}$ should satisfy 
\begin{eqnarray}
& 
\hat{H}_0^{(\alpha)}\Psi_{\kappa,\ell}^{(\alpha)}=\left(\frac{\hbar^2 
\kappa^2}{2m\alpha^2}+\hbar\alpha\Omega\ell\right) \Psi_{\kappa,\ell}^{(\alpha)}\,,\qquad
\hat{p}_\varphi \Psi_{\kappa,\ell}^{(\alpha)}=\hbar \ell  \Psi_{\kappa,\ell}^{(\alpha)}\,,\qquad 
\hat{H}_\Omega^{(\alpha)}\Psi_{\kappa,\ell}^{(\alpha)}=\frac{\hbar^2 
\kappa^2}{2m\alpha^2}\Psi_{\kappa,\ell}^{(\alpha)}\,.\qquad
&
\end{eqnarray} 
From these expressions we conclude that the transformed states 
have to be
 of the form
\begin{eqnarray}
&\label{transformFree}
\Psi_{\kappa,\ell}^{(\alpha)}\sim\psi_{\kappa_\ell,\ell}^{(\alpha)}\,,\qquad
\kappa_\ell=\kappa\sqrt{1+\frac{2m\alpha^3\Omega \ell}{\hbar \kappa^2}}\,.
&
\end{eqnarray}
However, 
in this way 
non-physical eigenstates can be obtained for negative values of
$\ell$ for $\Omega>0$.
For this reason we continue using states (\ref{FreeParticleE}) instead of 
 (\ref{transformFree}).
To construct the
symmetry operators associated with degeneracy of energy eigenvalues,
we apply the similarity transformation to the integral 
operators
of the free particle in a conical geometry in the reference frame at rest, 
$\Omega=0$. 
It is known 
from the previous work  \cite{InzPly7} that such symmetry operators 
exist only when $\alpha=q$
(otherwise, they can produce non-physical eigenstates),
and are given by 
\begin{eqnarray}\label{Pi+-}
&(\hat{\Pi}_\pm^{(q)})^{q}\,,\qquad \text{where}\qquad
\hat{\Pi}_\pm^{(q)}=-i\hbar
e^{\pm i\frac{\varphi}{q}}\left(\frac{1}{q}\frac{\partial}{\partial \rho}\pm i \frac{1}{\rho}\frac{\partial}{\partial \varphi}\right)\,. 
&
\end{eqnarray}
After applying the unitary transformation to them  one gets 
the mutually Hermitian conjugate operators
\be
\label{Iq}
(\hat{I}_+^{(q)})^{q}=\hat{W}_-^{(q)}
(\hat{\Pi}_+^{(q)})^{q}
\hat{W}_+^{(q)}\,,\qquad
 (\hat{I}_-^{(q)})^{q}=((\hat{I}_+^{(q)})^{q})^\dagger\,.
\ee
These operators
 commute with the Hamiltonian $\hat{H}_\Omega^{(\alpha)}$ and can be applied 
 to all physical states
 of the form (\ref{FreeParticleE}). 
 Their action can be deduced by means of  the commutation relations 
\be
[\hat{H}_0^{(q)},(\hat{I}_\pm^{(q)})^{q}]= \pm q \Omega\hbar (\hat{I}_\pm^{(q)})^{q}\,,\qquad
[\hat{p}_\varphi,(\hat{I}_\pm^{(q)})^{q}]= \pm  \hbar (\hat{I}_\pm^{(q)})^{q}\,,
\ee
from which one obtains 
\be
\label{Iq2}
(\hat{I}_\pm^{(q)})^{q}\psi_{\kappa,\ell}^{(\alpha)}\sim  \psi_{\kappa_\pm,\ell\pm 1}^{(\alpha)}\,,
\ee
where $\kappa_\pm$ correspond to (\ref{kappapm}). 
The operators $(\hat{I}_\pm^{(q)})^{q}$ produce exactly
the necessary 
change of the index in the wave functions which does not change the
corresponding energy eigenvalue. Nevertheless, in the cases when $\kappa$  satisfies
the condition    $\hbar \kappa^2<2m q^3\Omega$, the  operator 
 $(\hat{I}_-^{(q)})^{q}$ produces non-physical eigenstates.
 To avoid this problem,  we can take 
 the regularized operators
 \begin{eqnarray}\label{SymmetryOp}
& 
 \hat{\mathscr{I}}_-=(\hat{I}_-^{(q)})^{q}\Theta\left(\frac{1}{\hbar \Omega}\hat{H}_0^{(q)}-q\right)\,,\quad
  \hat{\mathscr{I}}_+=\Theta\left( \frac{1}{\hbar\Omega}\hat{H}_0^{(q)}-q \right)(\hat{I}_+^{(q)})^{q}\,,
 &
 \end{eqnarray}
  where $\Theta(\eta)$ is the Heaviside step function. 
  When $ \hat{\mathscr{I}}_-$
  acts on the eigenstates, the Heaviside 
   function annihilates all those physical eigenstates, 
  which can  potentiality be transformed into non-physical
   states by the action of $(\hat{I}_-^{(q)})^{q}$.
  On the other hand,  $(\hat{I}_+^{(q)})^{q}$ cannot produce non-physical states, and if 
   $\hbar \kappa^2<2m q^3\Omega$, then   $\hbar \kappa_+^2=\hbar \kappa^2+ 2m q^3\Omega\geq 2m q^3\Omega$. As
   a consequence, 
   the action of the operator $\Theta\left( \frac{1}{\hbar\Omega}\hat{H}_0^{(q)}-q \right)$
   after the application of   $(\hat{I}_+^{(q)})^q$  
   reduces to multiplication by one.
  The action of the symmetry operators (\ref{SymmetryOp})
   is well defined on all energy  eigenstates including 
   the constant wave function corresponding to zero eigenvalue
   of the operator $\hat{H}^{(\alpha)}_0$.

In conclusion of this section we note that as 
classically the system reveals 
a hidden symmetry in the case of rational values of the conical background 
parameter $\alpha$, while at the quantum level 
we are able to identify  the
hidden symmetry described by 
non-local operators in the case of only integer values of $\alpha$, 
we face a kind of a quantum anomaly here.

A rather natural  question is if the anomaly related to the geometry of the system can be 
``cured''
 by a dynamic extension of the system.
We  investigate  such a kind of 
extension of the system in the next section.

\section{Classical ERIHO system in a conical space}
\label{SecERIHOClasic}

Recently, in  \cite{ERIHO}
we considered the exotic rotationally invariant harmonic oscillator
generated 
from the planar free particle system
by application   
of the conformal bridge transformation \cite{InzPlyWipf1,PTrev}.
The model is described by the Hamiltonian~\footnote{In \cite{ERIHO}, the notation $g$ is used
instead of the parameter $\gamma$ here.}
\begin{eqnarray}
\label{PlanarHERIHO}
&H_\gamma=\frac{1}{2m}p_{i}p_{i}+\frac{1}{2}m\omega^2 x_ix_i +
\gamma\omega\epsilon_{ij}x_ip_j=\frac{1}{2m}\left(p_\rho^2+\rho^{-2}p_\varphi^2 \right)+
\frac{1}{2}m\omega^2 \rho^2+\gamma\omega p_\varphi\,,&
\end{eqnarray}
and its peculiar properties were studied 
by us from the perspective of symmetries.   
The solutions 
of the equations of motion of the system correspond to  closed trajectories 
only when $\gamma$ is a rational number. This property is reflected 
in the existence  of the 
additional integrals of motion in the case $\gamma\in \Q$, which  define the orientation of the orbit
like the Laplace-Runge-Lenz vector in Kepler's problem.
At the quantum level, these additional integrals explain the degeneracy of the spectrum of the 
model that only appears  when   $\gamma$ takes rational values.
 In spite of the explicit rotational invariance 
of (\ref{PlanarHERIHO}), the classical and quantum properties of the model resemble those of the
planar anisotropic Euclidean and Minkowskian harmonic oscillator systems. 
This is not accidental since both systems 
turn out to be related by a non-trivial unitary  $\mathfrak{so}(2)$-non-invariant transform 
composed from the special $\mathfrak{su}(2)$ rotation  considered 
by us earlier in \cite{InzPly7} and supplemented by an additional 
 Bogolyubov transformation.

After changing $\omega\rightarrow \Omega=-\gamma\omega$, one finds that the  Lagrangian of the system 
(\ref{PlanarHERIHO})  
can be presented in the form 
\be
\label{ERIHOLag}
L_\gamma=L_{\Omega}-\frac{1}{2}m(\gamma^{-1}\Omega)^2 x^ix^i=
\frac{m}{2}\dot{x}^i\dot{x}^i-\frac{m}{2}(\gamma^{-2}-1)\Omega^2 x^ix^i+
\frac{q_{GM}}{c}\dot{x}^iA^i\,,
\ee
where $L_{\Omega}$ is given by Eq. (\ref{flatLagrangian}). 
As a consequence,
 the model (\ref{PlanarHERIHO})  admits two other physical  interpretations  \cite{ERIHO}. 
 First, it corresponds to a two-dimensional isotropic  harmonic oscillator
  in a non-inertial reference frame rotating 
with angular velocity $\Omega$. Alternatively, it can be considered as 
 the  Landau problem in the presence of  an additional external 
harmonic potential 
$V_\gamma=\frac{m}{2}(\gamma^{-2}-1)\Omega^2 x^ix^i$. 
When $\gamma^2<1$ ($\gamma^2>1$), the sign of this quadratic 
potential is positive (negative), so it produces an attractive 
(repulsive) 
harmonic force
on the particle. 
The intermediate case  $\gamma^2=1$ corresponds to 
 the two Landau phases of the system different in  two possible orientations 
of the magnetic field orthogonal to the plane, which are defined 
by the  sign of
$\gamma$.

In this section we explore the direct classical and quantum 
generalization of the ERIHO system to a conical spacetime. 
As we will see, in this case
one has to take into account both  parameters $\gamma$ and $\alpha$ 
to find closed trajectories with which the integrals of motion of the hidden symmetry
are associated. 
Before the detailed analysis, one would expect that the presence of an additional parameter 
$ \gamma $, which  is associated here  with an additional ``gravitoelectric" harmonic potential, 
could ``cure" the quantum anomaly related with the background geometry, 
that  was described in the previous section.
In particular case of $\gamma^2=1$, this also will allow us to investigate
the Landau problem 
in conical background from the perspective of dynamics and symmetries.

\subsection{Classical picture}
\label{SecClasPic}
To find the analogue of the  ERIHO system in a conical background,  we start by changing
the implicit flat metric tensor $\delta_{ij}$ in Lagrangian (\ref{ERIHOLag}) 
by the conical metric tensor (\ref{conicalmetricCar}), that yields 
 \be\label{Laggamal}
 L_{\gamma}^{(\alpha)}=L_\Omega^{(\alpha)}-\frac{1}{2} m
 \gamma^{-2}
 \Omega^2g_{ij}x^ix^j=
\frac{m}{2}g_{ij}\dot{x}^i\dot{x}^j-\frac{m}{2}(\gamma^{-2}-1)\Omega^2 g_{ij}x^i x^j
+
\frac{q_{GM}}{c}g_{ij}\dot{x}^i A^{(\alpha)j}\,,
 \ee
 where $ L_{\Omega}^{(\alpha)}$ and $A^{(\alpha)i}$ are given by
  (\ref{cartesianLag}). 
 This Lagrangian 
means that, again, the system can be interpreted 
as  an isotropic harmonic oscillator 
in a conical  background of a cosmic string 
in a non-inertial uniformly rotating reference frame. 
Alternatively,  we can consider system (\ref{Laggamal}) 
as the Landau problem in conical geometry, extended by an external quadratic potential  
 $V_\gamma^{(\alpha)}=\frac{m}{2}(\gamma^{-2}-1)\Omega^2 g_{ij}x^i x^j=
 \frac{m}{2}(\gamma^{-2}-1)\Omega^2 \alpha^2\rho^2$. 
 {}From the nature of this additional potential term it is clear that
one has 
 to distinguish the cases $\gamma^2<1$  and $\gamma^2>1$, 
 where this potential is attractive  and repulsive, 
 and the Landau phases at 
$\gamma=\pm 1$. 
 The case  $\gamma=\infty$ corresponds here to the 
 system considered in the previous section.
 After  performing
 the Legendre transformation
  and by identifying $\Omega=-\gamma\omega$, 
 one gets the Hamiltonian of the system, which takes  more simple form  in 
 polar coordinates,
 \begin{eqnarray}
\label{Hgcone0}
&H_\gamma^{(\alpha)}=\frac{1}{2m}\left(\alpha^{-2}p_\rho^2 +\rho^{-2}p_\varphi^2\right)+
\frac{1}{2} m\omega^2\alpha^2 \rho^2+\gamma\omega\alpha p_\varphi\,.&
\end{eqnarray}
We assume without loss of generality that $\omega>0$.
Comparing (\ref{Hgcone0}) with  
the Hamiltonian of the Euclidean 
planar ERIHO system (\ref{PlanarHERIHO}) \cite{ERIHO}, 
it is obvious that the former can be obtained from the latter
by means of the local canonical transformation 
\begin{eqnarray}\label{canonical}
&\rho\rightarrow \alpha \rho\,,\qquad
p_{\rho}\rightarrow \alpha^{-1}p_\rho\,, \qquad
\varphi\rightarrow\alpha^{-1}\varphi\,,\qquad
p_{\varphi}\rightarrow\alpha p_{\varphi}\,.&
\end{eqnarray}
So, 
 (\ref{Hgcone0}) is 
 the ERIHO system in a conical background. 

 In terms of Cartesian coordinates, canonical transformation (\ref{canonical})
 is
\begin{eqnarray}
&\label{TransformationX}
x_1\rightarrow X_{\alpha,1}=\alpha \rho \cos(\frac{\varphi}{\alpha})\,,\qquad
x_2\rightarrow X_{\alpha,2}=\alpha \rho \sin(\frac{\varphi}{\alpha})\,,
&\\&
p_1\rightarrow P_{\alpha,1}=\frac{p_\rho}{\alpha}\cos(\frac{\varphi}{\alpha})-\frac{p_\varphi}{\rho}\sin(\frac{\varphi}{\alpha})\,,\qquad
p_2\rightarrow P_{\alpha,2}=\frac{p_\rho}{\alpha}\sin(\frac{\varphi}{\alpha})+\frac{p_\varphi}{\rho}\cos(\frac{\varphi}{\alpha})\,, 
&\\&
\{X_{\alpha,i},X_{\alpha,j}\}=\{P_{\alpha,i},P_{\alpha,j}\}=0\,,\qquad
\{X_{\alpha,i},P_{\alpha,j}\}=\delta_{ij}\,,
&
\end{eqnarray}
while the conical metric $ds^{2}=g_{ij}dx^idx^j$ takes the form 
$
ds^2=dX_{\alpha,1}^2+dX_{\alpha,2}^2$ . 
In local canonical variables ($X_{\alpha,i}$, $P_{\alpha,i}$)
Hamiltonian (\ref{Hgcone0})
has the form of the ERIHO Hamiltonian in usual Cartesian coordinates, 
\begin{eqnarray}
\label{Hgcone}
&H_\gamma^{(\alpha)}= \frac{1}{2m}(P_{\alpha,1}^2+P_{\alpha,2}^2)+
\frac{m\omega^2}{2}(X_{\alpha,1}^2+X_{\alpha,2}^2)+\omega \gamma \epsilon_{ij}X_{i,\alpha}P_{j,\alpha}\,.&
\end{eqnarray}
To analyse the dynamics and symmetries of the system, it is convenient 
to pass over  
to the following complex combinations of the local canonical variables, 
\begin{eqnarray}
&
b_{\alpha,1}^-=\frac{1}{\sqrt{2}}(a_{\alpha,1}^--ia_{\alpha,2}^-)=\frac{1}{2}e^{-i\frac{\varphi}{\alpha}}
\left(
\alpha\sqrt{m\omega}\rho
+\frac{p_\varphi}{\sqrt{m\omega}\rho} +\frac{ip_\rho}{\alpha\sqrt{m\omega}}\right)\,,\quad 
b_{\alpha,1}^+=(b_{\alpha,1}^-)^*\,,
&\\&
b_{\alpha,2}^-= \frac{1}{\sqrt{2}}(a_{\alpha,1}^-+ia_{\alpha,2}^-)=\frac{1}{2}e^{i\frac{\varphi}{\alpha}}
\left(
\alpha\sqrt{m\omega}\rho
-\frac{p_\varphi}{\sqrt{m\omega}\rho} +\frac{ip_\rho}{\alpha\sqrt{m\omega}}\right)\,,
\quad 
b_{\alpha,2}^+=(b_{\alpha,2}^-)^*\,,
&\\ &
a_{\alpha,i}^\pm=\sqrt{\frac{m\omega}{2}}\left(\,X_{\alpha,i} \mp \frac{i}{m\omega}\,P_{\alpha,i}\right)\,.
&
\end{eqnarray}
They satisfy  the Poisson bracket relations 
$\{b_{\alpha,j}^-,b_{\alpha,k}^-\}=\{b_{\alpha,j}^+,b_{\alpha,k}^+\}=0$, 
$\{b_{\alpha,j}^-,b_{\alpha,k}^+\}=-i\delta_{jk}$,
and reduce  the Hamiltonian to the form like that
of the anisotropic harmonic oscillator, 
\begin{eqnarray}
\label{gHamilbb+}
H_\gamma^{(\alpha)}= \omega\left(\ell_1 b_{\alpha,1}^+b_{\alpha,1}^-+\ell_2b_{\alpha,2}^+b_{\alpha,2}^- \right)\,,\qquad
\ell_1=1+\gamma\,,\qquad \ell_2=1-\gamma\,.
\end{eqnarray}
However, here the angular momentum is
\be
p_\varphi=\alpha^{-1}(b_{\alpha,1}^+b_{\alpha,1}^--b_{\alpha,2}^+b_{\alpha,2}^-)\,,
\ee
and Hamiltonian  (\ref{gHamilbb+})  
is rotationally invariant, $\{p_\varphi,H_\gamma^{(\alpha)}\}=0$.

The   equations of motion can easily be solved in variables $b_{\alpha,i}^\pm$, 
\be
\label{E.Q.O.M}
\dot{b}_{\alpha,i}^\pm=\{b_{\alpha,i}^\pm,H_\gamma^{(\alpha)}\}=\pm i\omega \ell_i b_{\alpha,i}^\pm\quad
\Rightarrow \quad b_{\alpha,i}^\pm(t)= e^{\pm i\omega \ell_i t }b_{\alpha,i}^\pm(0):= b_{\alpha,i}^\pm
\,.
\ee
Using the relation 
$\sqrt{m\omega}(X_{\alpha,1}+iX_{\alpha,2})=b^+_{\alpha,1}+b^-_{\alpha,2}$, 
one gets 
\be\label{traj}
X_{\alpha,+}(t)=X_{\alpha,1}(t)+iX_{\alpha,2}(t)=R_1e^{i\vartheta_1}e^{i\omega \ell_1 t}+
R_2e^{-i\vartheta_2}e^{-i\omega \ell_2 t}\,,
\ee
where $R_{1,2}\geq 0$, $R^2_1+R_2^2>0$, and $\vartheta_{1,2}\in \R$
are the constants of integration, $b_{\alpha,i}^\pm(0)=R_ie^{\pm i\vartheta_i}/\sqrt{m\omega}$.
From here one sees that $X_{\alpha,+}(t)$ is a periodic in time $t$ function 
only when $(1+\gamma)/(1-\gamma) \in \mathbb{Q}$, 
that implies rational values for the parameter $\gamma$. 
This, however,  does not mean periodicity of the
conical  trajectories 
as
$X_{\alpha,i}$ are not globally well defined functions of polar coordinates
for general values of the parameter $\alpha$\,:
$X_{\alpha,i} (\rho,\varphi=0 )\neq  X_{\alpha,i}(\rho,\varphi=2\pi)$
for $\alpha\neq 1/n$, $n=2,3,\ldots$.
To find the evolution of the polar coordinates   $\rho(t)$ and $\varphi(t)$, 
we use the relations 
\begin{eqnarray}
&
\label{chart}
\rho^2(t)=\frac{1}{\alpha^2}| X_{\alpha,+}(t)|^2\,,\qquad
e^{i\alpha^{-1}\varphi(t)} 
 =\left(\frac{X_{\alpha,+}(t)}{\alpha \rho(t)}\right)\,\Rightarrow\,
e^{i\varphi(t)}=\left(\frac{X_{\alpha,+}(t)}{\alpha \rho(t)}\right)^{\alpha}\,, &\\&
x_+(t)=x_1(t)+ix_2(t)=\rho(t)e^{i\varphi(t)}=\frac{1}{\alpha}| X_{\alpha,+}(t)|^{1-\alpha}
(X_{\alpha,+}(t))^{\alpha}\,.
&
\end{eqnarray}
Similar to the free rotation case from the previous section,
in order to visualize the trajectory, we should be careful with the  
complex function $f_\alpha=( X_{\alpha,+}(t))^{\alpha}$ in dependence on the value of $\alpha$. 
The $X_{q/k,+}(t)$ 
describes a curve that encircles  the origin of coordinates in 
the plane $(X_{\alpha,1},X_{\alpha,2})$ when $\gamma^2\neq 1$.
The only exception is possible in the Landau case 
$\gamma^2=1$  for 
certain selection of the initial data, that we discuss below. 
However, in general case these curves are closed only when $\gamma$ takes rational values. 
Then, for $\alpha=q/k$, 
 complex function $f_{q/k}$  is $k$-valued,  whose domain is a
 Riemann surface of $k$ sheets, 
 and we have to pay attention to the transitions from one sheet to another while we draw
  the trajectory.  
  When $k=1$, 
 the Riemann surface is the complex plane itself. 
For irrational 
 $\alpha$,
 function $f_\alpha$  is infinite-valued, 
and the path is not 
closed.

The closure of a trajectory implies the appearance of additional integrals of motion
corresponding to hidden symmetries.
From the structure of the Hamiltonian 
we distinguish the two obvious conserved quantities 
\be
\label{RealInt}
\mathcal{J}_{0}^{(\alpha)}=\frac{1}{2}(b_{\alpha,1}^+b_{\alpha,1}^-+b_{\alpha,2}^+b_{\alpha,2}^-)\,,\qquad
\mathcal{L}^{(\alpha)}=\frac{1}{2}(b_{\alpha,1}^+b_{\alpha,1}^--b_{\alpha,2}^+b_{\alpha,2}^-)=\frac{\alpha}{2}p_\varphi\,,
\ee
in terms of which 
\be\label{HgJ0La}
H_\gamma^{(\alpha)}=2\omega(\mathcal{J}_{0}^{(\alpha)}+\gamma\mathcal{L}^{(\alpha)})\,.
\ee
These integrals 
are well defined for arbitrary values of $\gamma$ and $\alpha$, and so,
 they separately cannot be 
 responsible 
for  the periodicity of the 
orbits. However, they enter into the Hamiltonian
(\ref{HgJ0La}) with a relative weight $\gamma$, and
one can expect  that the case of rational values of $\gamma$ 
could  be special analogously to  the ERIHO system with $\alpha=1$.
Assuming that $\gamma^2<1$ and representing it 
by an  irreducible fraction 
 \be
 \label{gmenos}
 \gamma=(s_2-s_1)/(s_2+s_1)\,,\qquad s_1,s_2=1,2,\ldots\,, 
 \ee
with the help of (\ref{E.Q.O.M}) we  find that the 
 quantities 
 \begin{eqnarray}
&\label{higherorderA}
\mathcal{L}_{\alpha,s_1,s_2}^{+}=(b_{\alpha,1}^+)^{s_1} (b_{\alpha,2}^-)^{s_2}\,,
\qquad \mathcal{L}_{\alpha,s_1,s_2}^{-}=(\mathcal{L}_{\alpha,s_1,s_2}^{+})^*\,,
&
\end{eqnarray}
are conserved in time.
 Indeed, 
 the time dependence of the quantities (\ref{higherorderA}) is 
given by  the exponential factors 
$e^{\pm  i\omega(\ell_1 s_1-\ell_2s_2)t}$, which for 
 rational $\gamma$  of the form  (\ref{gmenos}) reduce  to one. 
 However, the integrals   (\ref{higherorderA}) have the angular dependence 
$e^{\pm i\alpha^{-1}(s_1+s_2)\varphi}$
 and so,  they are not well defined phase space functions in general case. 
Instead of them, in dependence on the value of $\alpha=q/k$,
we can construct the 
integrals
\begin{eqnarray}
\label{Lintegralsande}
\mathscr{L}_{\alpha,s_1,s_2}^{(\epsilon)\pm}=(\mathcal{L}_{\alpha,s_1,s_2}^{\pm})^\epsilon\,,\qquad
\epsilon=
\left\{
\begin{array}{lll}
r & \text{if} 
 \quad
 q=r( s_1+s_2)\,,\quad r=1,2,\ldots,& \\
\\
q & \text{if} \quad q\neq r( s_1+s_2)\,, & 
 \\
\end{array}
\right.
\end{eqnarray}
to be the well defined in the phase space functions.
They
satisfy the Poisson brackets  relations 
\begin{eqnarray}
\label{Palpha1}
&\{\mathcal{J}_{0}^{(\alpha)},\mathscr{L}_{\alpha,s_1,s_2}^{(\epsilon)\pm}\}=
\mp i\frac{\epsilon}{2}(s_1-s_2)\mathscr{L}_{\alpha,s_1,s_2}^{(\epsilon)\pm}\,,\quad
\{\mathcal{L}^{(\alpha)},\mathscr{L}_{\alpha,s_1,s_2}^{(\epsilon)\pm}\}
=\mp i\frac{\epsilon}{2}(s_1+s_2)\mathscr{L}_{\alpha,s_1,s_2}^{(\epsilon)\pm}\,,\qquad&
\\
&\{\mathscr{L}_{\alpha,s_1,s_2}^{(\epsilon)+},
\mathscr{L}_{\alpha,s_1,s_2}^{(\epsilon)-}\}=P_{\alpha,s_1,s_2}(H_\gamma^{(\alpha)},p_\varphi)\,,&
\label{Palpha2}
\end{eqnarray} 
where $P_{\alpha,s_1,s_2}$ is a polynomial of 
the indicated arguments.
Relations (\ref{Palpha1}), (\ref{Palpha2}) together with 
$\{\mathcal{J}_{0}^{(\alpha)},\mathcal{L}^{(\alpha)}\}=0$ 
are identified as a non-linear deformation of the 
$\mathfrak{u}(2)\cong\mathfrak{su}(2)\oplus \mathfrak{u}(1)$ Lie algebra,
where the $\mathfrak{u}(1)$ sub-algebra is generated by   $H_\gamma^{(\alpha)}$.
The $\mathfrak{u}(2)$
algebra  corresponds to 
 the symmetry  of the 
two-dimensional Euclidean isotropic harmonic oscillator with    $\alpha=1$ and $\gamma=0$.

In the case of $\gamma^2<1$ with $\gamma$ of the form (\ref{gmenos}),  
one can also consider the  formal complex quantities 
\begin{eqnarray}
&
\label{higherorderB}
\mathcal{J}_{\alpha,s_1,s_2}^+=(b_{\alpha,1}^+)^{s_1} (b_{\alpha,2}^+)^{s_2}\,,\qquad
 \mathcal{J}_{\alpha,s_1,s_2}^{-}=(\mathcal{J}_{\alpha,s_1,s_2}^{+})^*\,,
&
\end{eqnarray}
which are not conserved but have  
the dependence on time of the form $e^{\pm i\omega(s_1\ell_1+s_2\ell_2)t}$. These objects can be 
promoted to the dynamical integrals by  multiplying them with  the corresponding time dependent inverse
phase factors.  
The angular dependence of these quantities 
is of the form
$e^{\pm i\alpha^{-1}(s_1-s_2)\varphi}$.  
As in the previous case, for rational value of $\alpha$ presented by an irreducible fraction  $\alpha=q/k$, 
the dynamical integrals
\begin{eqnarray}
\label{Jintegralsandd}
\mathscr{J}_{\alpha,s_1,s_2}^{(\delta)\pm}=(\mathcal{J}_{\alpha,s_1,s_2}^{\pm})^\delta\,,\qquad
\delta=
\left\{
\begin{array}{lll}
r' & \text{if}  
\quad
q=r'|s_2-s_1|\,,\quad r'=1,2,\ldots, &\\
\\
q  &\text{if}  
\quad
q\neq r'|s_2-s_1|\,,&
\\
\end{array}
\right.
\end{eqnarray}
are the well defined phase space functions.
These quantities   satisfy the Poisson brackets  relations
\begin{eqnarray}
&\{\mathcal{J}_{0}^{(\alpha)},\mathscr{J}_{\alpha,s_1,s_2}^{(\delta)\pm}\}=
\mp i\frac{\delta}{2}(s_1+s_2)\mathscr{J}_{\alpha,s_1,s_2}^{(\delta)\pm}\,,\quad
\{\mathcal{L}^{(\alpha)},\mathscr{J}_{\alpha,s_1,s_2}^{(\delta)\pm}\}=
\mp i\frac{\delta}{2}(s_1-s_2)\mathscr{J}_{\alpha,s_1,s_2}^{(\delta)\pm}\,,\qquad&
\\
&\{\mathscr{J}_{\alpha,s_1,s_2}^{(\delta)+},\mathscr{J}_{\alpha,s_1,s_2}^{(\delta)-}\}=
Q_{\alpha,s_1,s_2}(H_\gamma^{(\alpha)},p_\varphi)\,,&
\end{eqnarray} 
where $Q_{\alpha,s_1,s_2}$ is a polynomial of its arguments.
This algebra corresponds to 
 a non-linear deformation of the 
$\mathfrak{gl}(2,\R)\cong \mathfrak{sl}(2,\R)\oplus \mathfrak{u}(1)$  
dynamical symmetry.
 We notice here that the $\mathfrak{gl}(2,\R)$ Lie algebra  is the true symmetry 
of the Minkowskian isotropic oscillator described by the rescaled Hamiltonian
$\gamma^{-1}H_\gamma$
 with 
 $\gamma^{-1}=\infty$ $(s_1=s_2=1)$ and $\alpha=1$ \cite{ERIHO}.
 With the help of these relations one can see that
\be
\label{dualH1/g}
\{H_{1/\gamma}^{(\alpha)},\mathscr{J}_{\alpha,s_1,s_2}^{(\delta)\pm} \}=0\,,
\ee
i.e. the quantities (\ref{Jintegralsandd}) are the true integrals for the system $H_{1/\gamma}^{(\alpha)}$ 
with rational  $\gamma^{-1}= (s_2+s_1)/(s_2-s_1)\Rightarrow (\gamma^{-1})^2>1$, and $\alpha=q/k$.
For such system, the quantities (\ref{Lintegralsande}) multiplied by the
corresponding time-dependent factors are the  dynamical integrals of motion.
In correspondence with this, we note that in the limit $\gamma\rightarrow 0$, the quantity 
$\gamma H_{1/\gamma}^{(\alpha)}$ takes the form 
$\omega \alpha p_\varphi=\omega \alpha b_{\alpha,i}^+\eta_{ij} b_{\alpha, j}^-$ 
with $\eta=\text{diag}(1,-1)$, which can be interpreted as the Minkowskian isotropic 
harmonic oscillator  Hamiltonian
in a conical geometry.

For a given rational  value (\ref{gmenos}) of $\gamma$,  
the true integrals $\mathscr{L}_{\alpha,s_1,s_2}^{(\epsilon),\pm}$ and
the dynamical quantities 
$\mathscr{J}_{\alpha,s_1,s_2}^{(\delta),\pm}$  
generate
 a non-linear algebra, which in the Euclidean isotropic case
$\alpha=1$,  $\gamma=0$  reduces to the linear $\mathfrak{sp}(4,\R)$ algebra
\cite{InzPly7,ERIHO}. 
As in the Euclidean case, the transformation $\gamma\rightarrow {1}/{\gamma}$ changes 
the system $H_{\gamma}^{(\alpha)}$ into $H_{1/\gamma}^{(\alpha)}$, and the  
described additional dynamical integrals of the former system  are transmuted into the 
true integrals of motion of the latter,  and vise-versa
in accordance with  equation (\ref{dualH1/g}). 
This means that the inversion  transformation
$\gamma\rightarrow 1/\gamma$ with $\gamma$ given by Eq. (\ref{gmenos})
generates a kind of duality that interchanges the 
role of the additional true and dynamical symmetries.
The 
difference with the Euclidean case here is that the order in momenta of the true and dynamical integrals 
is different in general, and it depends on the values of both parameters 
$\alpha$ and $\gamma$ (more precisely, on the value of $q$ and 
$|s_1\pm s_2|$). 
Then one can hope that the dependence on additional rational parameter $\gamma$ 
could allow us to ``cure"  the problem of the quantum anomaly
in the cosmic string background which occurs  
in the case of the rational values of the parameter $\alpha$ different from integer values
$\alpha=q$. We analyze in detail this issue in the next section. 

With regard to  the dynamical symmetry algebra, we can also construct the complex dynamical integrals 
\be
\label{Jalphapmcla}
\mathcal{J}_{\pm}^{(\alpha)}= e^{\mp 2i\omega t}b_{\alpha,1}^\pm b_{\alpha,2}^\pm\,,
\ee 
which do not have the phase problem. 
Together with real integrals (\ref{RealInt}) they produce the 
$\mathfrak{gl}(2,\R)\cong\mathfrak{sl}(2,\R)\oplus\mathfrak{u}(1)$
 algebra
\begin{eqnarray}
&
\{\mathcal{J}_{0}^{(\alpha)},\mathcal{J}^{(\alpha)}_{\pm}\}=\mp i \mathcal{J}^{(\alpha)}_{\pm}\,,\qquad
\{\mathcal{J}_{\alpha,-},\mathcal{J}^{(\alpha)}_{+}\}=-2i \mathcal{J}_{0}^{(\alpha)}\,, &\\&
\{\mathcal{L}^{(\alpha)},\mathcal{J}_{0}^{(\alpha)}\}=\{\mathcal{L}^{(\alpha)},\mathcal{J}^{(\alpha)}_{\pm}\}=0\,,
&
\end{eqnarray}
where $\mathcal{J}_{0}^{(\alpha)}$, $\mathcal{J}^{(\alpha)}_{+}$ and $\mathcal{J}^{(\alpha)}_{-}$
are generators of the conformal $\mathfrak{sl}(2,\R)$ subalgebra.
As in the case of the free particle,
these generators of the $\mathfrak{sl}(2,\R)$  control the dynamics of  
the radial variable $\rho(t)$. As we will see, they  can be related to generators of conformal symmetry
of the free particle  system by conformal bridge transformation,  
and  play an  important role 
at the quantum level. 

 The case of Landau phases corresponding to $\gamma=1$  ($\gamma=-1$)
 can be obtained from  the  cases $\gamma^2<1$ or $\gamma^2>1$
 with rational $\gamma$  by setting in (\ref{gmenos}) $s_1=0$, $s_2=1$, ($s_1=1$, $s_2=0$).
 In Landau phases,  
the trajectory 
in the $(X_{\alpha,1},X_{\alpha,2})$ plane reduces to a circular path, 
\begin{eqnarray}
&\gamma=1\,: &\qquad
X_{\alpha,+}(t)=X_{\alpha,1}(t)+iX_{\alpha,2}(t)=R_1e^{i\vartheta_1}e^{i2 \omega t}+
R_2e^{-i\vartheta_2}\,,
\\&
\gamma=-1\,:& \qquad
X_{\alpha,+}(t)=X_{\alpha,1}(t)+iX_{\alpha,2}(t)=R_1e^{i\vartheta_1}+
R_2e^{-i\vartheta_2}e^{-i2\omega t}\,.
\end{eqnarray}
When $\gamma=1$ ($\gamma=-1$), $R_1$ ($R_2$) is
the radius and $R_2$ ($R_1$) corresponds to the radial 
coordinate 
 of the center of the  orbit.  
When the radius is bigger (smaller) than the radial position of the center, 
the path encircles (does not encircle)
the origin of the system of coordinates.
When $R_1=R_2$, 
the path
goes through  
 the origin, however this 
case corresponds to the ``falling to the center", which is excluded  from our analysis. 
Using this information, one can draw the trajectory of the particle in the  $(x_1,x_2)$ plane
when $\alpha=q$ by 
following the geometric arguments similar to those 
for the free particle in  conical background of Ref \cite{InzPly7}.

\begin{figure}[H]
\begin{center}
\begin{subfigure}[c]{0.28\linewidth}
\includegraphics[scale=0.3]{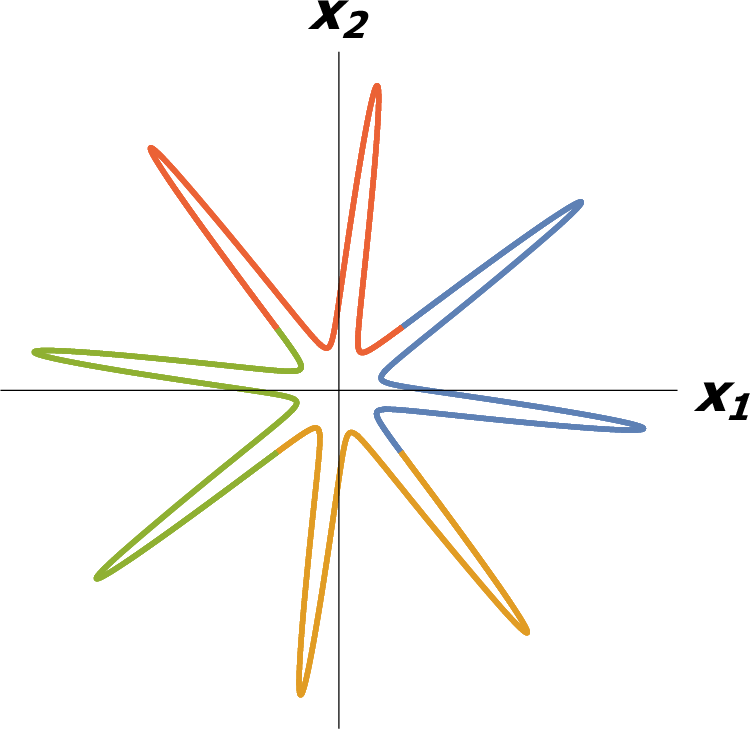}
\caption{\small{$\gamma=0$, $\alpha=1/4$.}}
\end{subfigure}
\begin{subfigure}[c]{0.28\linewidth}
\includegraphics[scale=0.3]{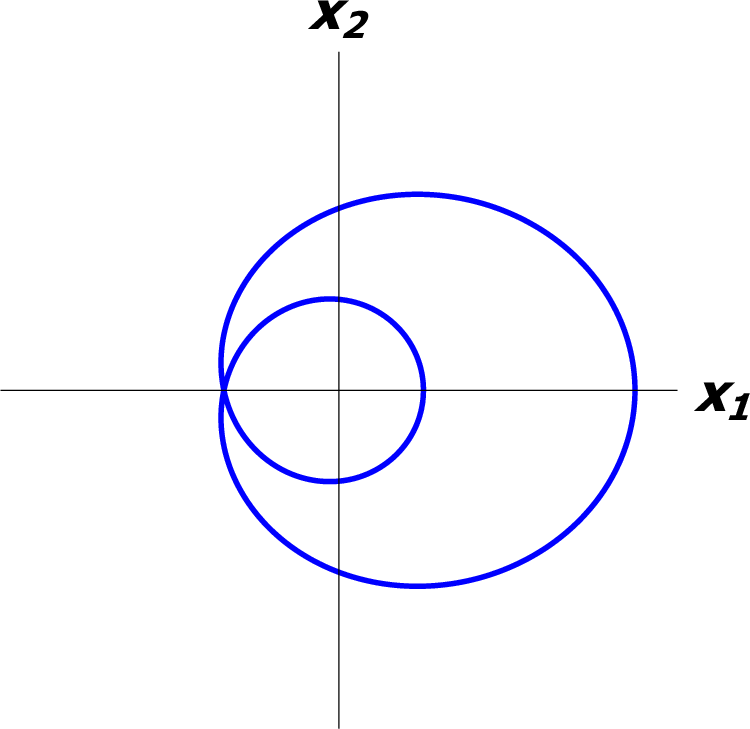}
\caption{\small{$\gamma=0$, $\alpha=4$.}}
\end{subfigure}
\begin{subfigure}[c]{0.28\linewidth}
\includegraphics[scale=0.3]{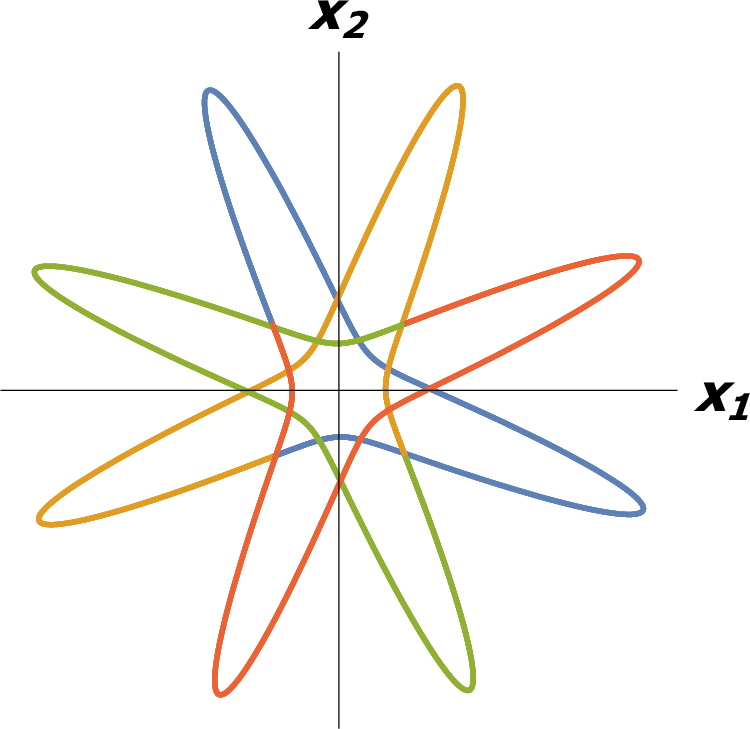}
\caption{\small{$\gamma=0$, $\alpha=3/4$.}}
\end{subfigure}
\vskip0.25cm
\begin{subfigure}[c]{0.28\linewidth}
\includegraphics[scale=0.3]{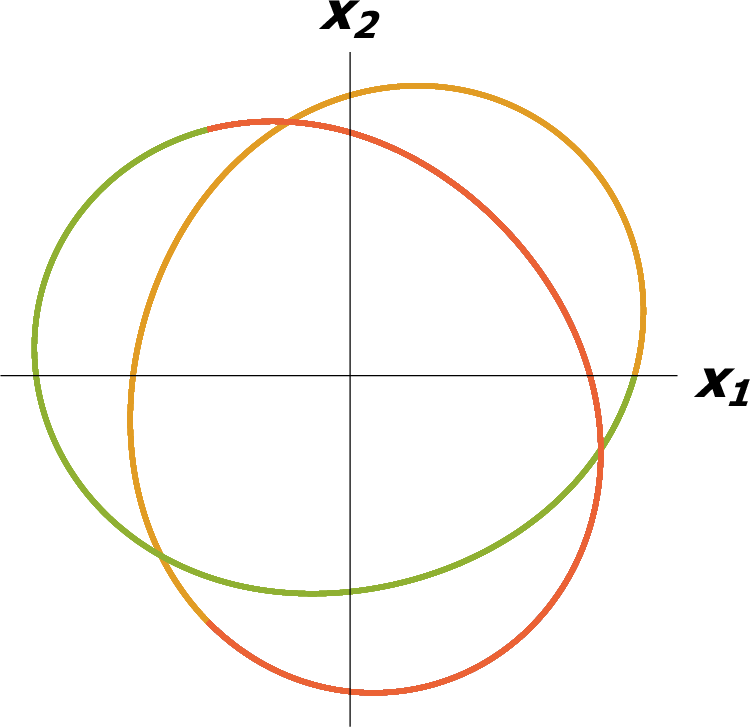}
\caption{\small{$\gamma=1$, $\alpha=2/3$.}}
\label{landau1}
\end{subfigure}
\begin{subfigure}[c]{0.28\linewidth}
\includegraphics[scale=0.3]{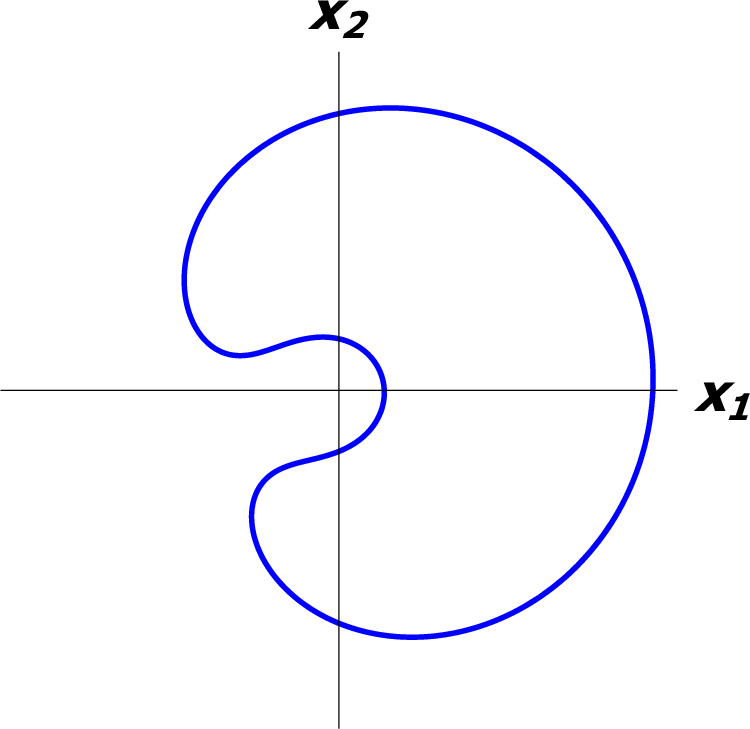}
\caption{\small{$\gamma=1$, $\alpha=3$.}}
\label{landau2}
\end{subfigure}
\begin{subfigure}[c]{0.28\linewidth}
\includegraphics[scale=0.3]{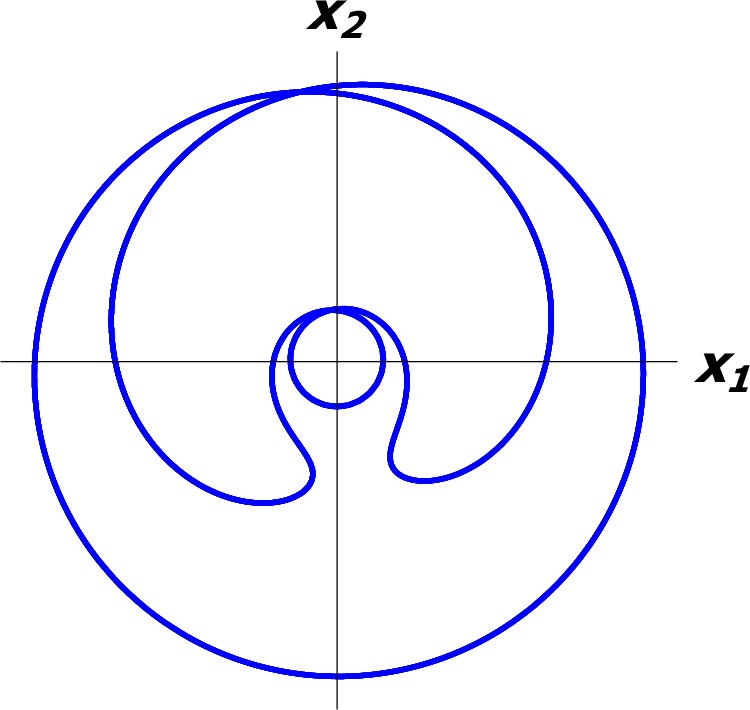}
\caption{\small{$\gamma=1$, $\alpha=7$.}}
\label{landau3}
\label{Fig(f)}
\end{subfigure}
\begin{subfigure}[c]{0.28\linewidth}
\includegraphics[scale=0.3]{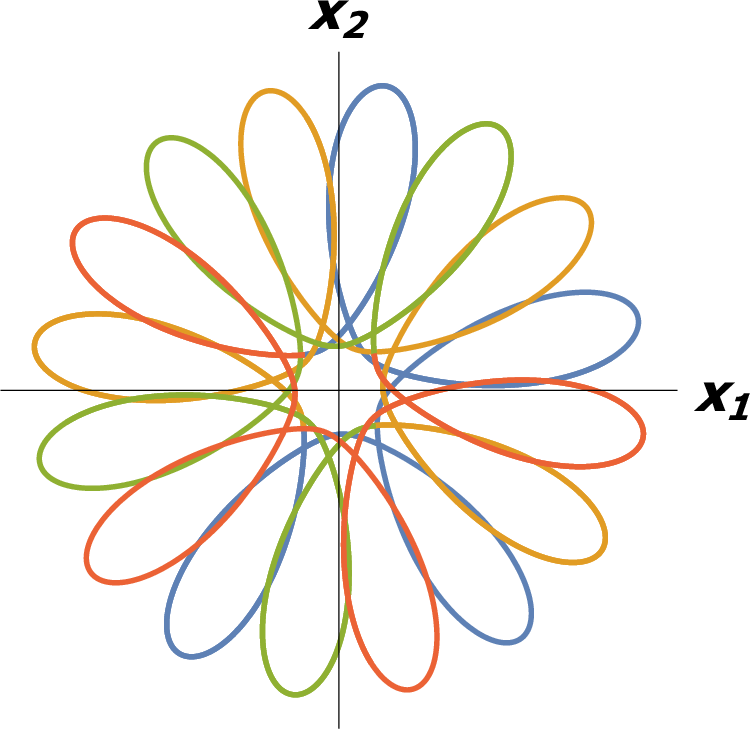}
\caption{\small{$\gamma=1/2$, $\alpha=3/4$.}}
\end{subfigure}
\begin{subfigure}[c]{0.28\linewidth}
\includegraphics[scale=0.3]{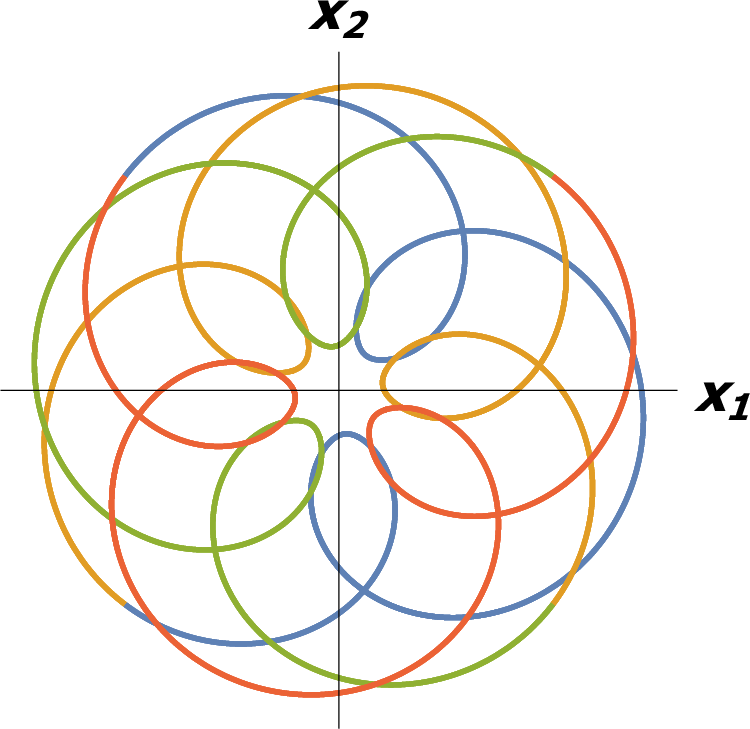}
\caption{\small{$\gamma=2$, $\alpha=3/4$.}}
\end{subfigure}
\begin{subfigure}[c]{0.28\linewidth}
\includegraphics[scale=0.3]{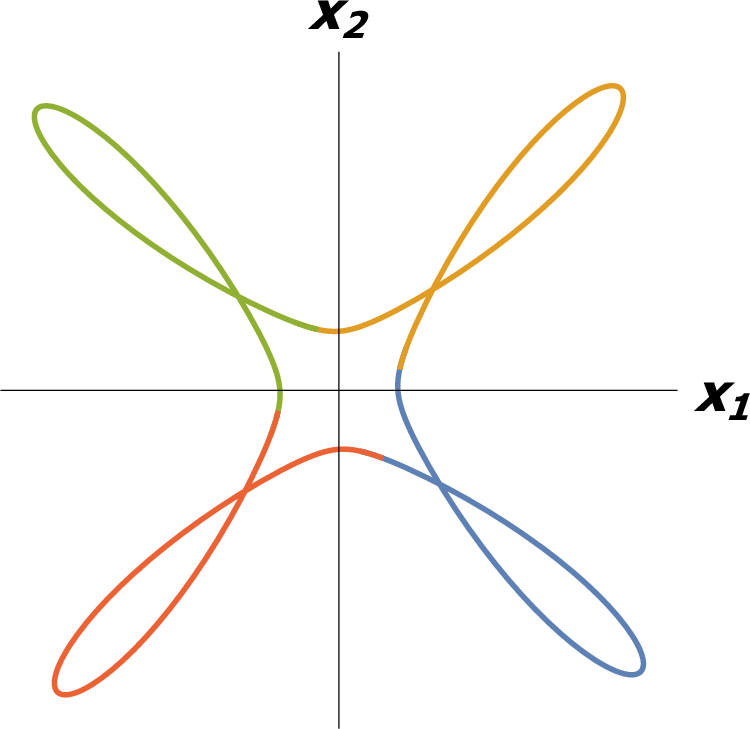}
\caption{\small{$\gamma=1/3$, $\alpha=3/4$.}}
\end{subfigure}
\vskip0.25cm
\begin{subfigure}[c]{0.28\linewidth}
\includegraphics[scale=0.3]{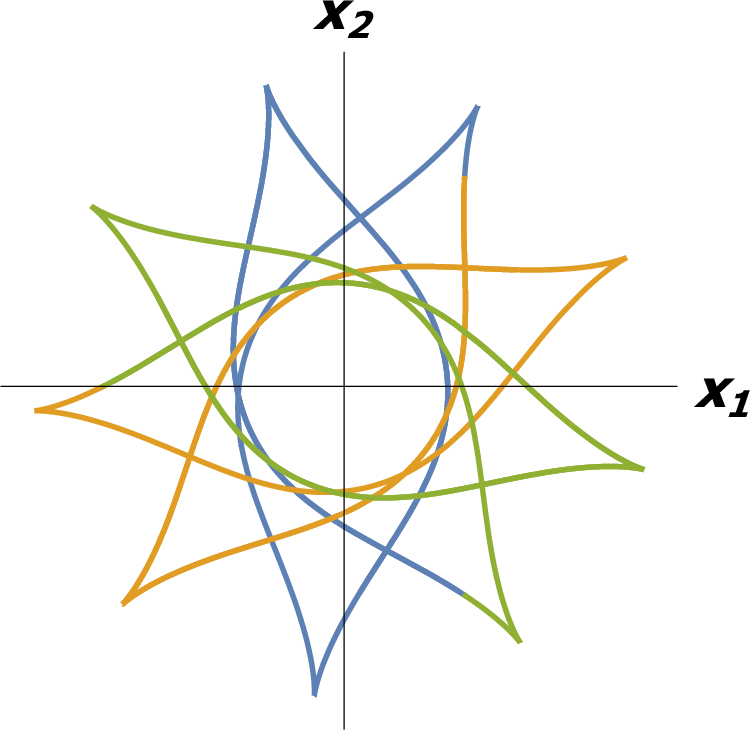}
\caption{\small{$\gamma=1/3$, $\alpha=5/3$.}}
\end{subfigure}
\begin{subfigure}[c]{0.28\linewidth}
\includegraphics[scale=0.3]{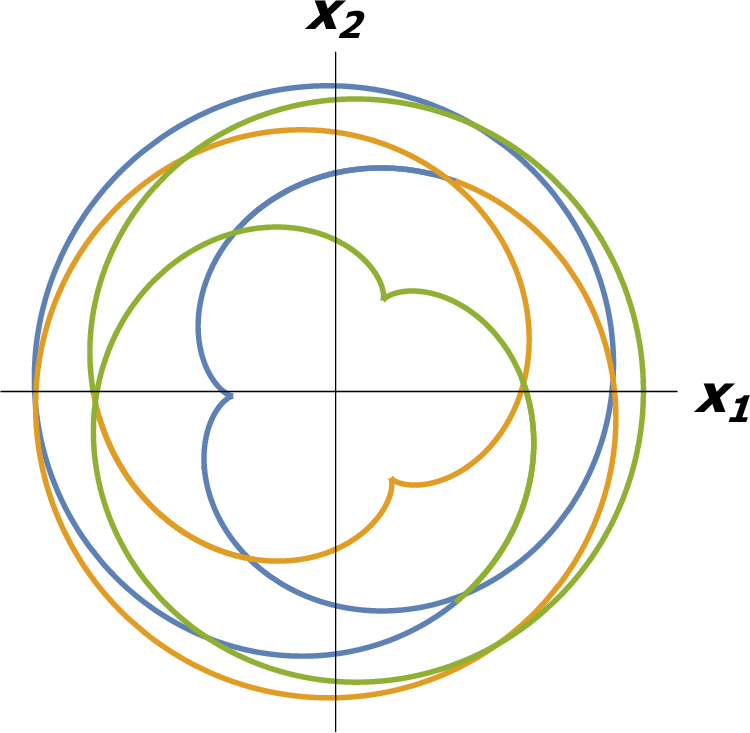}
\caption{\small{$\gamma=3$, $\alpha=5/3$.}}
\end{subfigure}
\begin{subfigure}[c]{0.28\linewidth}
\includegraphics[scale=0.3]{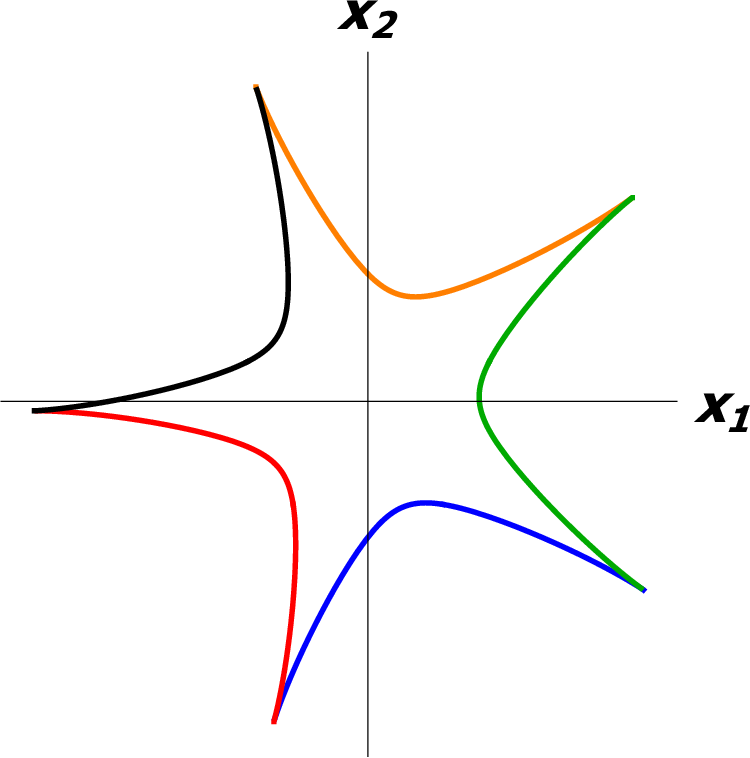}
\caption{\small{$\gamma=1/3$, $\alpha=3/5$.}}
\end{subfigure}
\end{center}
\caption{\small{Trajectories for some values of the parameters $\alpha$ and $\gamma$. 
 Different colours in the same orbit indicate different branches for the angular 
 variable $\varphi(t)$ according to  (\ref{chart}). Figures 
 \ref{landau1}, \ref{landau2} and \ref{landau3} correspond to 
 examples of the Landau phase $\gamma=1$. In Fig.  \ref{landau1}
 the trajectory encloses the origin of the coordinate system (here, 
$R_1>R_2$ and the total change
 for the period  is $\Delta\varphi=4\pi$), contrary to the cases of 
 \ref{landau2} and \ref{landau3} (in which $\Delta\varphi=0$ and $R_1<R_2$).   
 The last three
 figures correspond to the trajectories  with cusps
 that occur in the case  $R_1|\ell_1|=R_2|\ell_2|$.
}}
\label{Fig1}
\end{figure}

\begin{figure}[H]
\begin{center}

\begin{subfigure}[c]{0.45\linewidth}
\includegraphics[scale=0.3]{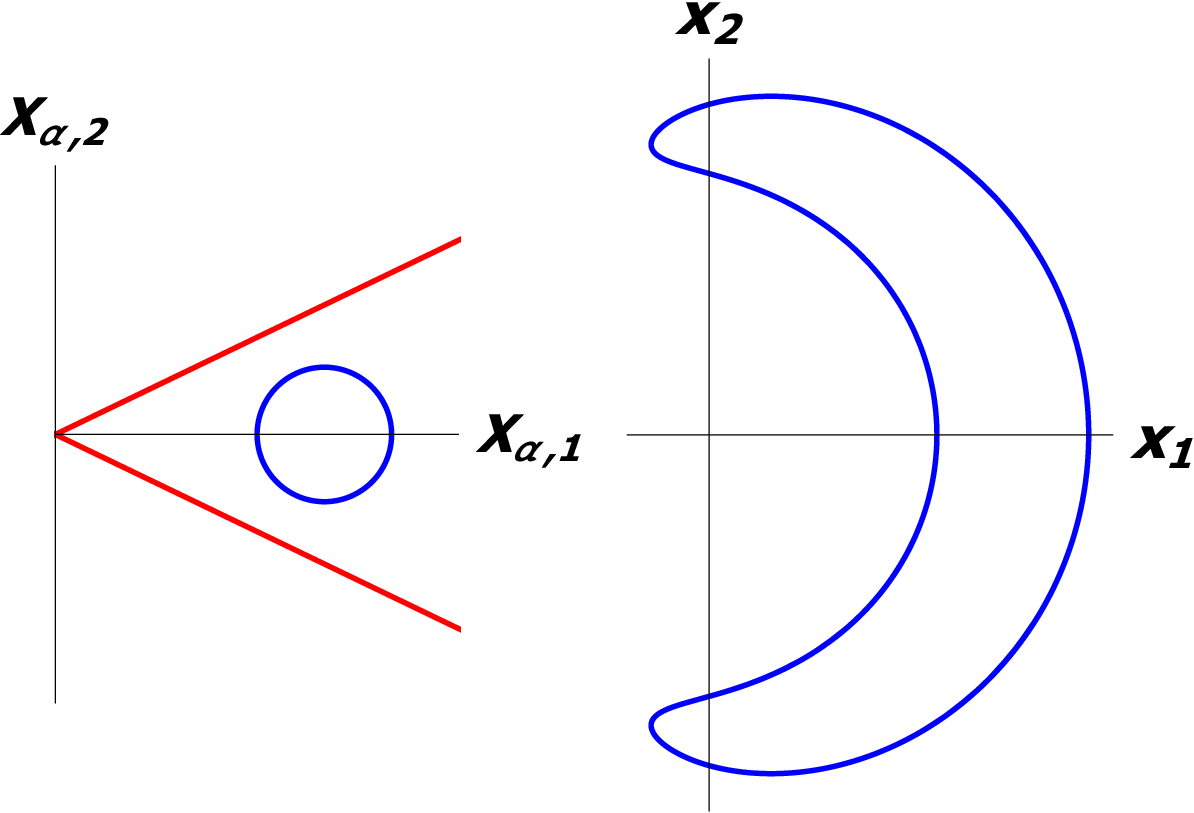}
\caption{\small{$\alpha=7$, $\gamma=1$,  $R_1<R_2\sin(\pi/\alpha)$.}}
\label{FigLandau1}
\end{subfigure}
\begin{subfigure}[c]{0.45\linewidth}
\includegraphics[scale=0.3]{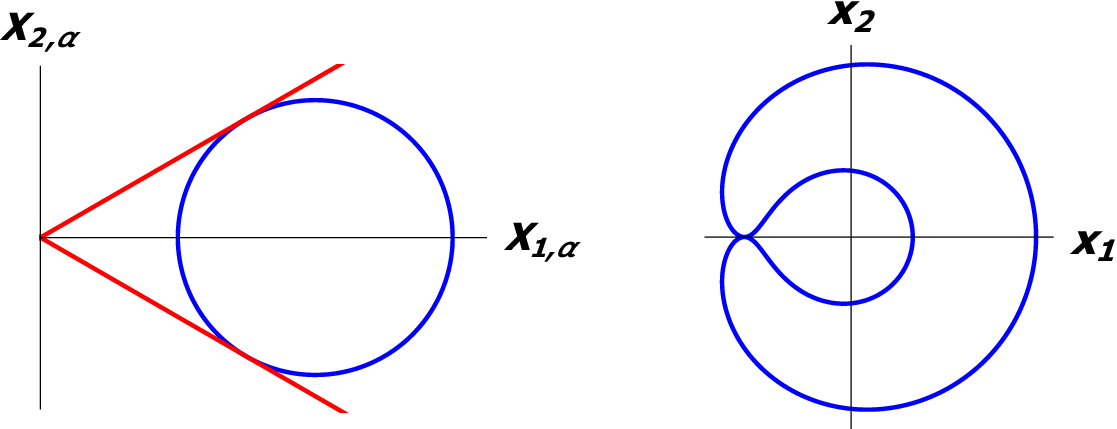}
\caption{\small{$\alpha=7$, $\gamma=1$, $R_1=R_2\sin(\pi/\alpha)$.}}
\label{FigLandau2}
\end{subfigure}
\end{center}
\vskip0.25cm
\hskip1.5cm
\begin{subfigure}[c]{0.6\linewidth}
\includegraphics[scale=0.14]{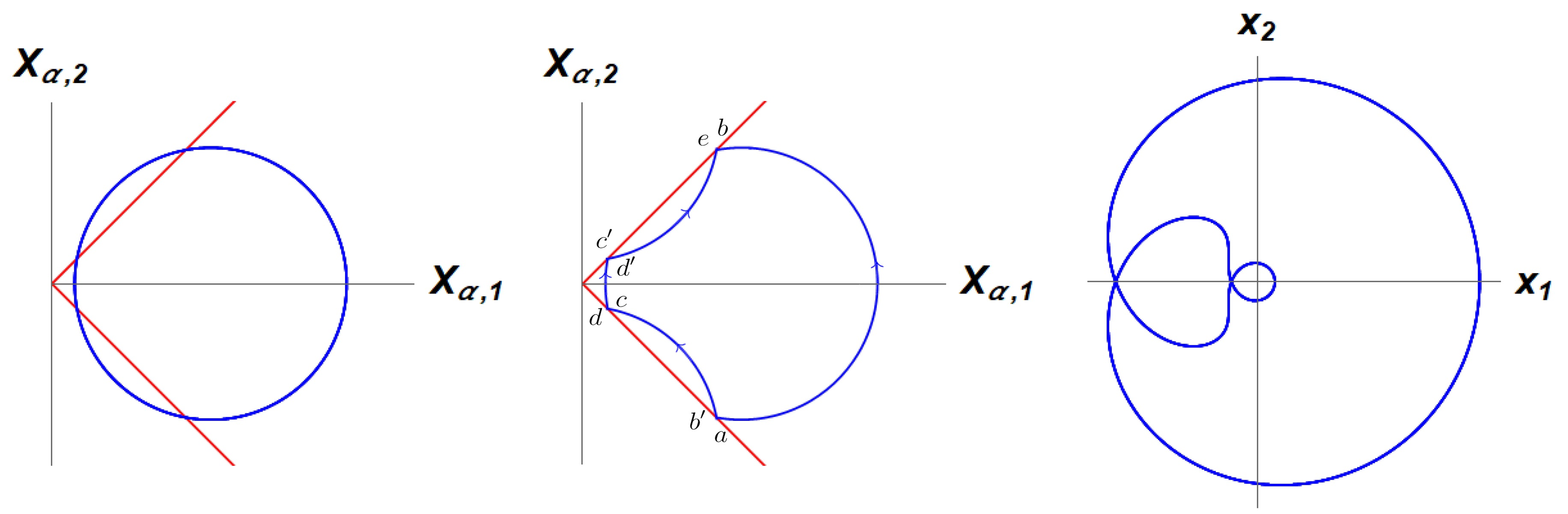}
\caption{\small{$\alpha=4$, $\gamma=1$, $R_2\sin(\pi/\alpha)<R_1<R_2$.}}
\label{FigLandau3}
\end{subfigure}

\caption{\small{
In Figs. \ref{FigLandau1} and \ref{FigLandau2} 
the 
trajectories of the particle in the $(X_{\alpha,1},X_{\alpha,2})$ plane (left) 
and $(x_1,x_2)$ plane (right) are presented.
The two red lines indicate the sector of the angle  $2\pi/\alpha$, which correspond
to the edges of the ``cut and flattened" cone space  that
have to be identified to each other.
In Fig. \ref{FigLandau1} the trajectory   does not touch these lines, and therefore the particle's orbit in
 $(x_1,x_2)$ plane  is a simple loop. In Fig. \ref{FigLandau2} 
 the trajectory  touches the boundaries in the same  (identified)  point, 
 implying that there is a single point    
 in the $ (x_1, x_2) $
  plane through which the particle passes twice within the 
   period of motion. 
 Finally, the trajectory in the plane  $(X_{\alpha,1},X_{\alpha,2})$ 
 crosses  the boundaries 
 in Fig. \ref{FigLandau3} (left), and to construct the real trajectory in the  
``flattened cone" space 
we use  the identity of the edges,
and the velocity vector in corresponding points 
at the edges
 is constructed by a parallel transport.
So when the particle moves from point $ a $ to point $ b $, 
it ``reappears"  at point $ b '$ on the opposite edge,
 and moves in the direction of point $ c $, and so on. 
 As a result, there are two points on the path in the plane $ (x_1, x_2) $ through which the particle passes 
 twice within the period of motion.
 The same logic  has to be employed in the analysis of  the cases in which the origin 
 of the plane $(X_{\alpha,1},X_{\alpha,2})$ is encircled  by the corresponding trajectory.
 }}
\label{FiGrid}
\end{figure}

In this context, 
the complex integrals 
$b_{\alpha,2}^\mp(0)=\sqrt{m\omega}R_2^{\pm i\vartheta_2}$
($b_{\alpha,1}^\pm(0)=\sqrt{m\omega}R_1^{\pm i\vartheta_1}$)
appear as 
 complex linear combinations of the coordinates of the
center of a circle in the plane $(X_{\alpha,1},X_{\alpha,2})$.
As a consequence, 
when $\alpha=q/k$, 
the well defined integrals 
$
\mathscr{L}_{\frac{q}{k},0,1}^{(q),\pm}=(b_{\frac{q}{k},2}^\mp)^{q}$ 
($
\mathscr{L}_{\frac{q}{k},1,0}^{(q),\pm}=(b_{\frac{q}{k},1}^\pm)^{q}$) correspond,
according to the transformation (\ref{chart}),
to the 
image of  this center in 
the $(x_1,x_2)$ plane.
In Fig. \ref{Fig1} some examples of the 
trajectories are shown for different rational values of
the parameters $\alpha$ and $\gamma$. 
Particularly, in Figs. \ref{landau1}, \ref{landau2} and \ref{landau3} we present  different 
examples of the trajectories corresponding to
the system in the Landau phases, which are  characterized 
by different values of the center position of a circular orbit
(in the plane $(X_{\alpha,1},X_{\alpha,2})$) and 
$\alpha$.  Figure \ref{FiGrid} 
illustrates the details corresponding to the construction of the trajectories
in the Landau phase $\gamma=+1$ for different 
integer values of $\alpha$ and different choices of the
constants $R_1$ and $R_2$.

\section{Quantum ERIHO in a conical background}
\label{SecERIHOQuantum}

To solve the quantum problem of the conical ERIHO system, 
we employ the conformal bridge transformation (CBT)  in this
geometry.
Earlier, this technique 
was used 
 in the study of the quantum isotropic harmonic oscillator 
in the cosmic string  spacetime background
\cite{InzPly7}. 
Here, we just very briefly describe the main elements of the construction 
of 
 ref. \cite{InzPly7}, in order 
to apply it then  for the study of 
 the quantum conical ERIHO system.

\subsection{CBT in a cosmic string background}
Consider the quantum version of the conformal symmetry 
generators (\ref{H0alpha}),  (\ref{Halpha0})
and 
(\ref{Halpha1}),  
which at $t=0$ are given  by
\begin{eqnarray}
\label{Qsl(2R)gen}
& \hat{H}^{(\alpha)}_0= -\frac{\hbar^2}{2m}
\left(\frac{1}{\alpha^2\rho}\frac{\partial}{\partial \rho}\left(\rho\frac{\partial}{\partial \rho}\right) +
\frac{1}{\rho^2}\frac{\partial^2}{\partial\varphi^2}\right)\,,\quad 
\hat{D}^{(\alpha)}=\frac{-i\hbar}{2}(\rho\frac{\partial}{\partial \rho}+1)\,,\quad
\hat{K}^{(\alpha)}=\frac{m\alpha^2}{2}\rho^2\,.\quad
&
\end{eqnarray}
They satisfy the  $\mathfrak{so}(2,1)$  algebra,
\be
[\hat{D}^{(\alpha)},\hat{H}^{(\alpha)}_0 ]=i\hbar \hat{H}^{(\alpha)}_0\,,\qquad
[\hat{D}^{(\alpha)},\hat{K}^{(\alpha)}]=i\hbar\hat{K}^{(\alpha)}\,,\qquad
[\hat{K}^{(\alpha)},\hat{H}^{(\alpha)}_0]=i2\hbar\hat{D}^{(\alpha)}\,.
\ee
Introduce the non-unitary operators
\be
\hat{\mathfrak{S}}=e^{-\frac{\omega}{\hbar}\hat{K}^{(\alpha)}}
e^{\frac{1}{2\hbar\omega}\hat{H}^{(\alpha)}_0}e^{\frac{i}{\hbar}\ln(2)\hat{D}^{(\alpha)}}\,,\qquad
\hat{\mathfrak{S}}^{-1}=e^{-\frac{i}{\hbar}\ln(2)\hat{D}^{(\alpha)}}
e^{-\frac{1}{2\hbar\omega}\hat{H}^{(\alpha)}_0} e^{\frac{\omega}{\hbar}\hat{K}^{(\alpha)}}\,.
\ee
They produce a 
similarity transformation, whose application to generators (\ref{Qsl(2R)gen}) yields
\be
\hat{\mathfrak{S}}
(\hat{H}^{(\alpha)}_0,\hat{D}^{(\alpha)},\hat{K}^{(\alpha)})\hat{\mathfrak{S}}^{-1}=
(-\hbar\omega \hat{\mathcal{J}}^{(\alpha)}_{-},-i\hbar\hat{\mathcal{J}}_{0}^{(\alpha)},\hbar\omega^{-1}
\hat{\mathcal{J}}^{(\alpha)}_{+})\,, 
\ee
where 
\begin{eqnarray}
&
\label{JJJ}
\hat{\mathcal{J}}_{0}^{(\alpha)}=\frac{1}{2\omega\hbar}(\hat{H}^{(\alpha)}_0 
+\omega^2\hat{K}^{(\alpha)})\,,\qquad
\hat{\mathcal{J}}^{(\alpha)}_{\pm}=-\frac{1}{2\omega\hbar}(\hat{H}^{(\alpha)}_0  -
\omega^2\hat{K}^{(\alpha)}\pm2i\hat{D}^{(\alpha)})\,
&
\end{eqnarray}
coincide with the dimensionless quantum version of the
 integrals $(\mathcal{J}_{0}^{(\alpha)},\mathcal{J}^{(\alpha)}_{\pm})$ 
taken at $t=0$,  which are given by Eqs.
(\ref{RealInt}) and (\ref{Jalphapmcla}). 
These operators   generate the quantum $\mathfrak{sl}(2,\R)$ conformal algebra 
\be
[\hat{\mathcal{J}}_{0}^{(\alpha)},\hat{\mathcal{J}}^{(\alpha)}_{\pm}]=\pm\hat{\mathcal{J}}^{(\alpha)}_{\pm}\,,\qquad
[\hat{\mathcal{J}}^{(\alpha)}_{-},\hat{\mathcal{J}}^{(\alpha)}_{+}]=2 \hat{\mathcal{J}}_{0}^{(\alpha)}\,.
\ee
The dimensionless quantum version of the integral  $\mathcal{L}^{(\alpha)}$
is 
invariant under this transformation, 
\begin{eqnarray}
&
\hat{\mathfrak{S}}(\hat{\mathcal{L}}^{(\alpha)})\hat{\mathfrak{S}}^{-1}=\hat{\mathcal{L}}^{(\alpha)}\,,\qquad
\hat{\mathcal{L}}^{(\alpha)}=-i\frac{\alpha}{2}\frac{\partial}{\partial \varphi}\,. 
&
\end{eqnarray}
The described  
 \emph{isotropic}  similarity transformation   
in a cosmic string background
is useful 
to solve the quantum problem of an isotropic harmonic oscillator in this geometry \cite{InzPly7}. 
This is because its 
Hamiltonian 
 is given by
\begin{eqnarray}
&
\hat{H}_{\text{osc}}^{(\alpha)}=2\omega\hbar\hat{\mathcal{J}}_{0}^{(\alpha)}=
-\frac{\hbar^2}{2m}
\left(\frac{1}{\alpha^2\rho}\frac{\partial}{\partial \rho}\left(\rho\frac{\partial}{\partial \rho}\right) +
\frac{1}{\rho^2}\frac{\partial^2}{\partial\varphi^2}\right)+\frac{1}{2} m \omega^2 \alpha^2\rho^2\,,
&
\end{eqnarray}
and $\hat{\mathcal{J}}^{(\alpha)}_{\pm}$ are identified as the second-order radial ladder operators. 
To find the corresponding energy eigenstates,  we apply the operator $\hat{\mathfrak{S}}$ to the states 
\be
\Omega_{n_\rho,\ell}^{(\alpha)}=\rho^{2n_\rho+\alpha
|\ell|}e^{i\ell \varphi}\,, \qquad n_\rho=0,2,\ldots\,,\qquad \ell=0,\pm 1,\pm 2\,,\ldots\,,
\ee
which  simultaneously are 
the Jordan states of zero energy of the free particle Hamiltonian, and the formal 
eigenstates of the $\mathcal{PT}$-symmetric operator $2i\hat{D}^{(\alpha)}$ with real eigenvalues \cite{PTrev}. 
These functions  satisfy the  relations
\begin{eqnarray}
&
\hat{H}^{(\alpha)}_0 \Omega_{n_\rho,\ell}^{(\alpha)}=
-2\frac{\hbar^2}{m\alpha^2}n_\rho(n_\rho+\alpha |\ell|)
\Omega_{n_\rho-1,\ell}^{(\alpha)}\,,\qquad
\hat{K}^{(\alpha)}\Omega_{n_\rho,\ell}^{(\alpha)}=\frac{m\alpha^2}{2}
\Omega_{n_\rho+1,\ell}^{(\alpha)}\,,
&\\&
2i\hat{D}^{(\alpha)}\Omega_{n_\rho,\ell}^{(\alpha)}=\hbar(2n_\rho+\alpha|\ell|+1)
\Omega_{n_\rho,\ell}^{(\alpha)}\,,\qquad 
\hat{\mathcal{L}}^{(\alpha)}\Omega_{n_\rho,\ell}^{(\alpha)}=\frac{\alpha\ell}{2}
\Omega_{n_\rho,\ell}^{(\alpha)}\,.
&
\end{eqnarray}
The application of the conformal bridge generator $\hat{\mathfrak{S}}$  from the left to
 these equations  yields 
\begin{eqnarray}
&
\hat{H}_{\text{osc}}^{(\alpha)}\psi_{n_\rho,\ell}^{(\alpha)}=\hbar\omega(2n_\rho+\alpha|\ell|+1)\psi_{n_\rho,\ell}^{(\alpha)}\,,\qquad
\hat{\mathcal{L}}^{(\alpha)}\psi_{n_\rho,\ell}^{(\alpha)}=
\frac{1}{2}\alpha\ell\psi_{n_\rho,\ell}^{(\alpha)}\,,
&\\
&\label{J1}
\hat{\mathcal{J}}^{(\alpha)}_{+}\psi_{n_\rho,\ell}^{(\alpha)}(r,\varphi)=- \sqrt{(n_\rho+1)
(n_\rho+\alpha |\ell|+1)} \psi_{n_\rho+1,\ell}^{(\alpha)}(r,\varphi)\,,&\\&
\hat{\mathcal{J}}^{(\alpha)}_{-}\psi_{n_\rho,\ell}^{(\alpha)}(r,\varphi)=-\sqrt{n_\rho
(n_\rho+\alpha |\ell|)} \psi_{n_\rho-1,\ell}^{(\alpha)} (r,\varphi)\,,\label{J2}&
\end{eqnarray}
where we have used relations
\begin{eqnarray}
&\label{CBTonstates}
\hat{\mathfrak{S}}\Omega_{n_\rho,\ell}^{(\alpha)}(r,\varphi)= \mathcal{N}_{n_\rho,\ell}^{(\alpha)}
\psi_{n_\rho,\ell}^{(\alpha)}(r,\varphi)\,,&\\&
\mathcal{N}_{n_\rho,\ell}^{(\alpha)}=(-1)^{n_\rho}
 \left(\frac{2\hbar}{m\omega \alpha^2}\right)^{n_\rho+\frac{\alpha |\ell|}{2}}\sqrt{\alpha\pi n_\rho!\Gamma(n_\rho+\alpha |\ell|+1)}\,,
&
\end{eqnarray}
and 
\begin{eqnarray}
&\label{EigenstatesHar}
\psi_{n_\rho,\ell}^{(\alpha)}(\rho,\varphi) =\left(\frac{m\omega\alpha^2}{\hbar}\right)^{\frac{1}{2}}
\sqrt{\frac{n_\rho!}{
\pi\alpha\Gamma(n_\rho+\alpha |\ell|+1)}}\,
\zeta^{\alpha |l|}L_{n_\rho}^{(\alpha |\ell|)}(\zeta^2)
e^{-\frac{\zeta^{2}}{2}  + i \ell \varphi}\,,\quad 
\zeta=\sqrt{\frac{m\alpha^2\omega}{\hbar}}\rho\,, \quad&
\end{eqnarray} 
with $L_{n}^{(\nu)}(z)$ to be the generalized Laguerre polynomials.
For more details on application of this technique to other models, see \cite{InzPly7,ERIHO,PTrev}. 
Here, we just take into account 
 that the Hamiltonian of our conical 
ERIHO system is a linear combination of the integrals $\mathcal{J}_{0}^{(\alpha)}$ and $\mathcal{L}^{(\alpha)}$,
the quantum versions of which are simultaneously diagonalized 
by  wave functions  (\ref{EigenstatesHar}). 
\subsection{Quantum conical ERIHO system}

The Hamiltonian operator of the quantum conical ERIHO system is given by
\begin{eqnarray}
\label{QuantumLandauH}
\begin{array}{lll}
\hat{H}_\gamma^{(\alpha)}&=& 2\hbar\omega(\hat{\mathcal{J}}_{0}^{(\alpha)}+\gamma\hat{\mathcal{L}}^{(\alpha)})\\
&=&
-\frac{\hbar^2}{2m}
\left(\frac{1}{\alpha^2\rho}\frac{\partial}{\partial \rho}\left(\rho\frac{\partial}{\partial \rho}\right) +
\frac{1}{\rho^2}\frac{\partial^2}{\partial\varphi^2}\right)+\frac{ m \omega^2 \alpha^2}{2}\rho^2-\omega \gamma \alpha i\hbar \frac{\partial}{\partial \varphi}\,.
\end{array}
\end{eqnarray}
Since the  structure of this operator is a
linear combination of the Hamiltonian of the  isotropic harmonic oscillator in the cone \cite{InzPly7} and the angular momentum operator,
 it is
easy to show that the eigenstates correspond to (\ref{EigenstatesHar}),
 and the spectrum is 
\begin{eqnarray}
&
E_{n_\rho,\ell}^{(\alpha,\gamma)}=\hbar\omega(2n_\rho+\alpha(1 + \gamma\,\text{sign}(\ell))|\ell|+1)\,,
&\\&
n_\rho=0,1,\ldots\,,
\qquad 
\ell=0,\pm 1,\pm2,\ldots\,.\qquad&
\end{eqnarray} 
Indeed,  operator (\ref{QuantumLandauH}) can be presented as 
\be
\hat{H}_\gamma^{(\alpha)}=\hat{\mathfrak{S}}2\omega(i\hat{D}^{(\alpha)}+\gamma\hbar
\hat{\mathcal{L}}^{(\alpha)})\hat{\mathfrak{S}}^{-1}\,,
\ee
and then
 $2\omega(i\hat{D}^{(\alpha)}+\gamma
\hbar\hat{\mathcal{L}}^{(\alpha)})
 \Omega_{n_\rho,\ell}^{(\alpha)}=E_{n_\rho,\ell}^{(\alpha,\gamma)}\Omega_{n_\rho,\ell}^{(\alpha)}$. 
Note that the spectrum depends on the two parameters $\alpha$ and $\gamma$, 
but the eigenfunctions depend only on the parameter $\alpha$. 
When $|\gamma|<1$, 
the spectrum is always positive and bounded from below. 
Meanwhile, the system with  $|\gamma|>1$
has  negative energy levels, and the spectrum is not bounded from below.
In the Landau phases 
$\gamma=\pm 1$, $E\geq \hbar\omega>0$,  
and each energy level 
is infinite degenerate since $(1+\gamma\text{sign}(\ell))$ can vanish for 
an infinite set of different  eigenstates.
 We will return to this point in the next subsection.

We also find  that,  according to equations (\ref{J1}), (\ref{J2}),
 $\hat{\mathcal{J}}^{(\alpha)}_{\pm}$ are still 
radial ladder operators. 
Let us look now  for other 
well defined operators.
To this aim, we introduce a new notation for the eigenstates and the spectrum,
\begin{eqnarray}
&\label{Egennotate}
\psi_{n_\rho,l}^{(\alpha,+)}=\psi_{n_\rho,l}^{(\alpha)}\,,\qquad
\psi_{n_\rho,l}^{(\alpha,-)}=\psi_{n_\rho,-l}^{(\alpha)}\,,\qquad
E_{n_\rho,l}^{(\alpha,\gamma,\pm)}=E_{n_\rho,\pm l}^{(\alpha,\gamma)}\,,\qquad
l=|\ell|\,.
\end{eqnarray}
With this notation, it is clear that the discrete formal transformation   $\sigma: l\rightarrow -l$
produces the functions 
\begin{eqnarray}&
\sigma(\psi_{n_\rho,l}^{(\alpha,\pm)}) =\left(\frac{m\omega\alpha^2}{\hbar}\right)^{\frac{1}{2}}
\sqrt{\frac{n_\rho!}{\pi\alpha\Gamma(n_\rho-\alpha l+1)}}\,
\zeta^{-\alpha l}L_{n_\rho}^{(-\alpha l)}(\zeta^2)
e^{-\frac{\zeta^{2}}{2}  \mp i l \varphi}\,,&
\end{eqnarray}
which 
cannot be physical eigenstates in the general case. 
Only when  $\alpha=q=1,2,\ldots$, the identity 
\be
\label{LagerIdentity}
\frac{(-\eta)^{i}}{i!}L_{n}^{(i-n)}(\eta)=
\frac{(-\eta)^{n}}{n!}L_{i}^{(n-i)}(\eta)\,,\qquad i,n=0,1,\ldots,
\ee
allows us to show that  
\be
\label{sigmatrans}
\sigma(\psi_{n_\rho,l}^{(q,\pm)})=
\left\{
\begin{array}{lll}
(-1)^{lq}\psi_{n_\rho-lq,l}^{(q,\mp)}  & \text{when}& n_\rho\geq q l \\
\\
0 & \text{otherwise } 
\end{array}
\right.\,,\qquad
\sigma(E_{n_\rho,l}^{(q,\gamma,\pm)})=E_{n_\rho-ql,l}^{(q,\gamma,\mp)}\,.
\ee
Therefore, if the action of a certain operator on a particular eigenstate in the physical Hilbert space can
 produce an  eigenstate with negative 
lower second index, we conclude that  such   operator 
is well defined only when $\alpha$ is an integer number. This helps us to see the quantum anomaly.

Now, consider the formal operators
\begin{eqnarray}
&\label{bhatalpha1}
\hat{b}_{\alpha,1}^-=\frac{1}{2}e^{-i\frac{\varphi}{\alpha}}\sqrt{\frac{m \omega }{\hbar }}
\left(
\alpha \rho+
\frac{\hbar }{m\omega\alpha}\left(\frac{\partial}{\partial \rho}
-\frac{i\alpha}{\rho}\frac{\partial}{\partial \varphi} \right)\right)\,, \qquad \hat{b}_{\alpha,1}^+=(\hat{b}_{\alpha,1})^\dagger\,, &\\&
\hat{b}_{\alpha,2}^-= \frac{1}{2}e^{i\frac{\varphi}{\alpha}}\sqrt{\frac{m \omega }{\hbar }}
\left(
\alpha \rho+
\frac{\hbar  }{m\omega\alpha}\left(\frac{\partial}{\partial \rho}
+\frac{i \alpha}{\rho}\frac{\partial}{\partial \varphi}\right) \right) \,, \qquad \hat{b}_{\alpha,2}^+=(\hat{b}_{\alpha,2}^-)^\dagger\,.
&\label{bhatalpha2}
\end{eqnarray}
They are obtained via the conformal bridge transformation, 
\begin{eqnarray}
&\label{CBTinb}
\mathfrak{\hat{S}}
 (\hat{\Xi}_+^{(\alpha)}, \hat{\Xi}_-^{(\alpha)},
 \hat{\Pi}_+^{(\alpha)},\hat{\Pi}_-^{(\alpha)})\mathfrak{\hat{S}}^{-1}=  
(\sqrt{\frac{2m\hbar}{\omega}}\hat{b}_{\alpha,1}^+,\sqrt{\frac{2m\hbar}{\omega}}
\hat{b}_{\alpha,2}^+,-i\sqrt{2m\hbar\omega}\hat{b}_{\alpha,2}^-,-i\sqrt{2m\hbar\omega}\hat{b}_{\alpha,1}^-)\,,\qquad 
&
\end{eqnarray}
applied to the free particle formal momenta
$\hat{\Pi}_\pm^{(q)}$ from  (\ref{Pi+-}) 
with $q$ changed for $\alpha$,  and Galilean boost
generators  taken at $t=0$,
see ref. \cite{InzPly7},
\begin{eqnarray}&
\label{Xi}
 \hat{\Xi}_\pm^{(\alpha)}=\alpha m \rho e^{i\frac{\varphi}{\alpha}}\,.&
\end{eqnarray}
To find the action  of operators (\ref{bhatalpha1}) and  (\ref{bhatalpha2})  and  some  powers of them
 on the eigenstates $\psi_{n_\rho,\l}^{(\alpha,\pm)}$, we first compute the action of (\ref{Pi+-}) and (\ref{Xi}) 
 on functions $\Omega_{n_\rho,\ell}^{(\alpha)}$, 
and then we use the relations (\ref{CBTonstates}) and (\ref{CBTinb}). As a result we get 
\begin{eqnarray}
&
(\hat{b}_{\alpha,1}^{\pm})^j\,\psi_{n_\rho,l}^{(\alpha,+)}(\rho,\varphi)=
\sqrt{\frac{\Gamma(n_\rho+\alpha l +1 +\beta_\pm j)}{\Gamma(n_\rho+\alpha l+1 -\beta_\mp j)}}\,
\psi_{n_\rho,l\pm \frac{j}{\alpha}}^{(\alpha,+)}(\rho,\varphi)\,,\qquad \beta_\pm=\frac{1\pm 1}{2}\,,&\\&
(\hat{b}_{\alpha,1}^{\pm})^j\,\psi_{n_\rho,l}^{(\alpha, -)}(\rho,\varphi)=
(-1)^j\sqrt{\frac{\Gamma(n_\rho+1+\beta_\pm j)}{\Gamma(n_\rho+1-\beta_\mp j)}}\,
\psi_{n_\rho\pm j,l\mp \frac{j}{\alpha}}^{(\alpha,-)}(\rho,\varphi)\,,
&\\&
(\hat{b}_{\alpha,2}^{\pm})^j\,\psi_{n_\rho,l}^{(\alpha,-)}(\rho,\varphi)=
\sqrt{\frac{\Gamma(n_\rho+\alpha l +1 +\beta_\pm j)}{\Gamma(n_\rho+\alpha l+1 -\beta_\mp j)}}\,
\psi_{n_\rho,l\pm \frac{j}{\alpha}}^{(\alpha,-)}(\rho,\varphi)\,,&\\&
(\hat{b}_{\alpha,2}^{\pm})^j\,\psi_{n_\rho,l}^{(\alpha, +)}(\rho,\varphi)=
(-1)^j\sqrt{\frac{\Gamma(n_\rho+1+\beta_\pm j)}{\Gamma(n_\rho+1-\beta_\mp j)}}\,
\psi_{n_\rho\pm j,l\mp \frac{j}{\alpha}}^{(\alpha,+)}(\rho,\varphi)\,.&
\end{eqnarray}
It is clear that when $j=1$, we obtain a function outside 
the Hilbert space in general case since the phase factor is multi-valued. 
When $\alpha=q/k$, we can select $j=q$ in these equations in order  to avoid problems with the phase factor, 
but we still have the problem of quantum anomaly
since when $k>l$, we can obtain the wave functions 
$$
\psi_{n_\rho,-|k-l|}^{(\frac{q}{k},\pm)}(\rho,\varphi)=\sigma(\psi_{n_\rho,|k-l|}^{(\frac{q}{k},\mp)}(\rho,\varphi))\,
$$
 on the right hand side of the corresponding equation.
They will be  physical eigenstates only when $k=1$, see Eq. (\ref{sigmatrans}). 

The action of operators $\hat{\mathcal{J}}^{(\alpha)}_{\pm}=\hat{b}_{1,\alpha}^{\pm}\hat{b}_{2,\alpha}^{\pm}$ 
(that is an  equivalent form of the second definition in  (\ref{JJJ}))
on eigenstates $\psi_{n_\rho,l}^{(\alpha,\pm)}$
produces a change of the radial quantum number $n_\rho$ (without touching $l$) for arbitrary values of $\alpha$. Besides, 
exceptionally   when $\alpha=q$, we have 
the well defined ladder operators $(\hat{b}_{q,i}^{\pm})^q$, which
allow to change the angular momentum quantum number. 
In conclusion, only in this last case we can have a spectrum generating set of the
ladder operators $(\hat{\mathcal{J}}^{(q)}_{\pm},(\hat{b}_{q,i}^{\pm})^q)$, whose combined action
allows us to generate from a given eigenstate all the 
physical eigenstates 
 that constitute a basis of the physical Hilbert space. 
On the other hand, operators $\hat{b}_{\alpha,i}^{\pm}$ are the building blocks for  the construction of any 
symmetry operator that correctly explains the emergent degeneracy of the spectrum 
in dependence on the values of the parameters $\alpha$ and $\gamma$. According to our
previous analysis,
this construction is only possible  
when $\alpha$  takes a positive integer value, that we imply from now on. 

Until now,
the construction is valid for arbitrary values of the parameter $\gamma$.

\subsection{Spectral degeneracy}

{\it Case $\gamma=\frac{s_2-s_1}{s_2+s_1}$}.
The quantum version of (\ref{Lintegralsande}) corresponds to the operators
\be
\label{QL}
\hat{\mathscr{L}}_{q,s_1,s_2}^{(\epsilon)+}=(\hat{b}_{1,q}^+)^{s_1\epsilon}(\hat{b}_{2,q}^-)^{s_2\epsilon}\,,\qquad
\hat{\mathscr{L}}_{q,s_1,s_2}^{(\epsilon)-}=(\hat{\mathscr{L}}_{q,s_1,s_2}^{(\epsilon)+})^\dagger\,. 
\ee
They are  obtained via conformal bridge transformation (up to an inessential
multiplicative constant),
from the higher order 
free particle symmetry operators 
\be
\hat{\mathscr{S}}_{q,s_1,s_2}^{(\epsilon)+}=(\hat{\Xi}_+^{(q)})^{s_1\epsilon}(\hat{\Pi}_+^{(q)})^{s_2\epsilon}\,,\qquad 
\hat{\mathscr{S}}_{q,s_1,s_2}^{(\epsilon)-}=(\hat{\Pi}_-^{(q)})^{s_1\epsilon}(\hat{\Xi}_-^{(q)})^{s_2\epsilon}\,,
\ee
which commute with the combination $2\omega(i\hat{D}^{(q)}+\gamma
\hbar \hat{\mathcal{L}}^{(q)})$ for the 
chosen value of the  parameter  $\gamma=(s_2-s_1)/(s_2+s_1)$. 

The action of these operators is given by 
\begin{eqnarray}
&\label{Lhatonstates1}
\hat{\mathscr{L}}_{q,s_1,s_2}^{(\epsilon)\pm}\psi_{n_\rho,l}^{(q,+)}=
(-1)^{\epsilon s_2}\sqrt{\frac{\Gamma(n_\rho+1+\beta_\mp \epsilon s_2)}{\Gamma(n_\rho+1-\beta_\pm\epsilon s_2)}
\frac{\Gamma(n_\rho+q l +1+\beta_\pm \epsilon s_1)}{\Gamma(n_\rho+q l+1-\beta_\mp\epsilon s_1)}}\psi_{n_\rho\mp \epsilon s_2,l
\pm\frac{\epsilon(s_1+s_2)}{q}}^{(q,+)}\,,
&\\&
\hat{\mathscr{L}}_{q,s_1,s_2}^{(\epsilon)\mp}\psi_{n_\rho,l}^{(q,-)}=
(-1)^{\epsilon s_1}\sqrt{\frac{\Gamma(n_\rho+1+\beta_\mp \epsilon s_1)}{\Gamma(n_\rho+1-\beta_\pm\epsilon s_1)}
\frac{\Gamma(n_\rho+q l +1+\beta_\pm\epsilon s_2)}{\Gamma
(n_\rho+q l+1-\beta_\mp \epsilon s_2)}}\psi_{n_\rho\mp \epsilon s_1,l\pm
\frac{\epsilon(s_1+s_2)}{q}}^{(q,-)}\,.
&
\label{Lhatonstates2}
\end{eqnarray}
In accordance with Eq. (\ref{Lintegralsande}),
the value of the parameter $\epsilon$ guarantees that  $\epsilon/q$ is integer. 
We conclude then  that 
these operators are responsible 
for the degeneracy of the system since the 
energy eigenvalues  satisfy the relations
\begin{eqnarray}
&
E_{n_\rho,l}^{ (q, \gamma,+)}=E_{n_\rho-\epsilon s_2,l+\frac{\epsilon(s_1+s_2)}{q}}^{ (q, \gamma,+)}=
E_{n_\rho+\epsilon s_2,l-\frac{\epsilon(s_1+s_2)}{q}}^{ (q, \gamma,+)}\,,&\\&
E_{n_\rho,-l}^{  (q, \gamma,-)}=E_{n_\rho-\epsilon s_1,-l+\frac{\epsilon(s_1+s_2)}{q}}^{ (q, \gamma,-)}=
E_{n_\rho+\epsilon s_1,l-\frac{\epsilon(s_1+s_2)}{q}}^{ (q, \gamma,-)}\,.
&
\end{eqnarray}
The poles of the Gamma functions and relation (\ref{sigmatrans}) guarantee 
 that only physical eigenstates are produced by the action of these operators. 

When $\gamma=1$ ($s_1=s_2=1$), 
equations (\ref{Lhatonstates1}), (\ref{Lhatonstates2}) are reduced to equations (6.57), (6.58) of \cite{InzPly7} when 
$q$ is even ($\epsilon=q/2$), and to (6.59), (6.60) when $q$ is odd ($\epsilon=q$). 
On the other hand, when $q=1$ ($\epsilon=1$), the integrals (\ref{QL}) are reduced to 
the integrals of the Euclidean quantum ERIHO system described in ref. \cite{ERIHO}. 
Precisely, another way to obtain the integrals
 (\ref{QL}) is first applying the quantum version of the canonical transformation (\ref{canonical}) 
 to the integrals of the Euclidean case, 
 and then taking the power $\epsilon$ of the resulting objects, 
 just as it was done at the classical level.

\vskip0.25cm
\noindent
{\it Case $\gamma=\frac{s_2+s_1}{s_2-s_1}$}. 
The quantum version of (\ref{Lintegralsande}) corresponds to 
\be
\hat{\mathscr{J}}_{q,s_1,s_2}^{(\delta)+}=(\hat{b}_{1,q}^+)^{s_1\delta}(\hat{b}_{2,q}^+)^{s_2\delta}\,,\qquad
\hat{\mathscr{J}}_{q,s_1,s_2}^{(\delta)-}=(\hat{\mathscr{J}}_{q,s_1,s_2}^{(\delta)+})^\dagger\,. 
\ee
These operators  are obtained (up to a numerical factor) 
after the application of the conformal bridge transformation  to the higher order operators 
\be
\hat{\mathscr{T}}_{q,s_1,s_2}^{(\delta)+}=(\hat{\Xi}_{+}^{(q)})^{s_1\delta}(\hat{\Xi}_-^{(q)})^{s_2\delta}\,,\qquad 
\hat{\mathscr{T}}_{q,s_1,s_2}^{(\delta)-}=(\hat{\Pi}_{-}^{(q)})^{s_1\delta}(\hat{\Pi}_+^{(q)})^{s_2\delta}\,,
\ee
which commute with $2\omega(i\hat{D}^{(q)}+\gamma\hbar\hat{\mathcal{L}}^{(q)})$ for the chosen 
value of $\gamma=(s_2+s_1)/(s_2-s_1)$. In the same way as for the previous case, 
these integrals can also be obtained from their version in the Euclidean case $(\alpha=1)$ \cite{ERIHO}
by the application of the quantum version of the canonical transformation (\ref{canonical})
and by taking then  the power $\epsilon$ of the resulting objects.

Their action on eigenstates is 
\begin{eqnarray}
&
\hat{\mathscr{J}}_{q,s_1,s_2}^{(\delta)\pm }\psi_{n_\rho,l}^{(q,+)}=
(-1)^{\delta s_2}\sqrt{\frac{\Gamma(n_\rho+1+\beta_\pm \delta s_2)}{\Gamma(n_\rho+1-\beta_\mp\delta s_2)}
\frac{\Gamma(n_\rho+q l +1+\beta_\pm\delta s_1)}{\Gamma(n_\rho+q l+1-\beta_\mp \delta s_1)}}\psi_{n_\rho\pm\delta s_2,l\mp 
\frac{\delta(s_2-s_1)}{q}}^{(q,+)}\,,
&\\&
\hat{\mathscr{J}}_{q,s_1,s_2}^{(\delta)\pm}\psi_{n_\rho,l}^{(q,-)}=
(-1)^{\delta s_1}\sqrt{\frac{\Gamma(n_\rho+1+\beta_\pm \delta s_1)}{\Gamma(n_\rho+1-\beta_\mp\delta s_1)}
\frac{\Gamma(n_\rho+q l +1+\beta_\pm\delta s_2)}{\Gamma(n_\rho+q l+1-\beta_\mp \delta s_2)}}
\psi_{n_\rho\pm 
\delta s_1,l\pm \frac{\delta(s_2-s_1)}{q}}^{(q,-)}\,.
&
\end{eqnarray}
Again, $\delta/q$ is an integer number here, 
and these operators reflect  the spectral degeneracy according to 
the relations
\begin{eqnarray}
&
E_{n_\rho,l}^{ (q, \gamma,+)}=E_{n_\rho+\delta s_2,l-\frac{\delta(s_2-s_1)}{q}}^{( q, \gamma,+)}=
E_{n_\rho-\delta s_2,l+\frac{\delta(s_2-s_1)}{q}}^{( q, \gamma,+)}\,,&\\&
E_{n_\rho,-l}^{ (q, \gamma,-)}=E^{(q, \gamma,-)}_{n_\rho+\delta s_1,l+\frac{\delta(s_2-s_1)}{q}}
=E^{ (q, \gamma,-)}_{n_\rho-\delta s_1,l-\frac{\delta(s_2-s_1)}{q}}\,.
&
\end{eqnarray}
Here, the poles in the Gamma function appear only in the action of the operator 
$\hat{\mathscr{J}}_{q,s_1,s_2}^{(\delta)-}$, 
that reflects
the infinite degeneracy of the system.

Now let us see  in detail how this 
scheme  works in the case of the Landau phases 
$\gamma=\pm 1$.  
First, in the case   $\gamma=-1$ ($s_1=1, s_2=0, \epsilon=q$) 
the equations (\ref{Lhatonstates1}), (\ref{Lhatonstates2})
reduce to 
\begin{eqnarray}
&\label{Lhatonstates3}
\hat{\mathscr{L}}_{q,1,0}^{(q)\pm }\psi_{n_\rho,l}^{(q,-)}=
(-1)^{q}\sqrt{\frac{\Gamma(n_\rho+1+q\beta_\pm)}{\Gamma(n_\rho+1-q\beta_\mp)}
}
\psi_{n_\rho\pm q,l\mp1}^{(q,-)}\,,
&\\&
\hat{\mathscr{L}}_{q,1,0}^{(q)\pm}\psi_{n_\rho,l}^{(q,+)}=\sqrt{
\frac{\Gamma(n_\rho+q (l+\beta_\pm) +1 )}{\Gamma(n_\rho+q (l-\beta_\mp)+1)}}
\psi_{n_\rho,l\pm 1}^{(q,+)}\,,&
\label{Lhatonstates4}
\end{eqnarray}
while for   $\gamma=1$ ($s_1=0, s_1=1, \epsilon=q$), we get 
\begin{eqnarray}
&\label{Lhatonstates5}
\hat{\mathscr{L}}_{q,0,1}^{(q)\pm }\psi_{n_\rho,l}^{(q,+)}=
(-1)^{q}\sqrt{\frac{\Gamma(n_\rho+1+q\beta_\mp)}{\Gamma(n_\rho+1-q\beta_\pm)}
}
\psi_{n_\rho\mp q,l\pm1}^{(q,+)}\,,
&\\&
\hat{\mathscr{L}}_{q,0,1}^{(q)\pm}\psi_{n_\rho,l}^{(q,-)}=\sqrt{
\frac{\Gamma(n_\rho+q (l+\beta_\mp) +1 )}{\Gamma(n_\rho+q (l-\beta_\pm)+1)}}
\psi_{n_\rho,l\mp 1}^{(q,-)}\,.&
\label{Lhatonstates6}
\end{eqnarray}
Then, for the case $\gamma=-1$ ($\gamma=1$), equation (\ref{Lhatonstates4})
(Eq. (\ref{Lhatonstates6})) shows us that 
the energy levels are infinitely degenerate, since the operator $\hat{\mathscr{L}}_{q,1,0}^{(q)+}$
($\hat{\mathscr{L}}_{q,0,1}^{(q)-}$) 
works 
as the raising
ladder operator for the quantum number $l$
when it acts on the states of
 the form $\psi_{n_\rho,l}^{(q,+)}$ ($\psi_{n_\rho,l}^{(q,-)})$ without annihilating any of them.

In this section  
we applied 
the conformal bridge 
transformation 
to identify the special characteristics of the integrals 
of the quantum conical ERIHO system in a 
 direct way.
In spite of our initial motivation  to solve the anomaly problem 
by introducing an additional parameter into  the dynamics,  
 the model still  carries 
the same quantum anomaly problem in 
the conical background in the case of rational non-integer values of
$\alpha$ \cite{InzPly7}.
We observed 
that there is an interplay between the 
geometric conical parameter $\alpha$ 
and the real parameter $ \gamma $, 
which characterizes the ratio of frequencies 
associated 
with the  isotropic harmonic potential term 
and 
the rotational speed of the corresponding cosmic string spacetime. 
This interplay is revealed first in the quantum spectrum of the model, 
that degenerates only in the case when 
 both parameters are rational and, 
secondly, in the integrals of motion, whose definition depends on both 
parameters,  and which can only be constructed when $ \alpha $ is a positive integer.
In the special case of $\gamma^2=1$ we obtained  the two quantum Landau phases in the conical 
background,
and identified their corresponding integrals of motion.
As these phases are  generated  by application of the conformal 
bridge transformation to the free particle in the conical background,
 the  quantum Landau problem on the cone is anomaly free 
only in the same cases of integer values of the parameter $\alpha=q$.

\section{
A  multi-particle generalization: vortices}\label{Vessel}

It is a theoretical and 
experimental fact that   
the  two dimensional 
 multi-particle bosonic
systems confined by (isotropic or anisotropic) harmonic  traps in a rotating 
reference frame 
can experience 
Bose-Einstein condensation at low-temperatures
\cite{Cooper,ALF}. In the case of isotropic harmonic traps, 
the Hamiltonian of such a system has the form 
of the   sum  
 of $N$ planar ERIHO Hamiltonians plus an interaction 
term which can represent the particles collision or a sort of electromagnetic interaction between them.
The condensation
occurs 
when this interaction potential is weak and 
the  harmonic trap  is stronger than the centrifugal force, i.e., when $|\gamma|\leq 1$ \cite{Cooper}. 
A special characteristics of the corresponding condensate wave function 
is 
 the emergence 
of quantum vortices, 
which in dependence on  
the form of the interaction term are expected to be distributed 
in regular lattices  
 \cite{Cooper,ALF}.  
Having noted that the shape of the classical and quantum solutions of the ERIHO 
system depends essentially on the geometric    
background  characterized by the cone
 parameter $\alpha$,
 it is interesting to see if there 
is any effect of it
at the level of   multi-particle   
systems. In particular, 
we would like to see what happens with the formation of vortices in the wave
 function of the condensate in this space. To this aim,  
  in this section 
 we first briefly  
 review
  the elements of Gross-Pitaevskii formalism to 
 apply it to  the condensate 
 wave function 
 in  conical geometry. Then,
 we analyze in detail the simplest case $\gamma^2=1$, 
 corresponding  to the Landau phases of the conical ERIHO model.

Let us start with the bosonic  Hamiltonian of the multi-particle conical ERIHO system 
\be
\hat{\mathcal{H}}=\sum_{n=1}^{N}H_{\gamma}^{(\alpha)}(\vr_i)+\mathcal{U}(\vr_1,\ldots, 
\vr_N)\,,\qquad \mathcal{U}(\vr_1,\ldots, \vr_N)=\sum_{i<j}
U(\vr_i,\vr_j)\,,
\ee 
where we have introduced a pairwise interaction between particles.
It could  be chosen,  
 for example, in the form of 
 the contact potential derived in the 
Hartree-Fock approximation for alkali  gases \cite{Leggett,Huang},
 \be
 \label{HFpotential}
 U(\vr_i,\vr_j)=\frac{4\pi\hbar a_s}{m}\delta(\vr_i-\vr_j)\,,
 \ee
 which we assume from now on. 
Here, $m$ is interpreted as
 the reduced mass and the scattering length parameter $a_s$  
  is assumed to be small \cite{Leggett,ALF,Huang}. 

To analyze this system we follow the Gross-Pitaevskii formalism \cite{Leggett,ALF,Huang,Cooper}. 
The first step is to perform a second quantization, obtaining as a result 
the Hamiltonian  
 \be
\hat{\mathcal{H}}=\int dV \hat{\Psi}^{\dagger}\left(\hat{H}_{\gamma}^{(\alpha)}(\vr)\right) \hat{\Psi} +
\frac{1}{2}\int dV dV'\hat{\Psi}^{\dagger} (r)\hat{\Psi}^{\dagger} (r') U(\vr,\vr')\hat{\Psi} (r)\hat{\Psi}(r') \,,\,\,\,\,
dV=\alpha \rho d\rho d\varphi\,,
\ee 
where the field operators $\hat{\Psi}(\vr)=\sum_{\lambda}\psi_\lambda(\vr)\hat{a}_{\lambda}$ 
and
$\hat{\Psi}^\dagger(\vr)=\sum_{\lambda}\psi_\lambda^*(\vr) \hat{a}_{\lambda}^\dagger $ 
are  constructed in terms of a single particle wave-function $\psi_\lambda(\vr)$.
Here $\lambda$ represents a set of quantum numbers, 
the annihilation and creation 
operators of the Fock space  are used to describe the quantum state of the system 
made up of a variable or indeterminate number of particles, and  
$[\hat{\Psi}(\vr),\hat{\Psi}^\dagger(\vr')]=\delta(\vr-\vr')$.
 The
 time evolution equation for  $\hat{\Psi}(\vr)$ is given by 
 \be
 i\hbar\frac{\partial}{\partial t}\hat{\Psi}=[\hat{\Psi},\hat{\mathcal{H}}]=
 \left(\hat{H}_\gamma^{(\alpha)}(\vr)+ \int dV' \hat{\Psi}^{\dagger} (r') 
 U (\vr,\vr')\hat{\Psi} (r')\right)\hat{\Psi}(\vr)\,.
  \ee

At very low temperature, the field operator can be approximated by its mean value 
$\psi=\langle \hat{\Psi} \rangle$,
which 
is called 
the order parameter, or 
 the condensate 
 wave-function \cite{Leggett,ALF,Cooper}.
By taking into account the potential (\ref{HFpotential}), 
we  arrive   
at  
a non-linear 
Schr\"odinger equation in a conical space
\be
\label{time-GP}
i\hbar\frac{\partial}{\partial t}\psi=\left(\hat{H}_\gamma^{(\alpha)}+ \frac{4\pi\hbar a_s}{m}|\psi|^2\right)\psi\,.
\ee
To have Bose-Einstein  
condensate at low temperature, 
the ground energy level  should  be equal to 
the chemical potential $\mu$, and function $\psi$ should satisfy $\int dV|\psi|^{2}=N$ (assuming 
that the number of particles in the condensate is $N_c\sim N$). Therefore 
$|\psi|^2=n(\vr)$ is understood as the particle density in the condensate, and  the 
anzats $\psi(\vr,t)=e^{-i\frac{\mu}{\hbar}t}\psi(\vr)$ gives us the stationary version of  (\ref{time-GP}),
\be
\label{es-GP}
\mu\psi(\vr)=\left(\hat{H}_\gamma^{(\alpha)}+ \frac{4\pi\hbar a_s}{m}|\psi|^2\right)\psi(\vr)\,.
\ee
Finally, for the case $a_s\sim 0$, one can use the variational method to 
approximate the ground state, which should be close
to the ground state of the single 
particle system. 

\begin{figure}[H]
\begin{center}
\begin{subfigure}[c]{0.32\linewidth}
\includegraphics[scale=0.4]{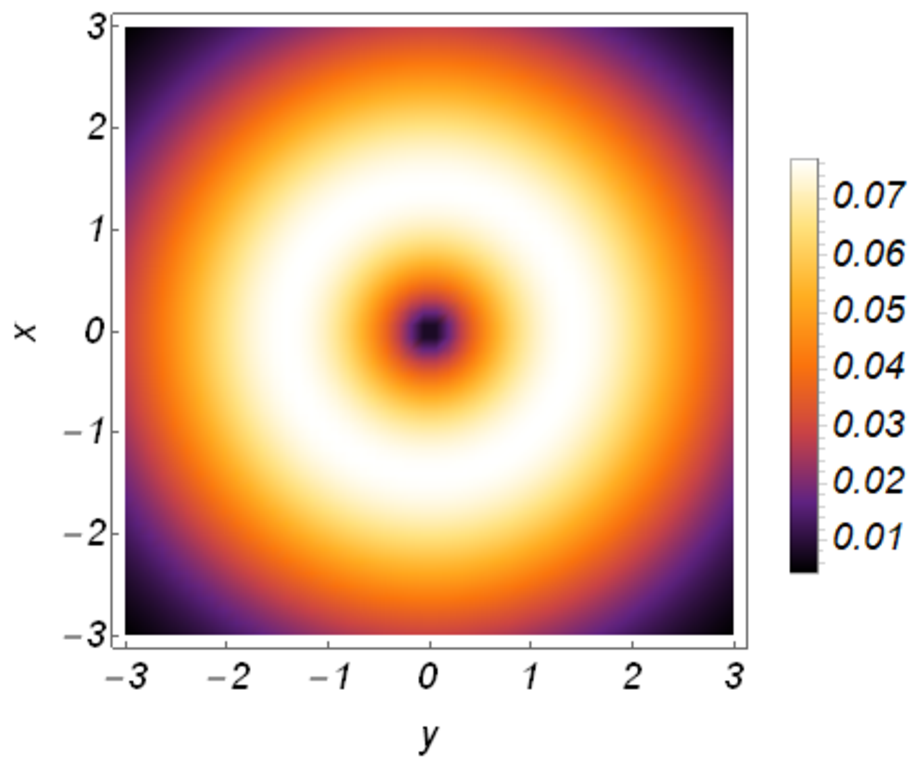}
\caption{\small{$\alpha=1/2$}}
\label{Fig(a)}
\end{subfigure}
\begin{subfigure}[c]{0.32\linewidth}
\includegraphics[scale=0.4]{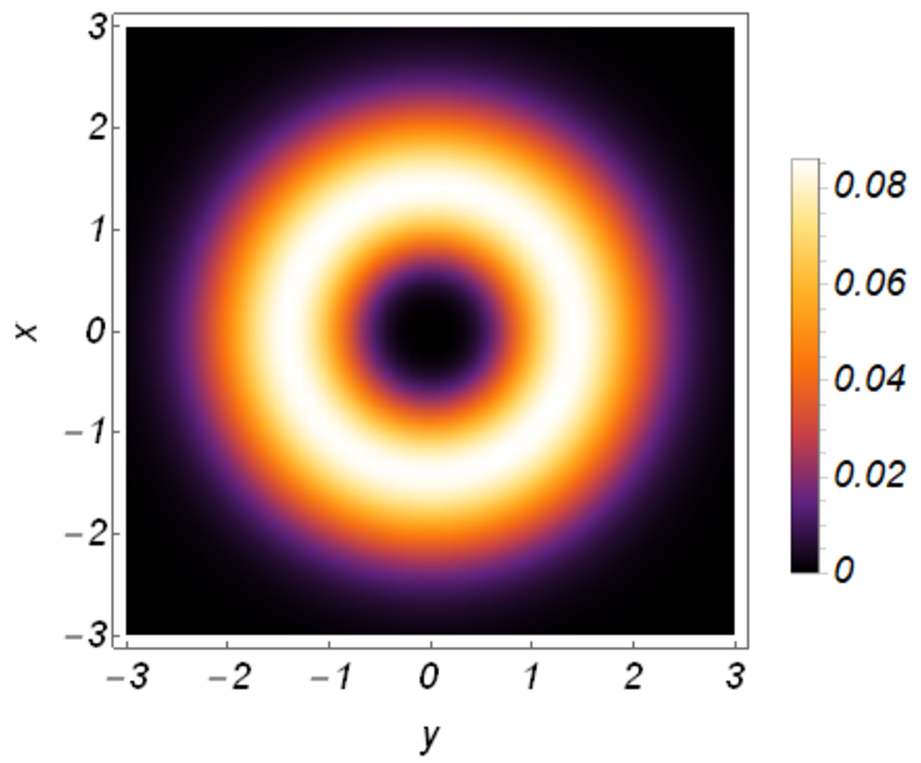}
\caption{\small{$\alpha=1$}}
\end{subfigure}
\begin{subfigure}[c]{0.32\linewidth}
\includegraphics[scale=0.4]{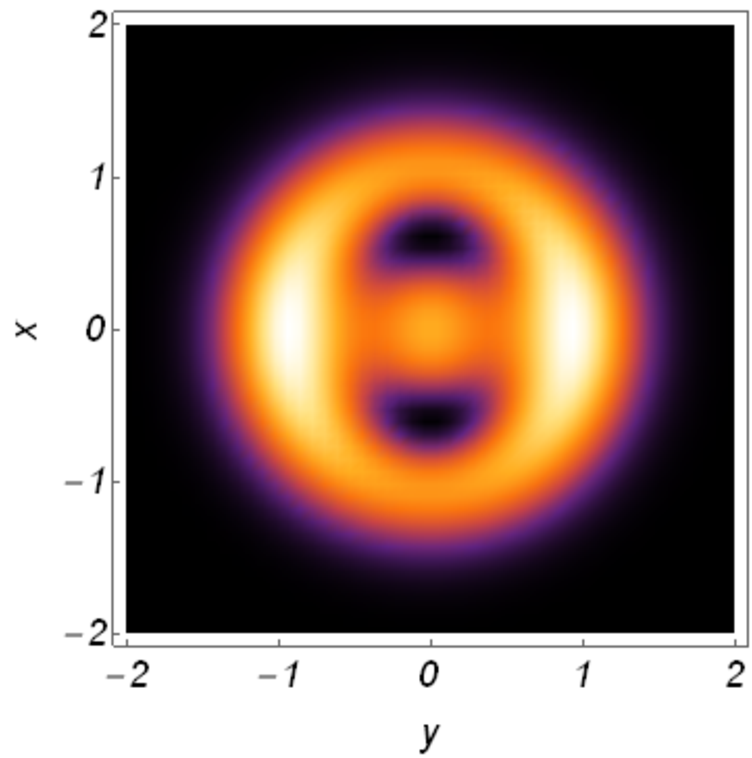}
\caption{\small{$\alpha=2$}}
\end{subfigure}
\vskip0.25cm
\begin{subfigure}[c]{0.32\linewidth}
\includegraphics[scale=0.4]{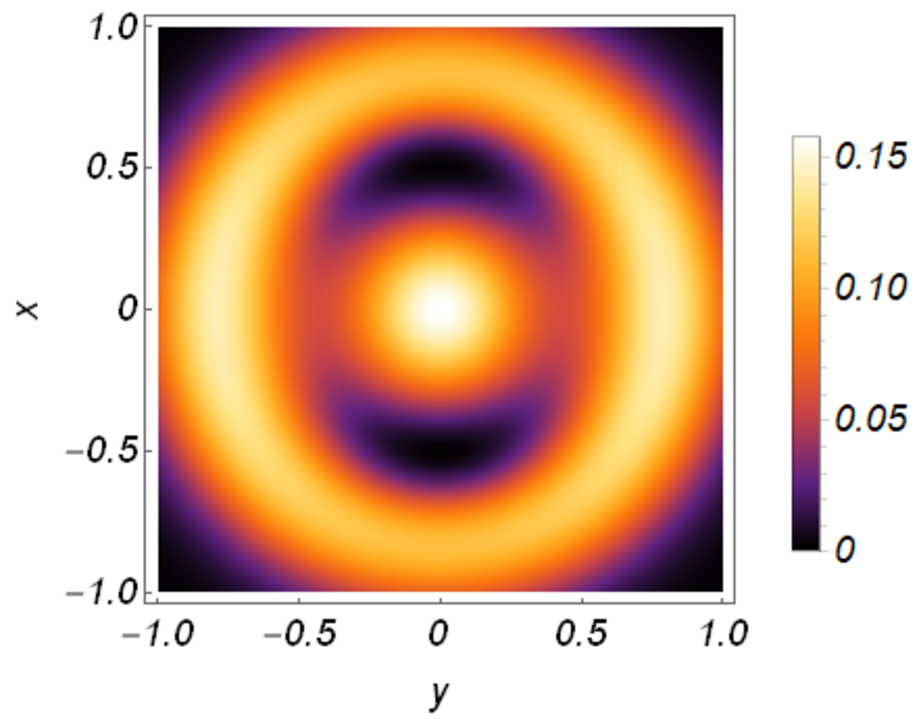}
\caption{\small{ $\alpha=3$}}
\end{subfigure}
\begin{subfigure}[c]{0.32\linewidth}
\includegraphics[scale=0.4]{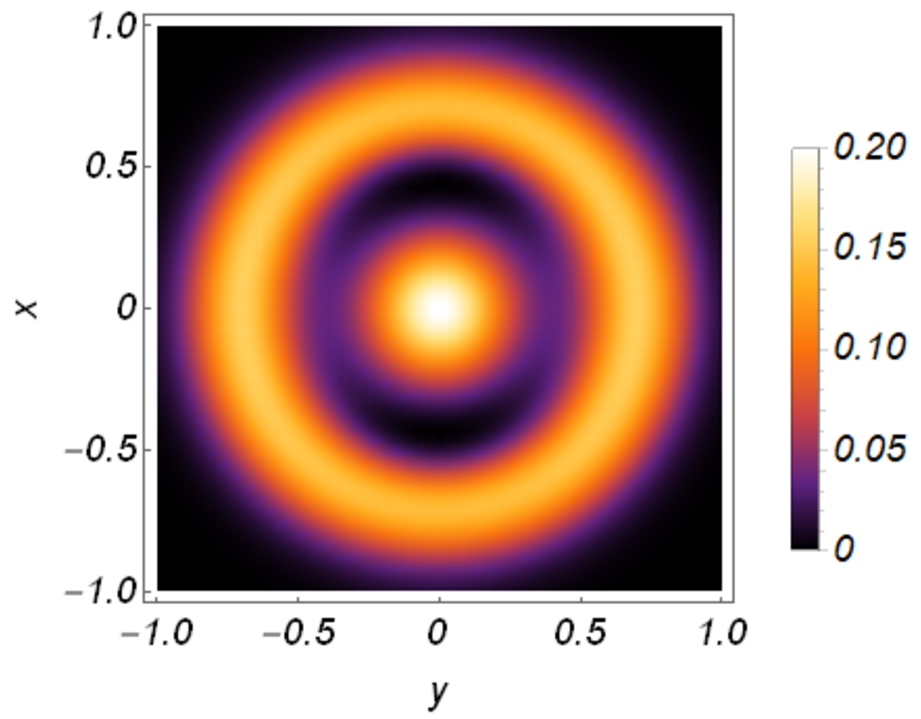}
\caption{\small{$\alpha=4$}}
\label{Fig(e)}
\end{subfigure}
\begin{subfigure}[c]{0.32\linewidth}
\includegraphics[scale=0.3]{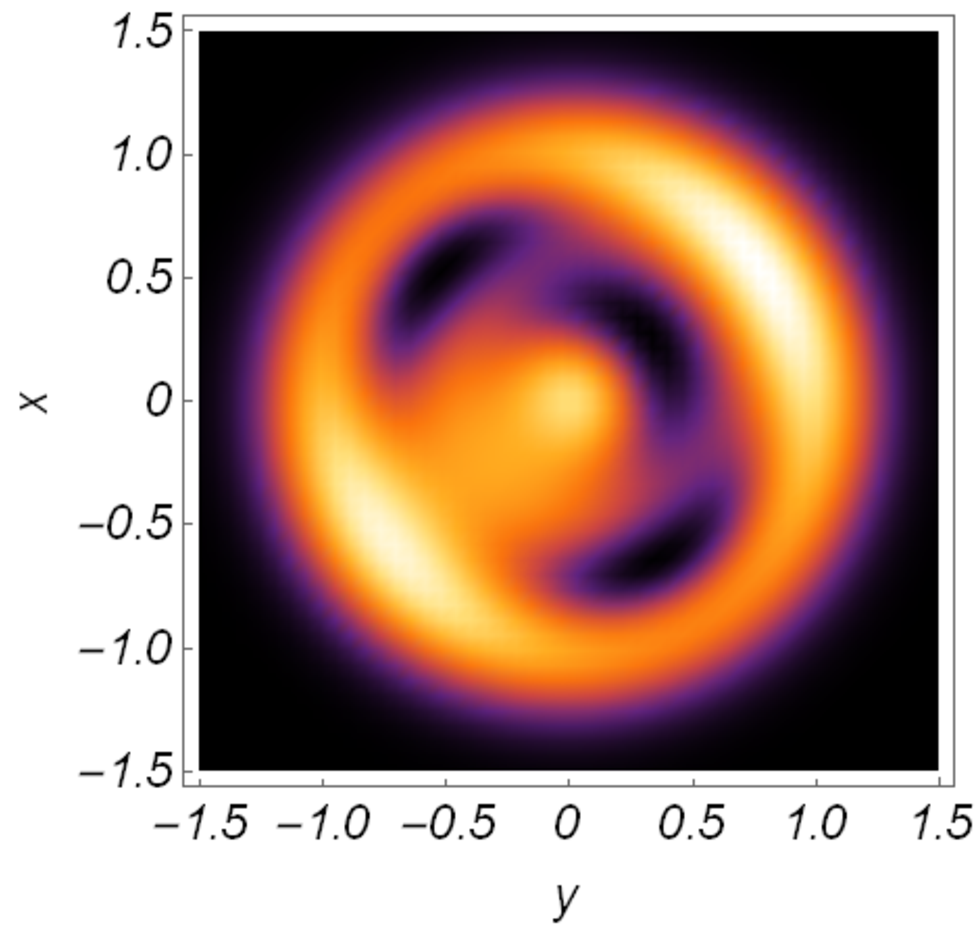}
\caption{\small{$\alpha=3$}}
\label{FigDen(f)}
\end{subfigure}

\end{center}
\caption{\small{Density plots
of the function  $|\psi_{a_0,a_1,a_2}^{(\alpha)}|^2/N$ for different values of $\alpha$ 
are shown in figures from 
 \ref{Fig(a)} to \ref{Fig(e)}, and the  
coefficients $a_j$ are chosen  according to
Table \ref{tab:coef}.
In Fig. \ref{FigDen(f)} we consider the density plot of the function 
$|\psi_{a_0,a_1,a_2,a_3}^{(\alpha)}|^2/N$ with 
$\alpha=3$, 
where the complex values $a_i$ with $i=0,\ldots,3$ are such that 
$I_{a_0,a_1,a_2,a_3}=0.073$.
Bright (dark)  colours  indicate maximum (minimum) values of this function.
Note that as 
 $\alpha$ grows from $0<\alpha\leq 1$ to $\alpha=2,3,\ldots,$
  a local  maximum  appears in the origin of coordinates and 
the two appearing  vortices  seem to degenerate in one single region without particles. 
}}
\label{Fig4}
\end{figure}

The interesting cases to see here are  
$\gamma=1$ and $\gamma=-1$,  both of them corresponding to the Landau phases of our 
conical ERIHO system, with different orientations of 
the homogeneous magnetic field. 
 Since the ground state is infinitely degenerate, 
in the case $\gamma_\pm=\pm 1$ the solution of  (\ref{es-GP}) 
is approximated by 
\be
\label{LLL}
\psi^{(\alpha,\gamma_\pm)}=\sqrt{N}\sum_{l=0}^{l_{\text{cut}}}c_l
\psi_{0,l}^{(\alpha,\pm)}\,,
\qquad \sum_{l=0}^{l_{\text{cut}}}c_lc_l^{*}=1\,,
\ee
where $l_{\text{cut}}$ is a cut-off  introduced
by hand, 
and the usual criteria to select the complex constants $c_l$ is by minimizing 
the non-linear potential term 
expectation value  $\frac{4\pi\hbar a_s}{m}\int \alpha r drd\varphi|\psi^{(\alpha)}|^4$.  
Note that the structure of this function is a polynomial in the variable $r^{\alpha}e^{i\varphi}$ 
(times a Gaussian term) of degree $l_{\text{cut}}$. 
In the planar case, the corresponding nodes of this function represent quantized vortices in a 
superfluid, 
and here this also can 
happen, but now 
the influence of  geometry of the space is expected to be 
essential.   
This is more clear by following the Onsager and  Feynman ideas \cite{Onsager,Feynman},
and selecting 
the solution of (\ref{time-GP}) in the form 
\be
\label{Feyn}
\psi=\sqrt{n(\vr,t)}e^{i\theta(\vr,t)}\,.
\ee
By introducing this anzats, and ignoring the non-linear term, we get the following two 
equations 
\begin{eqnarray}
&\label{Continity}
\frac{\partial}{\partial t} n(\vr,t)=\nabla_c\cdot(n(\vr,t) \vv)\,,&\\&
\label{Bernulli}
\frac{\partial}{\partial t}\Phi+\frac{1}{2}v^2+\frac{P}{m}=0\,,
&
\end{eqnarray}
where 
\be
\Phi=\frac{\hbar \theta}{m}\,,\qquad
\vv=\nabla_c \Phi-\frac{q}{m}\vA\,,\qquad
P=-\frac{\hbar^2}{2m}\frac{1}{\sqrt{n}} \nabla_c \cdot \left(\nabla_c \sqrt{n}\right)\,,
\ee
and $\nabla_c$ is the curvilinear gradient operator that acting on a scalar 
function $F$ and 
a vector field $\vG$ yields 
\begin{eqnarray}
&
\nabla_c F=\frac{1}{\alpha}\frac{\partial F}{\partial \rho}\ve_r+
\frac{1}{\rho}\frac{\partial F}{\partial \varphi}\ve_\varphi\,,\quad 
\nabla_c\cdot \vG=\frac{1}{\alpha\rho}
\frac{\partial (\rho G_\rho)}{\partial \rho}+\frac{1}{\rho}\frac{\partial G_{\varphi}}{\partial \varphi}\,,
&\\&
\nabla_c \cross \vG=\left[\frac{1}{\alpha\rho}
\frac{\partial (\rho G_\varphi)}{\partial \rho}-\frac{1}{\rho}\frac{\partial G_{\rho}}{\partial \varphi}\right]\ve_z\,,
\qquad 
\nabla_c \cdot (\nabla_c F)=\frac{1}{\alpha^2\rho}
\frac{\partial}{\partial \rho}\left(\rho\frac{\partial F}{\partial \rho}\right)+
\frac{1}{\rho^2}\frac{\partial^2 F}{\partial \varphi^2}\,.
&
\end{eqnarray}
Equation (\ref{Continity}) has the form of the continuity equation for the particle density, and if we interpret 
$\Phi$ as a fluid potential, then (\ref{Bernulli}) looks like  the Bernoulli equation for a charged fluid coupled to a 
constant magnetic field,  but now 
in a conical vessel space 
when $\alpha>1$, or for a rotating condensate in the presence of
radial dislocations  \cite{Volovik} for $\alpha<1$. 
Then, since in general case 
$\nabla_c \cross (\nabla_c F)=0$ for a continuous scalar function $F$, we can only have vortices when 
$\theta$ has singularities. Therefore, if we manage to write down the  function 
(\ref{LLL}) in the form (\ref{Feyn}), it is seen that the points where vortices can appear are related to 
the nodes of (\ref{LLL}), where phase can not be defined.  As we mentioned before, 
the polynomial structure of the wave function depends on $\alpha$, and this can affect the vortices 
location with respect to the origin. 
In order to see this, let us consider the case $l_{\text{cut}}=2$ in (\ref{LLL}), 
which gives
us  
the condensate wave-function (in  units in which $\frac{m\omega\alpha^2}{\hbar}=1$)
\begin{eqnarray}
&\label{2vertexsolution}
\psi_{a_0,a_1,a_2}^{(\alpha)}(\rho,\varphi)=
\sqrt{\frac{\alpha N}{\pi(|a_0|^2+|a_1|^2+|a_2|^2)}}
 \left(a_0+
 \frac{a_1(\alpha\rho)^{\alpha}}{\sqrt{\Gamma(\alpha+1)}} e^{i\varphi}+
\frac{a_2(\alpha\rho)^{2\alpha}}{\sqrt{\Gamma(\alpha+1)}} e^{i2\varphi}\right)e^{-\frac{\alpha^2\rho^2}{2}}\,.
\quad
&
\end{eqnarray}
Here,  
$\sum_{0}^{l_{\text{cut}}}c_lc_l^{*}=1$, where  $c_l=a_l(|a_{0}|^2+|a_{1}|^2+|a_{2}|^2)^{-\frac{1}{2}}$, $l=0,1,2$.
 In  Table \ref{tab:coef},   
 we present
 some numerical results for the coefficients  
 $a_l$, that were 
 obtained to minimize
 the integral
\begin{eqnarray}
&
\label{Ia0a1a2}
I_{a_0,a_1,a_2}=\int_{0}^{\infty}\int_{0}^{2\pi}\alpha 
\rho d\rho d\varphi |\psi_{a_0,a_1,a_2}^{(\alpha)}(\rho,\varphi)|^4\,.
&
\end{eqnarray}
\begin{table}[H]
\begin{center}
\begin{tabular}{| c || c | c |c|c|}
\hline
$\alpha$& $a_0$ & $a_1$ & $a_2$ & $I_{a_0,a_1,a_2}$\\ \hline
 \hline
$1/4$ & $0$ &$0$  &$1$ & $ 1/(4\pi)\approx 0.025$ \\
\hline
$1/3$ & $0$ &$0$ & $1$ & $\Gamma(7/6)/(6\pi^{\frac{3}{2}}\Gamma(5/3))\approx 0.031$\\
\hline
$1/2$ & $0$ &$0$ & $1$ & $1/(8\pi)\approx 0.040$\\\hline
$1$ &  $0$ &$0$ & $1$ & $3/(16\pi) \approx 0.060$\\
\hline
$2$ &  $1$ &$0$ & $\sqrt{112/19}\approx 2.428$ & $33/(131\pi)\approx 0.080$\\
\hline
$3$ & $1$ &$0$ & $\sqrt{992/199} \approx 2.233$ & $115/(397\pi) \approx 0.092$\\
\hline
$4$ & $1$ &$0$ & $\sqrt{32512/6179}\approx 2.294  $ & $12866/(38691\pi) \approx 0.106$\\
\hline
\end{tabular}
\caption{\small{The values of parameters
that minimize $I_{a_0,a_1,a_2}$ for different values
 of 
$\alpha$, where $a_i$ are
assumed to be real numbers
(the imaginary part only contributes with a global sign in this particular case).   
Note that the value of $I_{a_0,a_1,a_2}$ grows with the 
increasing value of $\alpha$.}}
\label{tab:coef}
\end{center}
\end{table}
\vskip-0.75cm
For the cases $ 0 <\alpha \leq 1 $, 
 two of the three parameters are  zero. So, in these cases the solution of the form (\ref{2vertexsolution}) that minimizes 
 $ I_{a_0,a_1,a_2} $ has only one vortex at the origin of coordinates. On the other hand, when $ \alpha> 1 $,
 two parameters are non-zero,   
 and as is seen  
from Figure \ref{Fig4},
two vortices
  appear in the corresponding solutions.

\vskip-0.5cm
If by hand we put $a_2=0$ in function (\ref{2vertexsolution}) and minimize
 the integral (\ref{Ia0a1a2}) for  integer values of
  $\alpha\geq 1$,  we get the 
results presented in Table \ref{tab:coef2}.
\begin{table}[H]
\begin{center}
\begin{tabular}{| c || c | c |c|}
\hline
$\alpha$& $a_0$ & $a_1$  & $I_{a_0,a_1,0}$\\ \hline
\hline
$1$ &  $0$ &$1$ &  $1/(4\pi) \approx 0.080$\\
\hline
$2$ &  $0$ &$1$ & $3/(8\pi) \approx 0.119$\\
\hline
$3$ & $1$ &$2\sqrt{3} \approx 3.464$ &  $6/(13\pi) \approx 0.147$\\
\hline
$4$ & $1$ &$4\sqrt{7/19}\approx 2.428$ & $66/(131\pi) \approx 0.160$\\
\hline
\end{tabular}
\caption{\small{ The values of the parameters
 that minimize    $I_{a_0,a_1,a_2=0}$   in dependence  on  $\alpha$.   
Note that the values of $I_{a_0,a_1,0}$ are    greater 
than  the values obtained
  in the case when $a_2\not=0$.
When $\alpha$ grows, 
 one vortex 
 has to
 appear far from the origin.}}
\label{tab:coef2}
\end{center}
\end{table}
\vskip-1cm
By comparison of Tables \ref{tab:coef} and \ref{tab:coef2} we learn that the 
condensate wave-function gives a better approximation to the ground state of 
(\ref{es-GP}) when more terms in the series are taken 
into account, i.e., by introducing  a higher number of vortices. 
Also it is observable that when $\alpha$ grows, the local maximum that appears at the origin of the coordinate
(the vertex of the cone) corresponds to an accumulation of particles
near that point,   
and this prohibits the formation of vortices there. 
Then, in contrast to the Euclidean case where there are 
no special points and the vortex network corresponds to a 
 triangular lattice
\cite{Cooper,ALF}, 
one can expect that in the conical case any underlying structure should  be built around this singular point.

\section{Discussion and outlook}
\label{SecDisOut}

In conclusion, we discuss several  open problems that seem 
 interesting from the perspective of the obtained results.

\textbf{1.} We 
began 
our consideration with
the $(3+1)$ 
dimensional Minkowski 
space subjected to a uniform rotation with angular velocity $\Omega$  
about a fixed axis 
and  
introduced 
the  conical singularity by imposing a geometrical constraint. 
 One can note that in the case 
 of $\Omega\not=0$, the spacetime metric
 (\ref{ConeGeo1}) 
 can be 
 related to  the metric of the  spinning cosmic string background  \cite{Volovik,DJtH,Mazur,Spinning2},
 \begin{eqnarray}
 &
 \label{spinning}
 ds_{\text{sp}}^{2}=-c^2(dt+ 4c^{-4} G J d\varphi)^2 +\alpha^2d\rho^2+\rho^2 d\varphi^2\,,
& 
 \end{eqnarray}
  where  $J$ is the linear angular momentum density of the string.
  This is 
  achieved by applying 
  the coordinate transformation 
$d\varphi \rightarrow \frac{1}{2}(d\varphi-2\alpha\Omega dt)$, 
$dt \rightarrow dt+\frac{1}{2\alpha\Omega}d\varphi$ 
 in (\ref{ConeGeo1}),
followed by the identification $J=\frac{c^{4}}{ 8 \alpha \Omega G}
$. 
The peculiarity of this
transformation is that 
it requires  the compactification of the time variable since 
$\varphi=0\cong \varphi=2\pi$ 
implies that the new time coordinate is periodic, $t\cong t+\pi/(\alpha\Omega)$ \cite{Mazur}.
The non-relativistic limit can also be applied to the geodesic motion 
in background of the spinning cosmic string. For this, we present the action 
corresponding to the metric 
(\ref{spinning}) in the  form
\begin{eqnarray}
&S_{\text{sp}}=-mc\int\sqrt{- ds_{\text{sp}}^{2}}
=-mc^2\int \sqrt{
1-\frac{1}{c^2}\left[
\alpha^2\dot{ \rho}^2+(\rho^2-\frac{c^2}{4\alpha^2\Omega^2})\dot{\varphi}^2-
\frac{c^2}{\alpha\Omega}\dot{\varphi}
\right]}\,dt\,,&
\end{eqnarray}
and take the limit $c\rightarrow \infty$, $\Omega\rightarrow\infty$ in such a way 
 that $c^2/\alpha\Omega=\theta=const$.
 As a result, we obtain Lagrangian of a non-relativistic anyon in the cone background,
 \begin{eqnarray}\label{Lany}
 &L_{\text{any}}=
 \frac{m}{2}\left(\alpha^2\dot{\rho}^2+\rho^2\dot{\varphi}^2-\theta \dot{\varphi}\right)\,,&
\end{eqnarray}
which also includes the planar case at $\alpha=1$. The non-relativistic limit we employed is analogous  
to the Jackiw-Nair procedure  considered earlier
in the context of relation of the physics of anyons to exotic extension of the planar Galilei group 
and non-commutative geometry  \cite{JackiwNair,HPany1,HPany2}.
It would be interesting to study non-relativistic system (\ref{Lany}) 
 from the perspective of hidden  symmetries and quantum
 anomaly.
 
 \textbf{2.} As we saw, 
 the trajectories 
 of  the two-dimensional non-relativistic model in a 
  conical spacetime subjected to a homogeneous  rotation 
  can be obtained  
  from the corresponding orbits in the  Euclidean plane
  in the inertial frame system ($\Omega=0$). This can be achieved 
   by applying  
  the time-dependent canonical transformation
  supplied with  an additional $\alpha$-dependent 
   local canonical transformation to produce 
  the cone geometry. 
  In this way we obtained 
the trajectories that lie on the corresponding  finite- or infinite- sheeted
Riemann surfaces in dependence on whether the cone parameter $\alpha$ 
takes a rational or irrational value. 
This procedure also helped us  to obtain the formal integrals,  
which are the building blocks for  the construction of the  
well defined in phase space symmetry generators
in the case of rational values of $\alpha$.
As the problem of geometric quantum anomaly 
arises when we employ the canonical quantization procedure,
one may wonder if the  Riemann surface structure should play 
a special  role at the quantum level. 
 It seems to be worthing to analyze the problem of the quantum 
 anomaly in more detail by using the method of geometric quantization with 
 taking into account the finite-sheeted Riemann surfaces
 structure.

 \textbf{3.} 
 In the case of geodesic dynamics in non-rotating ($\Omega=0$) cosmic string background with 
 an arbitrary integer value $\alpha=q$ of the geometric parameter 
 we were able to construct 
 the well defined higher order integrals (\ref{Pi+-})  
 that generalize the momenta operators of a free particle in Euclidean geometry,
 and which reflect correctly the degeneracy  of quantum energy levels
 of the system. 
 By means of the $\Omega$-dependent unitary transformation supplemented by 
 a `Heaviside step function regularization' procedure,
 we obtained the analogs (\ref{SymmetryOp}) of  these integrals,  which are non-local 
 operators, and do the same job in the quantum systems  with $\alpha=q$  and 
 arbitrary values of the rotation parameter $\Omega\neq 0$. 
 The conformal bridge transformation allowed then us
 to construct  on their basis the analogous  integrals 
 for the quantum conical ERIHO systems with 
 integer $\alpha=q$.  
   It is interesting to note here that in the case of the free particle
   the presence of the frequency parameter $\Omega$ 
   admits the construction of 
   a quantum combination $\hbar\Omega$ of the energy dimension, 
   with which it is possible to regularize by means of the Heaviside step function 
   (whose argument is  dimensionless). 
   In the classical theory, the construction of a combination of parameters
    of the same dimension would require the introduction by hand 
    of an additional constant of the dimension of length.
 
  It would be natural to try to modify somehow the  described approach  
  to ``cure" the quantum anomaly  that still persists 
   in the case of rational values of  $\alpha=q/k$ with $k>1$.
 For this, let us consider first the case of $\Omega=0$, 
 and introduce the alternative integrals
\begin{eqnarray}\label{Ladd1}
&
\mathfrak{L}_+^{(\frac{q}{k})}=\left(\hat{\Pi}_+^{(\frac{q}{k})}\right)^{q}\Theta\left(\frac{\hat{p}_\varphi}{\hbar}-k\right)\,,\qquad 
\mathfrak{L}_-^{(\frac{q}{k})}=\Theta\left(\frac{\hat{p}_\varphi}{\hbar}-k\right)\left(\hat{\Pi}_-^{(\frac{q}{k})}\right)^{q}\,.
&\\&
\mathfrak{J}_+^{(\frac{q}{k})}=\Theta\left(k -\frac{\hat{p}_\varphi}{\hbar}\right)\left(\hat{\Pi}_+^{(\frac{q}{k})}\right)^{q}\,,\qquad 
\mathfrak{J}_-^{(\frac{q}{k})}=\left(\hat{\Pi}_-^{(\frac{q}{k})}\right)^{q}\Theta\left(k-\frac{\hat{p}_\varphi}{\hbar}\right)\,.
\label{Ladd2}
&
\end{eqnarray}
Due to inclusion of  the Heaviside step  function, these operators do not produce 
non-physical eigenstates under their application to physical eigenstates.
Their explicit action is given by the following relations, 
\begin{eqnarray}
&
\label{RegIn1}
\left\{
\begin{array}{ll}
\mathfrak{L}_+^{(\frac{q}{k})}\psi_{\kappa,j}^{(\frac{q}{k})}=0\,,  & j\leq k-1\\
\\
\mathfrak{L}_+^{(\frac{q}{k})}\psi_{\kappa,j}^{(\frac{q}{k})}\sim \psi_{\kappa,j+k}^{(\frac{q}{k})}\,, & j>k-1
\end{array}
\right.\,,
\quad 
\left\{
\begin{array}{ll}
\mathfrak{L}_-^{(\frac{q}{k})}\psi_{\kappa,j}^{(\frac{q}{k})}=0\,,  & j\leq 2k-1\\
\\
\mathfrak{L}_-^{(\frac{q}{k})}\psi_{\kappa,j}^{(\frac{q}{k})}\sim \psi_{\kappa,j-k}^{(\frac{q}{k})} \,,& j>2k-1
\end{array}
\right.\,,&
\end{eqnarray}
\begin{eqnarray}
&
\label{RegIn1}
\left\{
\begin{array}{ll}
\mathfrak{J}_+^{(\frac{q}{k})}\psi_{\kappa,j}^{(\frac{q}{k})}=0\,,  & j\leq k-1\\
\\
\mathfrak{J}_+^{(\frac{q}{k})}\psi_{\kappa,j}^{(\frac{q}{k})}\sim \psi_{\kappa,j+k}^{(\frac{q}{k})}\,, & j<k-1
\end{array}
\right.\,,
\quad 
\left\{
\begin{array}{ll}
\mathfrak{J}_-^{(\frac{q}{k})}\psi_{\kappa,j}^{(\frac{q}{k})}=0\,,  & j\geq 2k-1\\
\\
\mathfrak{J}_-^{(\frac{q}{k})}\psi_{\kappa,j}^{(\frac{q}{k})}\sim \psi_{\kappa,j-k}^{(\frac{q}{k})}\,, & j\geq k-1
\end{array}
\right.\,,\quad j\in\Z\,.\qquad &
\end{eqnarray}
To generalize for the case with $\Omega\not=0$, we can follow the procedure described in 
Sec. \ref{SecQuantumFreeRot}.  With these two pairs of operators, we have access 
to all the states with a given energy eigenvalue, except the 
states of the form $\psi_{\kappa, \ell}^{(\frac{q}{k})}$ with $|\ell|<k-1$.   
Thus, the question on identifying the 
symmetry operators that would correctly explain the degeneracy  
of the energy spectrum  when $\alpha$ is rational remains to be open.
Perhaps,  the  operators  (\ref{Ladd1})
and (\ref{Ladd2}) still need to be  supplemented with some 
additional integrals, similar to what  happens under  the construction of the complete
set of the spectrum generating operators in rationally extended 
harmonic oscillator \cite{CariPlyu}
and  rational deformations of conformal mechanics \cite{CariInzPlyu}.

 \textbf{4.}  For the harmonic oscillator of frequency $\omega$
  in a uniformly rotating  cosmic string background,
 the numerical factor 
$ \gamma = - \Omega / \omega $ can be introduced. 
This allows us to identify the system as the conical version of the 
ERIHO model \cite{ERIHO,PTrev}, in which  
closed trajectories and the  associated 
integrals of the hidden symmetry appear
when  both parameters $ \alpha $ and $\gamma$ take rational values.
The model reveals a kind of duality under the inversion transformation
$\gamma\rightarrow 1/\gamma$ in the same way as this happens in  Euclidean version 
 \cite{ERIHO}. Namely,  
in the case   
$|\gamma|<1$, the system is characterized by a spectrum bounded from bellow,   
and
 its true (not depending explicitly on time) symmetry generators 
  satisfy a  non-linear version of the $\mathfrak{u}(2)$ algebra. Besides,  the system 
also is characterized  by a dynamical non-linear 
  $\mathfrak{gl}(2,\R)$ algebra.
Additionally, it also is characterized by 
 the conformal $\mathfrak{sl}(2,\R)$ symmetry.
 After applying the duality transformation,  we get a system with $|\gamma'|=|\gamma^{-1}|>1$, 
whose
 quantum spectrum is not bounded
 from below, 
 and the dynamical (explicitly depending on time) and true symmetries are interchanged,
while the conformal $\mathfrak{sl}(2,\R)$ symmetry is not changed.
  In the intermediate case $\gamma^2=1$, 
  we use the gravitoelectromagnetic interpretation to identify the model 
  with the Landau problem (in symmetric gauge)  in a conical background. 
  The sign of  $\gamma$ determines there orientation 
  of the  magnetic field. Such a system corresponding to our 
  case  $|\gamma|=1$ was considered earlier  in  \cite{DeAMarques,Muniz},
   where it
  was shown that some energy levels have  infinite degeneracy
  for arbitrary values of $\alpha$.
 However, the system was not investigated there  in the light of classical and quantum symmetries for
 special
 values of the parameter $\alpha$, and this is just a particular case in our analysis. 
 On the other hand, note 
that   the same classical trajectories as we obtained  for the 
 case $(0<\alpha<1,\Omega=0)$ were recently  observed for geodesics in 
the conical BTZ spacetimes \cite{BTZ1}, 
and it would be interesting to see if the trajectories 
in the case $\Omega\not=0$ can be related to the dynamics in 
such a 
 kind of spacetimes as well.

\textbf{5.} Finally, we studied the multi-particle conical ERIHO system in its Landau phases using 
the mean-field 
theory. Effectively, we have  
replaced
the pair-wise interaction  (that could  be a Coulomb one) by a contact potential. 
The numerical calculation in the 
simplest examples shows that in the case $ 0 <\alpha \leq1$ \footnote{ 
The cases with $0<\alpha<1$ can be associated, in particular,  with  a geometric background 
describing radial dislocations in superfluids \cite{Volovik}.} the best approximation 
of the wave function of the condensate is the one-vortex solution. In the case of $ \alpha> 1 $, 
the best approximation is obtained by considering 
multi-vortex configurations,
and  the accumulation of the particles is possible (while 
the formation of 
vortices
is prohibited) at the vertex of the cone, where the probability density  has a 
local maximum.
Naturally, one may ask if the approximation that we considered is good enough 
from the point of view of a 
 possible experimental realization,  
for example by means of the  Bose-Einstein condensate 
in  a rotating harmonic trap, or in the system of 
charged particles  subjected to a magnetic field.
 Anyway, our numerical results indicate 
 that the cone's vertex should necessarily 
 produce some effect. 
It would also be interesting to compare these results with a field theory approach, 
such as the collective field theories, where the large $N$ limit is applied \cite{ColFT1}. 
This problem also is  interesting since it can bring some insight 
into a generalization of the conformal bridge transformation for 
an infinite number of degrees of freedom.

\vskip0.3cm

\noindent {\large{\bf Acknowledgements} } 
\vskip0.1cm

The work was partially supported by the FONDECYT Project 1190842 
and the DICYT Project 042131P\_POSTDOC.

\appendix
\section{Time-dependent canonical transformation}
\label{AppCantime}
The system in a rotating flat geometry introduced 
at
the beginning of  Section \ref{SecFreeRotatingD} can be obtained 
from the non-relativistic 
two-dimensional free particle in the inertial reference frame ($\Omega=0$),
\begin{eqnarray}
&\label{Freaction}
S=\int Ldt\,, \qquad L=\frac{m}{2} \frac{d\chi_i}{dt} \frac{d\chi_i}{dt}\,,\quad \rightarrow \quad H=\frac{1}{2m}\pi_i\pi_i\,,\qquad
\pi_i=m \frac{d\chi_i}{dt}\,,& 
\end{eqnarray}
by
applying 
 the time dependent canonical transformation
$\chi_i\rightarrow x_i$, $\pi_i\rightarrow p_i$,
which can be presented in a 
form
\begin{eqnarray}\label{canxchi}
x_\pm= e^{\mp i\Omega t }\chi_\pm=\rho  e^{\pm i(\Omega t+\varphi)}\,,\qquad 
p_\pm =e^{\mp i \Omega t}\pi_\pm\,.
\end{eqnarray} 
Here $\chi_\pm=\chi_1\pm i\chi_2$, $\pi_\pm=\pi_1\pm i\pi_2$, etc.,  
 and $\{\chi_\pm,\pi_\pm\}=0$, $\{\chi_\pm,\pi_\mp\}=2$. 
Under the time-dependent  
 rotation transformation $\chi_i\rightarrow x_i$, 
 Lagrangian 
in (\ref{Freaction}) transforms into Lagrangian (\ref{flatLagrangian}), 
while the canonical transformation (\ref{canxchi}) applied to 
the action one-form 
$\theta=\pi_i d\chi^{i}-  H dt$ of  the system (\ref{Freaction}) 
yields $\theta=p_i  dx^{i}-  H_\Omega  dt$, 
where $H_\Omega$ is the Hamiltonian
 (\ref{FreeClasicH}).
 At the same time,
the angular momentum is invariant 
under transformation  (\ref{canxchi}), 
$\pi_\varphi=\epsilon_{ij}\chi_i\pi_j=\epsilon_{ij}x_ip_j=p_\varphi$. 

For the system (\ref{Freaction}),
canonical momenta are the  integrals of motion,
$\frac{d}{d\tau}\pi_\pm=\{H,\pi_\pm\}=0$.
On the other hand,    
according to
(\ref{canxchi}), the explicitly depending on time
phase space functions  
\be\label{P+-}
P_\pm:=e^{\pm  i\Omega t}p_\pm=\pi_\pm
\ee
are the dynamical integrals of motion for the system 
$H_\Omega$, $\frac{d}{dt}P_\pm=\frac{\partial}{\partial t}P_\pm +
\{P_\pm,H_\Omega\}=0$.
Analogously, the explicitly depending on time
phase space functions  
\begin{eqnarray}\label{X+-}
&X_\pm=e^{\pm i\Omega t}\left(\chi_\pm -\frac{1}{m}t\pi_\pm\right)&
\end{eqnarray}
are the dynamical integrals of motion
for the system $H_\Omega$, where
 $(\chi_\pm -\frac{1}{m}t\pi_\pm)$ are
  the 
 Galileo boosts generators
 of the system
(\ref{Freaction}).
Quadratic rotationally invariant polynomials of (\ref{P+-}) and  (\ref{X+-}) 
provide us with the true integrals, $p_\varphi=\frac{i}{2}(X_+P_--X_-P_+)$ and 
$H_\Omega=H_0-\Omega p_\varphi$, where 
$H_0=\frac{1}{2m}P_+P_-$,
and the dynamical integrals,  
\begin{eqnarray}
&D=\frac{1}{4 }(X_+P_-+X_-P_+)=\frac{1}{2}\rho p_\rho-H_0 t\,,&\\
&K=\frac{m}{2} X_+X_-=\frac{m}{2} \rho^2-2Dt-H_0 t^2\,,&
\end{eqnarray}
which correspond to the $\alpha=1$ case of 
(\ref{Halpha0}), (\ref{Halpha1}).
Multiplying dynamical integrals (\ref {P+-}) with 
dynamical integrals $\exp(\pm i\Omega D/H_0)$, we obtain
\begin{eqnarray}
&I_\pm^{(1)}= \exp\big (\pm i \Omega \frac{\rho p_\rho}{2H_0}\big)\cdot p_\pm\,.&
\end{eqnarray}
These are the well defined for $H_0 \neq 0$ true integrals of the system $H_\Omega$, 
to which the integrals (\ref{I1})  reduce 
at $\alpha=1$.

 To establish the relation between the quantum 
 Hamiltonians, 
 consider the time-dependent Schr\"odinger equation for the 
 quantum analog of the system 
(\ref{Freaction}),
 \begin{eqnarray}\label{Schrtchi}
 &
i\hbar \frac{\partial}{\partial t}
 \psi(\chi_1,\chi_2,t)= -\frac{\hbar^2}{2m}\left(\frac{\partial^2}{\partial \chi_1^2}+
 \frac{\partial^2}{\partial \chi_1^2}\right)\psi(\chi_1,\chi_2,t)\,.
&
 \end{eqnarray}
 Denoting $\Psi(x_1,x_2,t)= \Psi(x_1(t),x_2(t),t)=\psi(\chi_1,\chi_2,t)$,
 where $x_i(t)$ are given by the  first relation in (\ref{canxchi}), 
 we have $i\frac{\partial}{\partial t}\psi(\chi_1,\chi_2,t)=\left(i\hbar\frac{\partial}{\partial t}+\hbar 
 \Omega\hat{p}_\varphi \right)\Psi(x_1,x_2,t),
 $
 and
 equation (\ref{Schrtchi}) takes the form
 \begin{eqnarray}
&
i\hbar\frac{\partial}{\partial t}\Psi(x_1,x_2,t)=\hat{H}_\Omega \Psi(x_1,x_2,t)\,,
&\\&
\hat{H}_\Omega=-\hbar^2\left(\frac{\partial^2}{\partial x_1^2}+\frac{\partial^2}{\partial x_2^2}\right)-\Omega\hat{p}_{\varphi}\,,
\qquad 
\hat{p}_\varphi=-i\hbar\left(x_1\frac{\partial}{\partial x_2}-x_2\frac{\partial}{\partial x_2}\right)\,.& 
\end{eqnarray}

Applying the canonical transformation  (\ref{canonical}), 
the described procedure can   be
generalized  
for the system in conical geometry.

\end{document}